\RequirePackage{snapshot}
\documentclass[
    reprint,
    superscriptaddress,
    amsmath,amssymb,
    aps,
    pra,
]{revtex4-2}
\newif\iffigdraft
\figdraftfalse
\newif\ifgentoc
\newif\ifeditmode
\editmodefalse
\gentocfalse
\newif\ifshowparagraphinline
\showparagraphinlinefalse
\ifeditmode\else
\showparagraphinlinefalse
\fi
\usepackage{layouts}
\usepackage[procnames]{listings}
\usepackage{chngcntr}
\usepackage{amsfonts}
\usepackage{blindtext}
\let\oldblindtext\blindtext
\renewcommand{\blindtext}{\textcolor{gray}{\oldblindtext}}
\usepackage{cmap}
\usepackage{datetime}
\usepackage{environ}
\usepackage{etoolbox}
\usepackage[utf8]{inputenc}
\DeclareUnicodeCharacter{03BD}{\ensuremath{\nu}}
\DeclareUnicodeCharacter{03BC}{\textmu}
\DeclareUnicodeCharacter{00B0}{\ensuremath{^\circ}}
\DeclareUnicodeCharacter{00B1}{\ensuremath{\pm}}
\DeclareUnicodeCharacter{2192}{\ensuremath{\rightarrow}}
\iffigdraft
    \usepackage[draft]{graphicx}
\else
    \usepackage{graphicx}
\fi
\usepackage{hyphenat}
\usepackage{ifxetex}
\usepackage{mathrsfs}
\usepackage{calc}
\newcommand{\nts}[1]{{\color{red}#1}}
\newcommand{\ie}{\textit{i.e.}\xspace}
\newcommand{\vs}{\textit{vs.}\xspace}
\newcommand{\eg}{\textit{e.g.}\xspace}
\newcommand{\etc}{\textit{etc.}\xspace}
\newcommand{\via}{\textit{via}\xspace}

\def\us{\texorpdfstring{\ensuremath{\;\mbox{\textmu s}}}{us}\xspace}
\newcommand\pdfcommentG[1]{%
\pdfcomment[author=general,color=yellow]{#1}
}

\newcommand\pdfcommentJF[1]{%
\pdfcomment[author=John Franck,color=blue]{#1}
}
\newcommand\pdfcommentLater[1]{%
\pdfcomment[author=John Franck,color=gray]{#1}
}
\newcommand\pdfcommentAB[1]{%
\pdfcomment[author=Alec Beaton,color=red]{#1}
}

\ifeditmode
\usepackage{pdfcomment}
\newcommand{\sepcomment}{\\ \vspace*{3ex}}
\else
\newcommand{\pdfcomment}[2][]{}
\newcommand{\sepcomment}{}
\fi
\usepackage{hyperref}
\hypersetup{colorlinks=true, pdfstartview=FitV, linkcolor=linkcolor, citecolor=dbluecolor, urlcolor=dgreencolor, bookmarksdepth=subparagraph}
\usepackage{bookmark}
\usepackage{sidecap}
\usepackage{soul}
\usepackage{textcomp}
\usepackage{url}
\usepackage{xcolor}
\usepackage{xspace}
\usepackage{multirow}
\usepackage{array}
\newcolumntype{R}{>{\vspace{2ex}\displaystyle}{r}}
\newcolumntype{L}{>{\displaystyle}{l}}
\definecolor{SUgrey}{HTML}{6F777D}
\definecolor{SUorange}{HTML}{D44500}
\definecolor{dbluecolor}{rgb}{.01,.02,0.29}
\definecolor{dgraycolor}{rgb}{0.50,0.50,0.50}
\definecolor{dgreencolor}{rgb}{0.0,0.4,0}
\definecolor{linkcolor}{cmyk}{0,1,1,1}
\usepackage{accsupp}    
\newcommand{\noncopynumber}[1]{%
    \BeginAccSupp{method=escape,ActualText={}}%
    #1%
    \EndAccSupp{}%
}

\usepackage[compact]{titlesec}
\ifshowparagraphinline
\titleformat{\paragraph}[runin]{\color{gray}\normalfont\bfseries\footnotesize}{}{3pt}{\hspace{0.75em}\ul{\footnotesize\thesubsection\theparagraph)\;}}[:]
\else
\renewcommand{\paragraph}[1]{\par\phantomsection\addcontentsline{toc}{paragraph}{#1}}
\fi
\makeatletter
\@ifundefined{linelabel}{%
\newcommand{\linelabel}[1]{}}{}
\makeatother
\usepackage[acronyms]{glossaries}
\makeglossaries
\renewcommand{\glossarysection}[2][]{}
\newacronym{oec}{OEC}{oxygen-evolving complex}
\newacronym{psii}{PSII}{Photosystem II}
\newacronym{odnp}{ODNP}{Overhauser Dynamic Nuclear Polarization}
\newacronym{nmr}{NMR}{Nuclear Magnetic Resonance}
\newacronym{esr}{ESR}{Electron Spin Resonance}
\newacronym{deer}{DEER}{Double Electron-Electron Resonance}
\newacronym{pds}{PDS}{Pulse Dipolar Spectroscopy}
\newacronym{eseem}{ESEEM}{Electron Spin Echo Envelope Modulation}
\newacronym{sdsl}{SDSL}{Site-Directed Spin-Labeling}
\newacronym{pre}{PRE}{Paramagnetic Relaxation Enhancement}
\newacronym{awg}{AWG}{Arbitrary Waveform Generation}
\newacronym{br}{BR}{bacteriorhodopsin}
\newacronym{pr}{PR}{proteorhodopsin}
\newacronym{md}{MD}{molecular dynamics}
\newacronym{mtsl}{MTSL}{\textit{S}-(1-oxyl-2,2,5,5-tetramethyl-2,5-dihydro-1H-pyrrol-3-yl)methyl methanesulfonothioate)}
\newacronym{ednmr}{EDNMR}{electrically detected \gls{nmr}}
\newacronym{mmm}{MMM}{the open-source Matlab tool known as Multiscale Modeling of Macromolecules}
\newacronym{ddm}{DDM}{n-Dodecyl-β-D-Maltopyranoside}
\newacronym{dppc}{DPPC}{dipalmitoylphosphatidylcholine}
\newacronym{dopa}{DOPA}{dioleoylphosphatidic acid}
\newacronym{sdr}{SDR}{software-defined radio}
\newacronym{avhe}{AVHE}{Academic Virtual Hosting Environment}
\newacronym{pcet}{PCET}{proton-coupled electron transfer}
\newacronym{rnr}{RNR}{Ribonuclease reductase}
\newacronym{iq}{IQ}{quadrature}
\newacronym{sasa}{SASA}{solvent-accessible surface area}
\newacronym{mr}{MR}{Magetic Resonance}
\newacronym{dcct}{DCCT}{domain colored coherence transfer}
\newacronym{ct}{CT}{coherence transfer}
\newacronym{lp}{LP}{Linear Prediction}
\newacronym{mw}{μw}{microwave} 
\newacronym{fid}{FID}{free induction decay}
\newacronym{rm}{RM}{reverse micelle}
\newacronym{snr}{SNR}{signal to noise ratio}
\newacronym{fwhm}{FWHM}{full-width-half-maximum}
\newacronym{l2g}{L2G}{Lorentzian-to-Gaussian}

\renewcommand{\thesection}{\Roman{section}}
\renewcommand{\thesubsection}{\thesection.\arabic{subsection}}

\makeatletter
\renewcommand{\p@subsection}{}
\renewcommand{\p@subsubsection}{}
\makeatother
\newcounter{subfigure}[figure]
\newcounter{subfigurenonumber}
\newcounter{tempfigure}
\renewcommand\thesubfigurenonumber{(\alph{subfigurenonumber})}
\newcommand{\subfig}[2]{%
    \setcounter{tempfigure}{\value{figure}}%
    \addtocounter{tempfigure}{1}%
    \refstepcounter{subfigure}%
    \setcounter{subfigurenonumber}{\value{subfigure}}%
    \expandafter\edef\csname ref#2\endcsname{\thesubfigurenonumber}
    \label{#1}%
    }
\newif\ifpoormancref
\poormancreffalse
\ifpoormancref
\usepackage[poorman]{cleveref}
\else
\usepackage{cleveref}
\fi
\crefname{equation}{Eq.}{Eqs.}
\crefname{table}{Table}{Tables}
\crefname{figure}{Fig.}{Figs.}
\crefname{section}{Sec.}{Sec.}
\crefname{subfigure}{Fig.}{Figs.}
\crefname{lstlisting}{Appendix, example code \#}{Appendix, example codes\#}
\Crefname{lstlisting}{Listing}{Listings}
\usepackage{xr}
\begin{document}
\counterwithin{lstlisting}{section}
\newlength\myfigwidth
\setlength{\myfigwidth}{3.5in}
\newcommand\SUaffil{\affiliation{Department of Chemistry, Syracuse University, Syracuse, NY 13210, USA}}
\author{Alec A. Beaton}
\author{Alexandria Guinness}
\author{John M. Franck}
\SUaffil
\email{jmfranck@syr.edu}
\title{%
A Modernized View of Coherence Pathways\\
Applied to
Magnetic Resonance Experiments in Unstable,
Inhomogeneous Fields
}
\date{\today}

\begin{abstract}
Over recent decades, the value of conducting
    experiments at lower frequencies and in
    inhomogeneous and/or time-variable fields has
    grown.
For example, an interest in the nanoscale heterogeneities
    of hydration dynamics demands increasingly
    sophisticated and automated measurements
    deploying \gls{odnp} at low field.
The development of these methods poses
    various challenges that drove us to develop
    a standardized alternative to
    the traditional schema for acquiring and analyzing
    coherence pathway information
    employed by the overwhelming majority of
    contemporary \gls{nmr} research.
Specifically,
    on well-tested, stable \gls{nmr} systems running well-tested
    pulse sequences in highly optimized, homogeneous
    magnetic fields,
    traditional hardware and software quickly
    isolate a meaningful subset of data by
    averaging and discarding between 3/4 and 127/128
    of the digitized data.
In contrast, spurred by recent advances in the capabilities of
    open-source libraries, the \gls{dcct} schema
    implemented here builds
    on the long-extant concept of Fourier transformation along
    the pulse phase cycle dimension
    to enable data visualization that more fully
    reflects the rich physics underlying these \gls{nmr}
    experiments.
In addition to discussing the outline and
    implementation of the general \gls{dcct} schema and
    associated plotting methods, this manuscript
    presents a collection of algorithms that provide
    robust phasing, avoidance of baseline distortion,
    and the ability to realize relatively
    weak signals amidst background noise through
    signal-averaged correlation alignment.
The methods for visualizing the raw data,
    together with the processing routines whose development
    they guide should apply directly
    to or extend easily to other techniques facing
    similar challenges.

\end{abstract}

\keywords{ESR, ODNP, NMR, cross-relaxation}
\newcommand{\figStepByStep}{\begin{figure*}[tbp]
    \centering
    \subfig{fig:StepByStepInvRawSig}{RawSig}
    \subfig{fig:StepByStepInvPhaseCorr}{PhaseCorr}
    \subfig{fig:StepByStepInvAligned}{Aligned}
    \subfig{fig:StepByStepInvReal}{Real}
    \newlength{\templen}\setlength{\templen}{\dimexpr(0.25\linewidth-0.25em)\relax}%
    \begin{tabular}{cccc}
    \refRawSig & \refPhaseCorr &
    \refAligned & \refReal
    \\
    \includegraphics[height=3.5in]{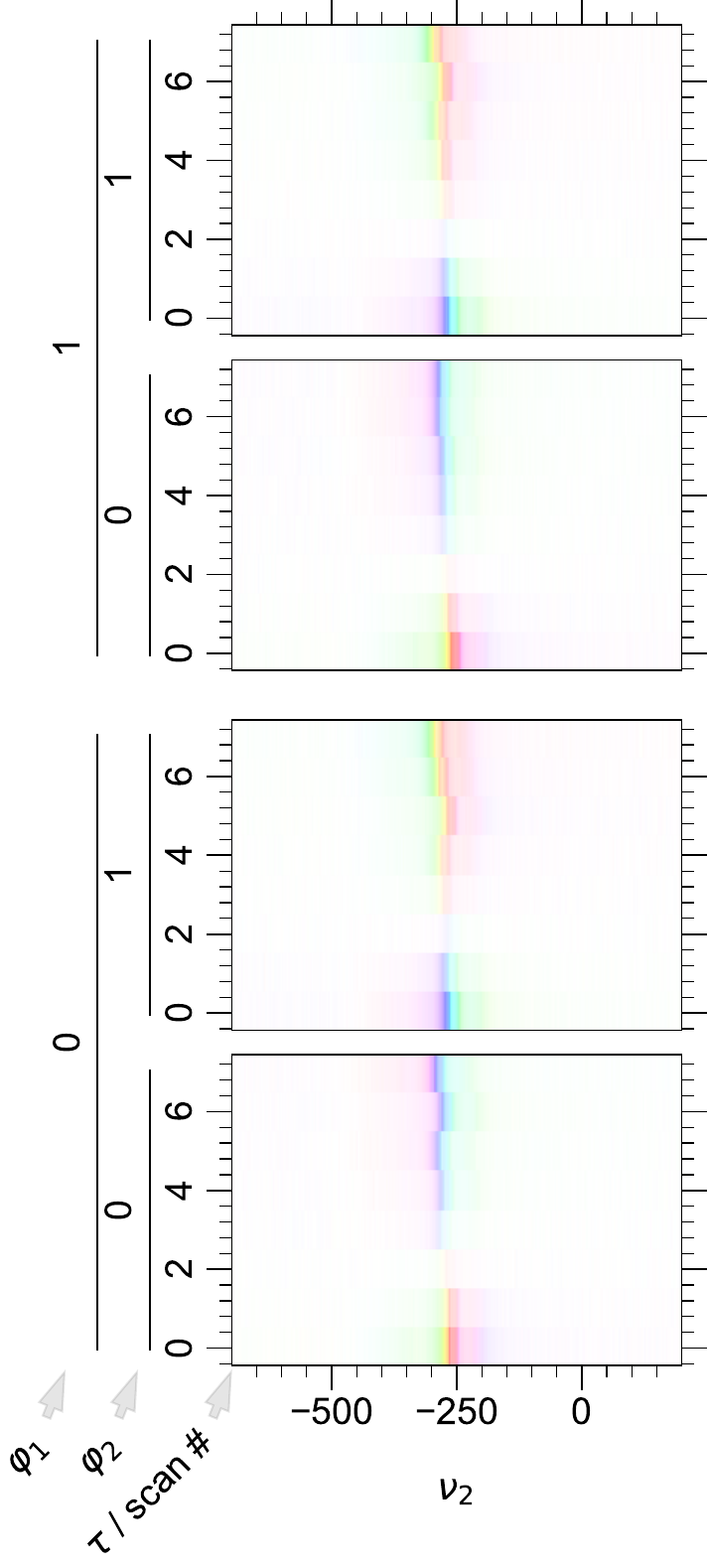}
    &
    \includegraphics[height=3.5in]{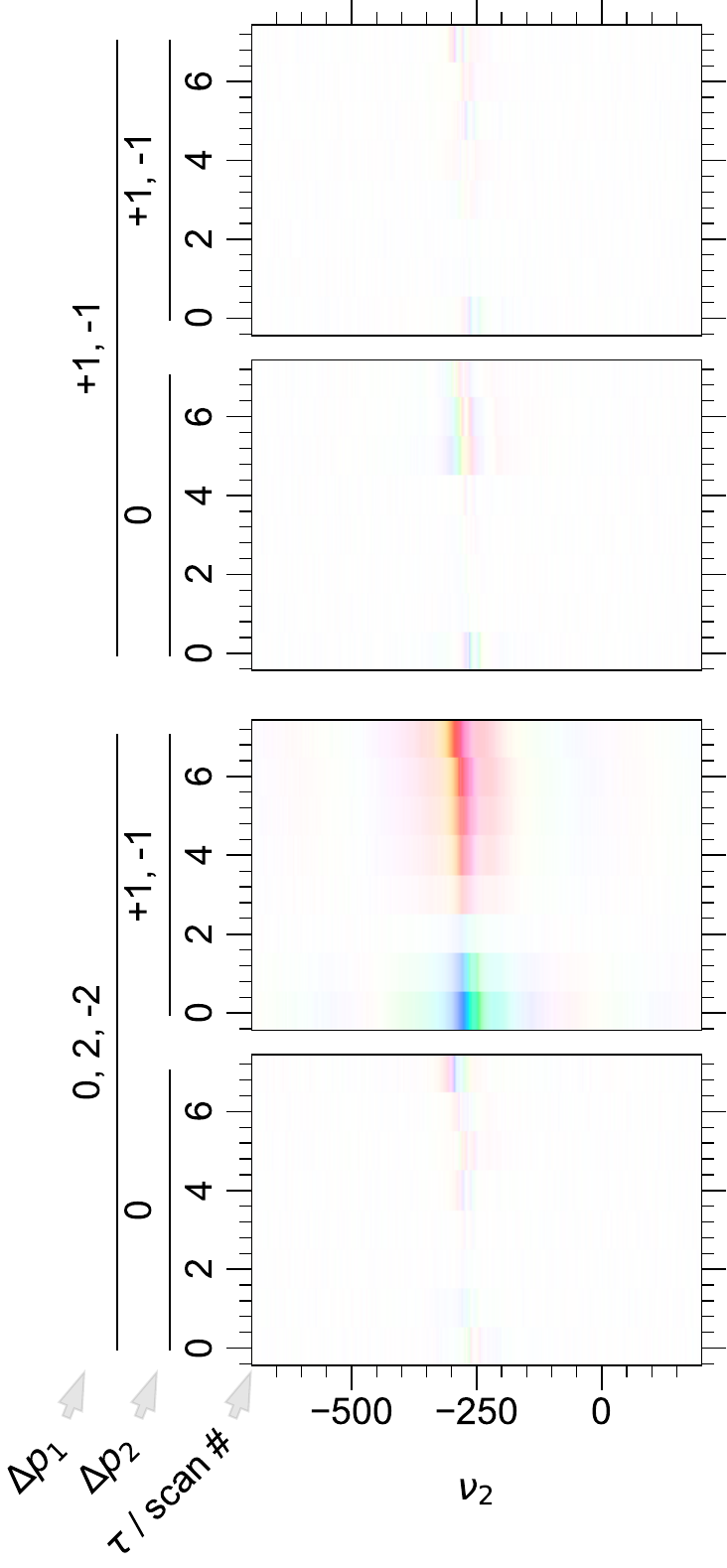}
    &
    \includegraphics[height=3.5in]{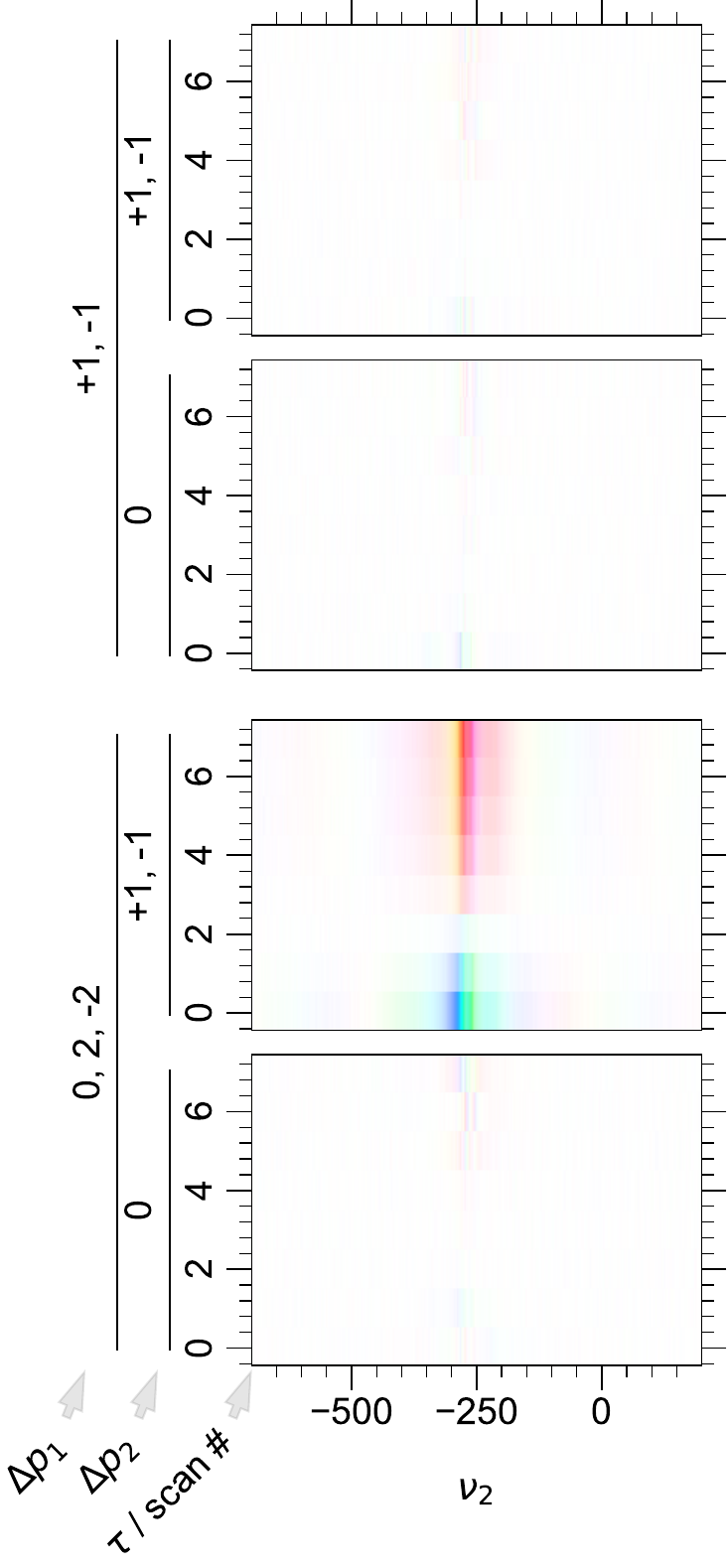}
    &
    \setlength{\templen}{\dimexpr(3.5in-5ex)\relax}
    \raisebox{4ex}{\includegraphics[height=\templen]{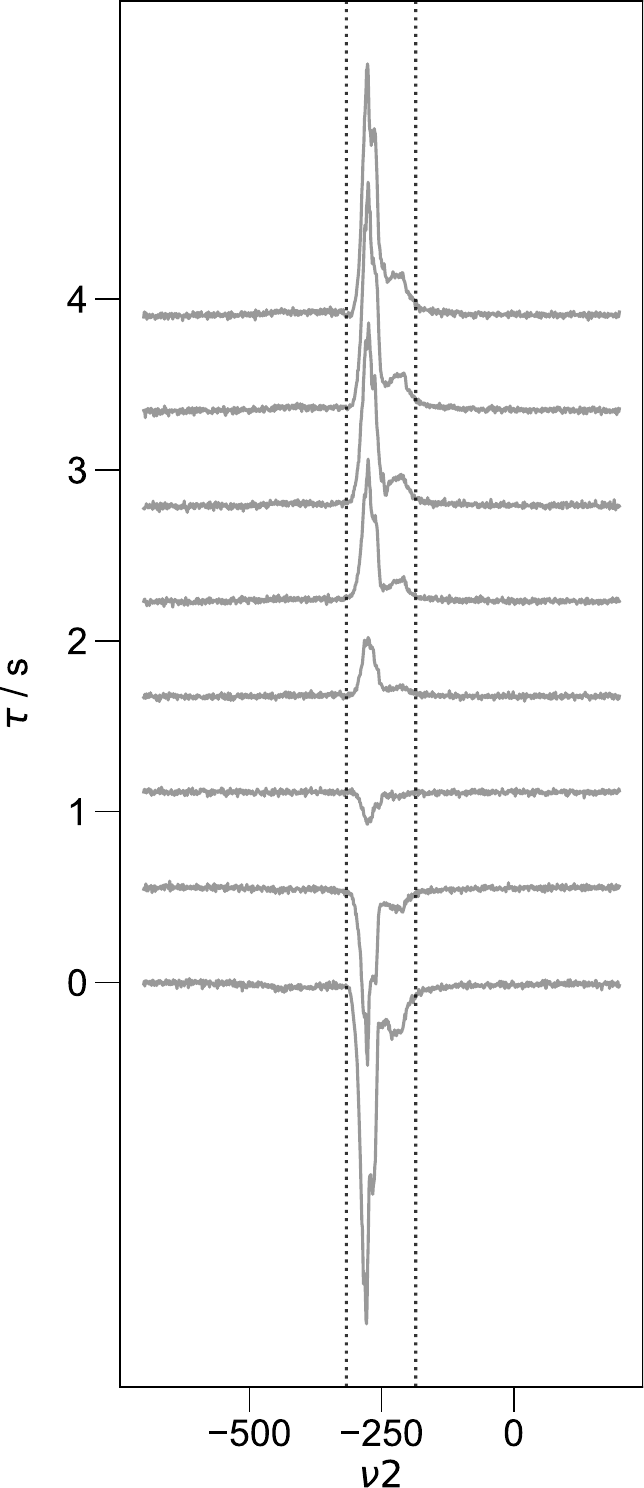}}\\
    \end{tabular}
    \caption{
        Application of the processing
        and visualization
        algorithms to an inversion recovery experiment
        produces a high-quality result,
        despite fluctuating fields and
        imperfections in phase cycling.
        In \refRawSig, the raw data
        has not been Fourier transformed along the
        phase cycling dimension.
        Here, because the $\tau$ values are not equally
        spaced,
        the labels on the indirect dimension
        correspond to the scan number.
        This data already clearly
        shows a null followed by phase inversion
        along the indirect dimension.
        Fourier transformation of  this data
        from the phase cycling to the coherence
        domain yields a \gls{dcct} map shown in
        \refPhaseCorr,
        where the Hermitian symmetry cost function
        \cref{eq:fastHermitian}
        determines the center of the spin echo,
        yielding phase-corrected signal.
        The alignment routine corrects the frequency
        drift, yielding
        \refAligned.
        Finally, after slicing out the \gls{fid}
        in the time domain,
        following \cref{eq:s_FID},
        \refReal\ provides the real part of the
        signal, free of baseline distortion,
        with automatically chosen integration bounds
        delineated by the dotted lines.
    }
    \label{fig:StepByStepInv}
\end{figure*}}

\newcommand{\figStepByStepEp}{\begin{figure*}[tbp]
    \centering
    \subfig{fig:stepOneEp}{RawSig}
    \subfig{fig:StepByStepEpCohDom}{CohDom}%
    \subfig{fig:stepTwoEp}{PhaseCorr}
    \subfig{fig:stepThreeEp}{Aligned}
    \setlength{\templen}{\dimexpr(0.25\linewidth-0.25em)\relax}%
    \begin{tabular}{cccc}
    \refRawSig & \refCohDom & \refPhaseCorr &
    \refAligned \\
    \includegraphics[height=3.5in]{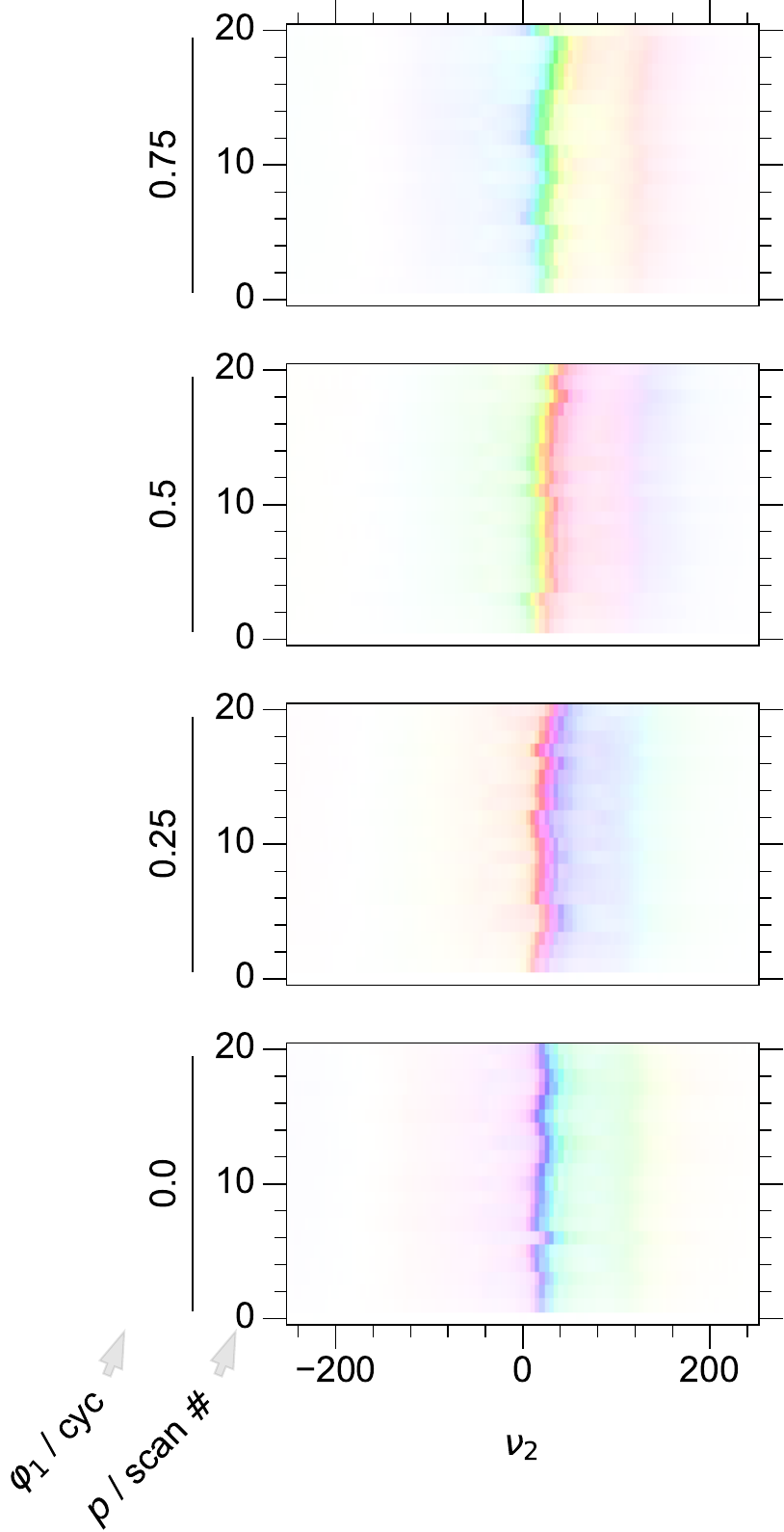}
    &
    \includegraphics[height=3.5in]{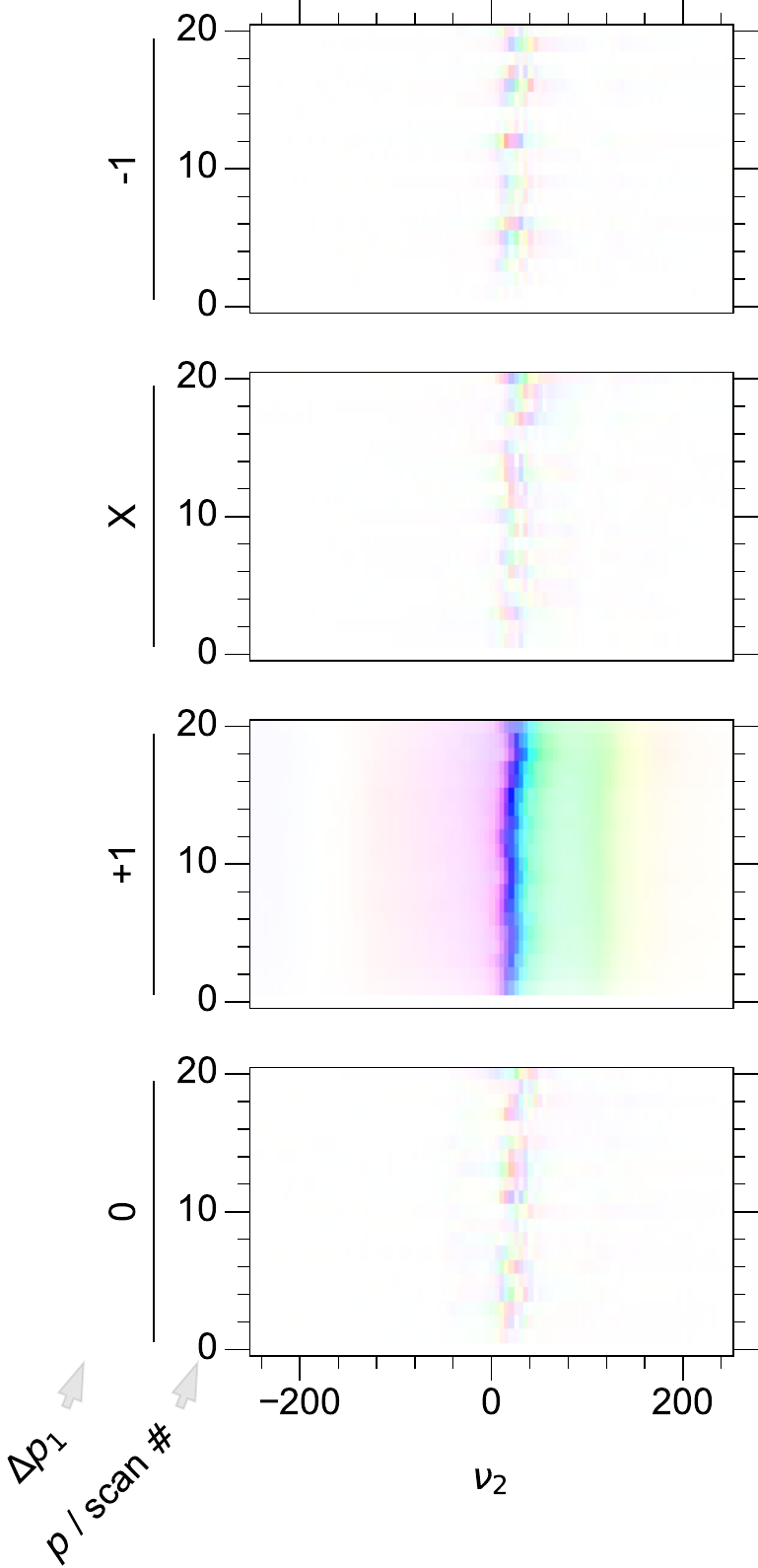}
    &
    \includegraphics[height=3.5in]{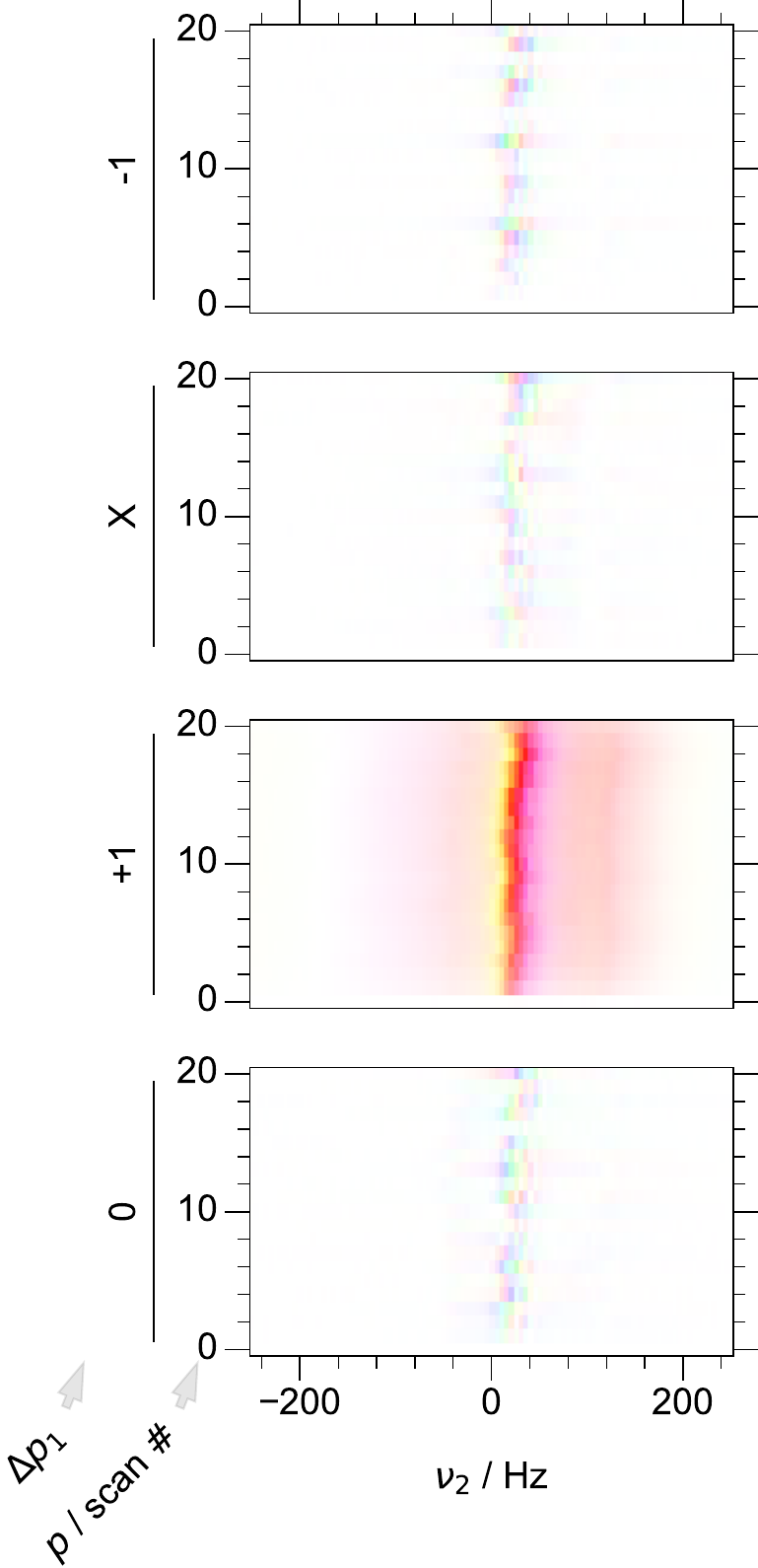}
    &
    \includegraphics[height=3.5in]{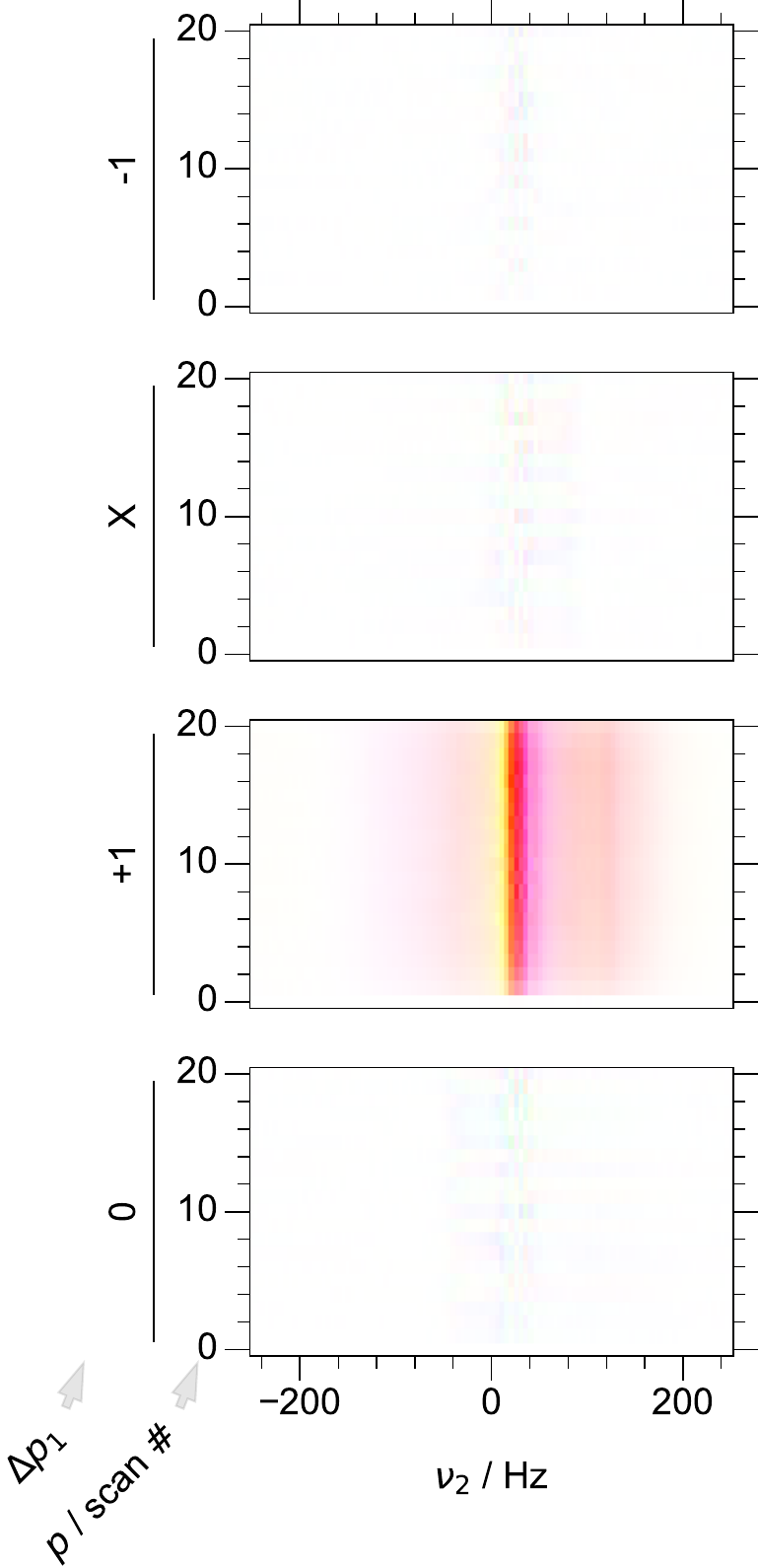}
    \\
    \end{tabular}
    \caption{
        Similar to \cref{fig:StepByStepInv} but
        applied to a progressive enhancement
        experiment.
        Here, because the microwave power ($p$) values are not equally
        spaced,
        the labels on the indirect dimension
        correspond to the scan number.
        \refRawSig: The raw data is
        Fourier transformed only along the direct
        dimension; \refCohDom: Fourier transformation 
        has been applied along the phase cycling dimension
        to yield coherence transfer dimension;
        \refPhaseCorr: A timing correction corresponding to
        $\Delta t_{min}/2$ from \cref{eq:fastHermitian}
        has been applied.
        \refAligned: The data after applying the
        correlation alignment.
        The 1D line plots for this data are shown in
        \cref{fig:AlignEpOverlay}.
    }
    \label{fig:StepByStepEp}
\end{figure*}}

\newcommand{\figMicelleProcessing}{\begin{figure}[tbp]
    \centering
    \textcolor{red}{show RM data (900 transients)
    -- both the 2D color plot for the various
    repeats and the 1D average}
    \caption{
        This figure presents the signal from a 
        challenging dataset -- a sample
        of reverse micelles.
        As this sample contains $30 \times$ less
        proton spins than a standard aqueous
        sample,
        $900\times$ signal averaging is required,
        posing unusual challenges in terms of low
        \gls{snr} per transient and field fluctuations
        between the various transients.
    }
    \label{fig:MicelleProcessing}
\end{figure}}

\newcommand{\figHermitianPhasing}{\begin{figure}[tbp]
    \centering
    \subfig{fig:HermitianPhasingCost}{cost}
    \subfig{fig:HermitianPhasingImaginary}{imaginary}
    \subfig{fig:HermitianPhasingFT}{hermitianFt}
    \begin{tabular}{cc}
        \refcost
        & \raisebox{-\height+1ex}{ 
            \includegraphics[width=0.9\linewidth]{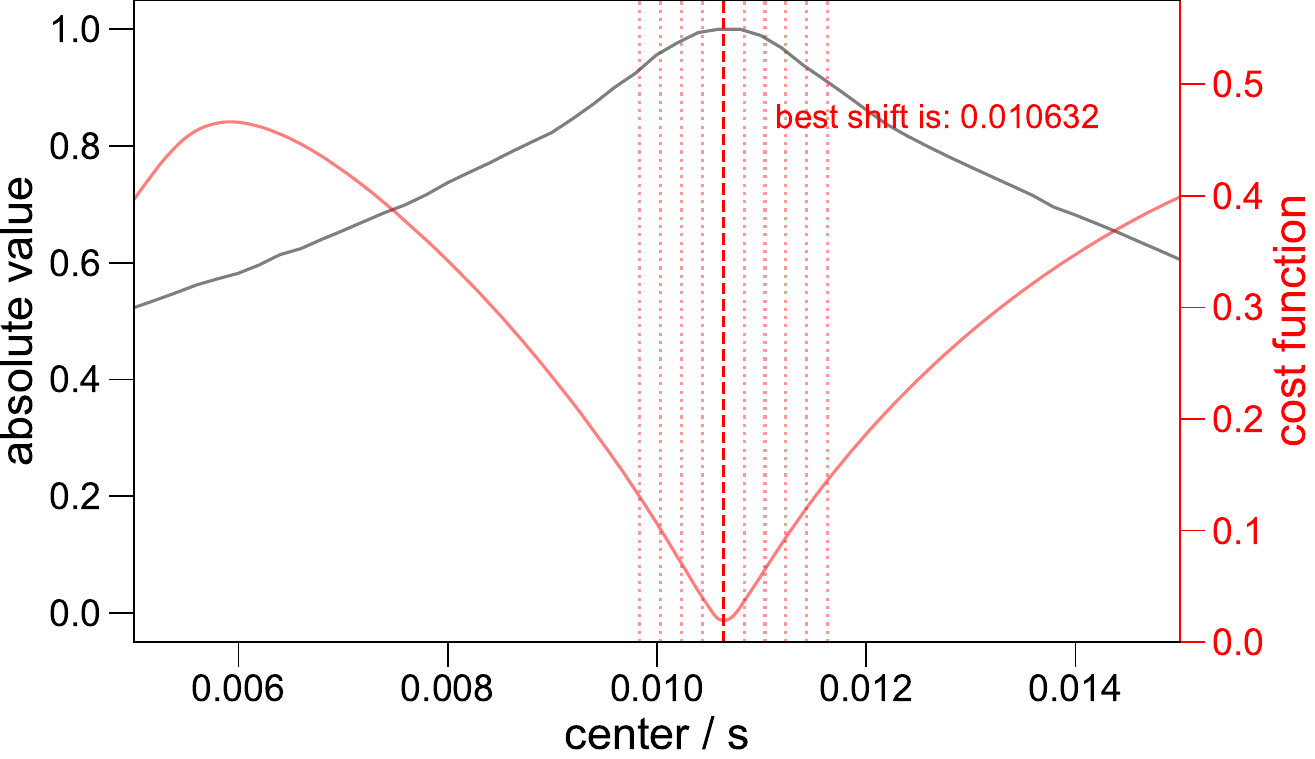}
        }
        \\
        \refimaginary
        & \raisebox{-\height+1ex}{
            \includegraphics[width=0.9\linewidth]{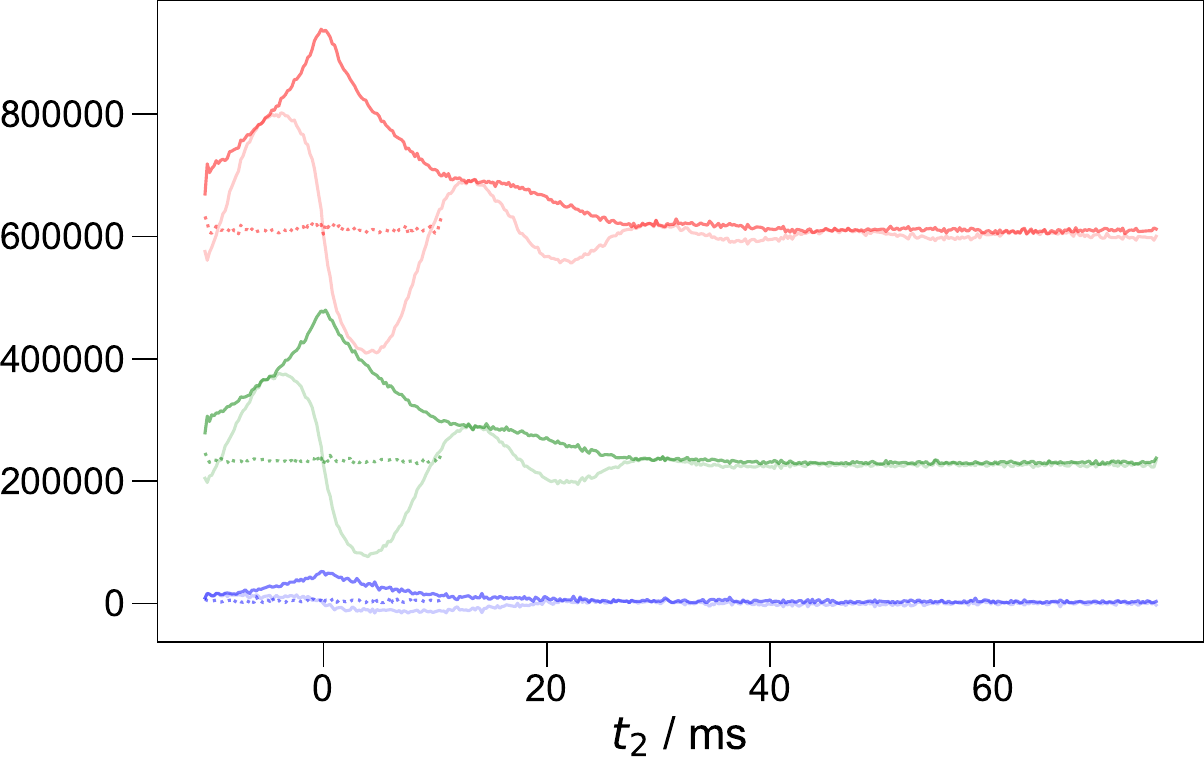}
        }
        \\
        \refhermitianFt
        & \raisebox{-\height+1ex}{
            \includegraphics[width=0.9\linewidth]{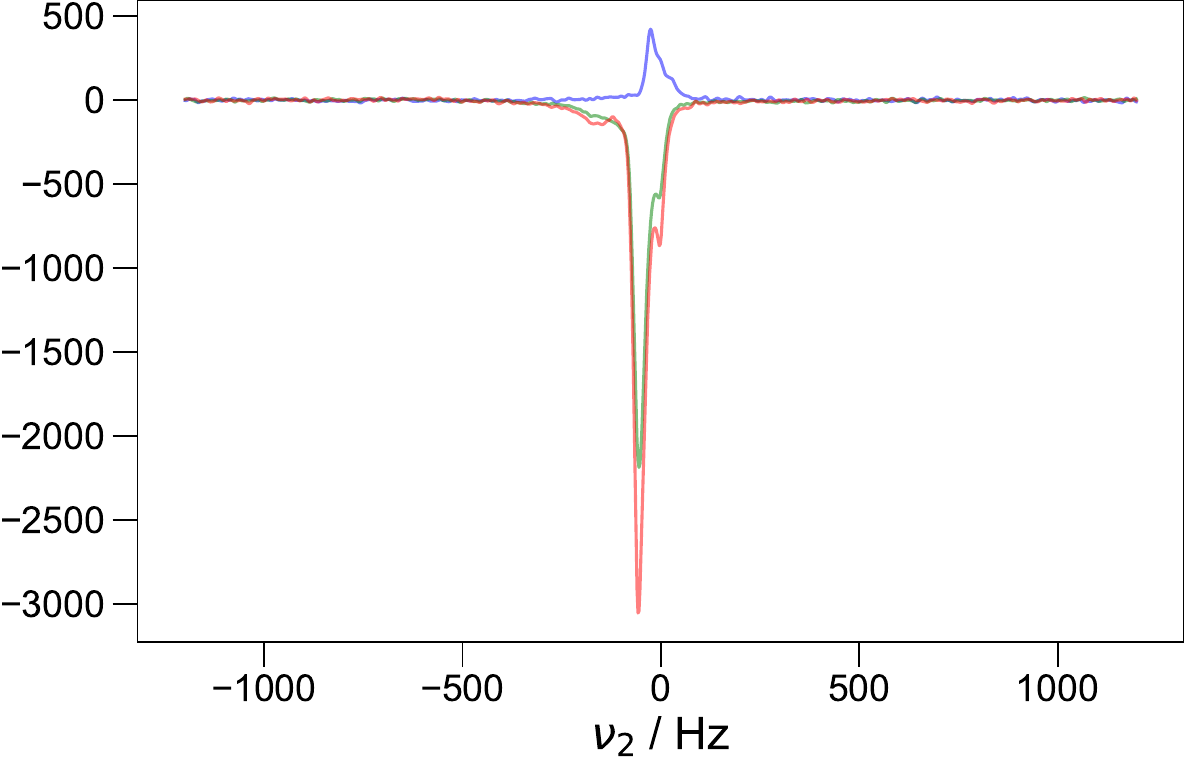}
        }
    \end{tabular}
    \centering
    \caption{\refcost: The cost function
        \cref{eq:fastHermitian}
        (where $\Delta t/2=$ ``center'' above)
        demonstrates a clear minimum
        at the optimum time shift, $\Delta t_{min}/2$,
        which represents the precise center of the
        spin echoes. 
        Subtraction of $\Delta t_{min}/2$ from the time axis
        centers the echoes at $t=0$.
        \refimaginary: After applying a uniform zeroth-order
        phase shift,
        the imaginary component of the
        echoes crosses zero at $t=0$
        for all microwave
        powers employed to measure \gls{odnp}
        enhancement.
        Subsequent slicing of the \gls{fid}
        from these echoes,
        following \cref{eq:s_FID},
        yields
        well-phased, absorptive, baseline-free
        signal,
        the real component of which is shown in
        \refhermitianFt.
    }
    \label{fig:HermitianPhasing}
\end{figure}}
\newcommand{\figCpmg}{%
\begin{figure}[tbp]
    \subfig{fig:CPMGsignal}{signal}
    \subfig{fig:CPMGbleeding}{bleeding}
    \begin{tabular}{cc}
        \refsignal
        & \raisebox{-\height+1ex}{ 
            \includegraphics[width=0.9\linewidth]{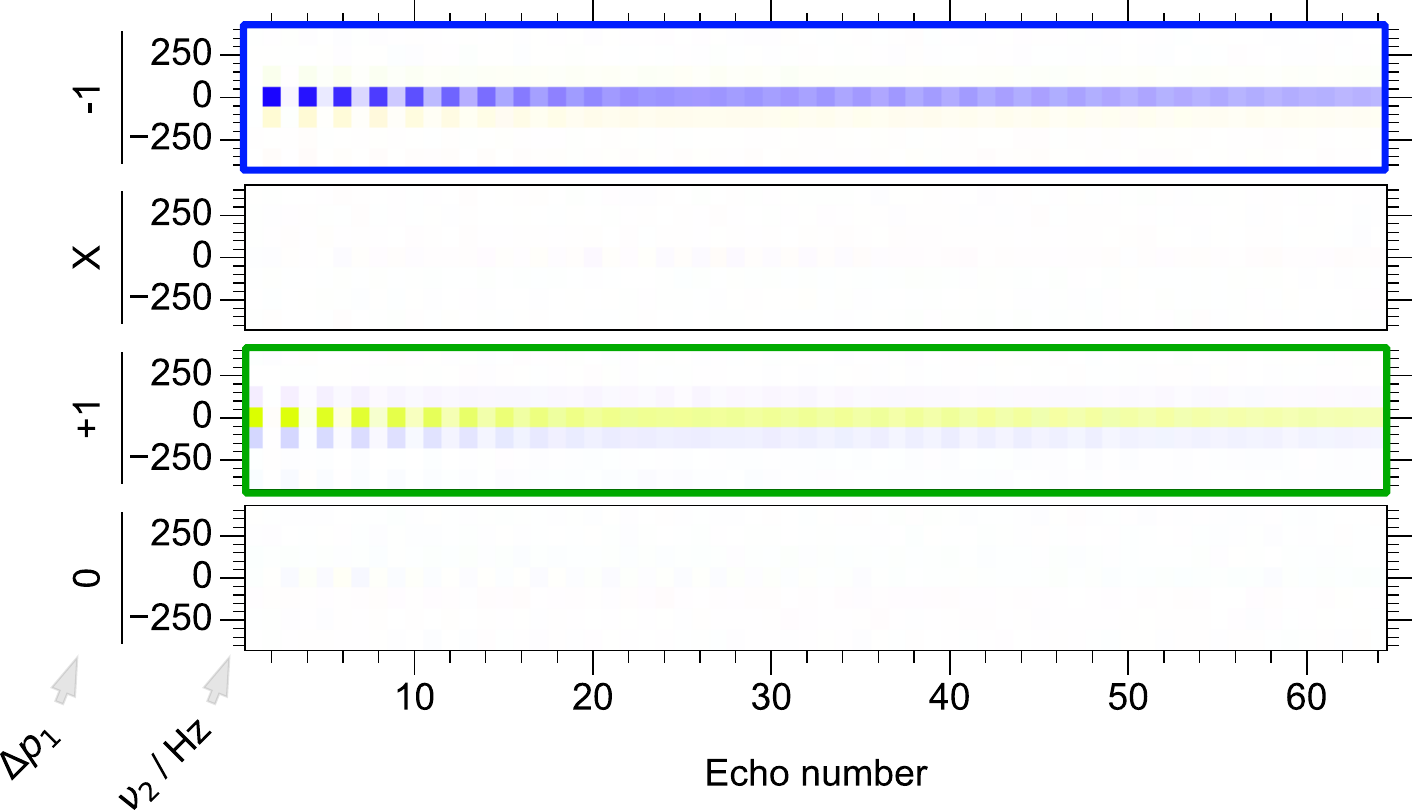}
        }
        \\
        \refbleeding
        & \raisebox{-\height+1ex}{ 
            \includegraphics[width=0.9\linewidth]{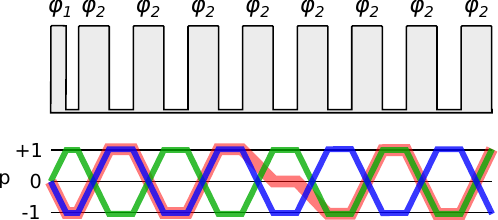}
        }
    \end{tabular}
    \caption{\gls{dcct} map presentation of CPMG data
    acquired at 15~MHz with a 4-step phase cycle of
    the initial excitation pulse \refsignal\ and
    an independent 4-step phase cycle of all 180° pulses in
    concert.
    Here, the $x$-axis corresponds to the echo number,
    while the direct dimension within each echo window
    appears as the innermost dimension along the
    $y$-direction.
    For simplicity,
    the single element of the $\Delta p_1 + \Delta p_2$ dimension
    that contains valid signal
    ($\Delta p_1 + \Delta p_2 = -1$)
    has been selected.
    Detected signal alternates between two coherence
    pathways $\Delta p_1 = 1$, $\Delta p_1 + \Delta p_2 = -1$ (green) and
    $\Delta p_1 = -1$, $\Delta p_1 + \Delta p_2 = -1$(blue)
    with each $\pi$ pulse.
    Beginning with the 6\textsuperscript{th} or
    7\textsuperscript{th} $\pi$ pulse,
    signal `bleeding' from the alternate pathway,
    as shown in \refbleeding,
    starts to become significant.
    The red line gives one example of the many pathways
    that, cumulatively, give rise to such ``bleeding.''
}
    \label{fig:cpmg}
\end{figure}
}
\newcommand{\figBrukerCpmg}{\begin{figure}[tbp]
    \centering
    \includegraphics[width=\linewidth]{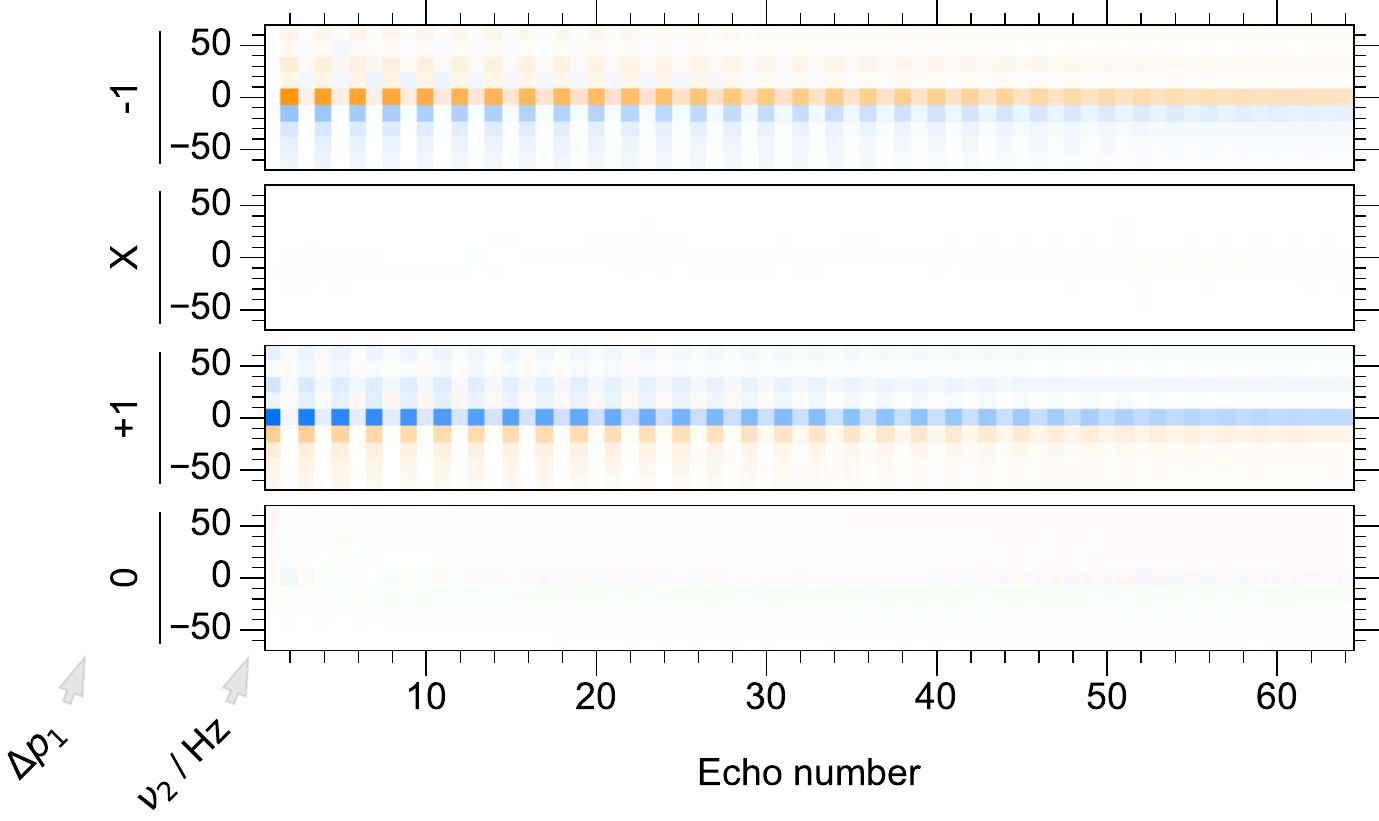}
    \caption{CPMG data was acquired on high-field (400
        MHz) Bruker spectrometer for 64 180°
        pulses. After each 180° pulse, the signal
        alternates between $\Delta p_1 = +1$ and
        $\Delta p_1 = -1$ until eventually
        bleeding is observed (by around the
            12\textsuperscript{th} echo).}
    \label{fig:BrukerCPMG}
\end{figure}}
\newcommand{\figScopeData}{\begin{figure}[tbp]
    \centering
    \subfig{fig:ScopeDataMisset}{misset}
    \subfig{fig:ScopeDataFixed}{fixed}
    \subfig{fig:ScopeDataCoh}{coh}
    \begin{tabular}{rc}
        \refmisset
        & \hspace*{-2em}\raisebox{-\height+1ex}{ 
            \includegraphics[width=\linewidth-2em]{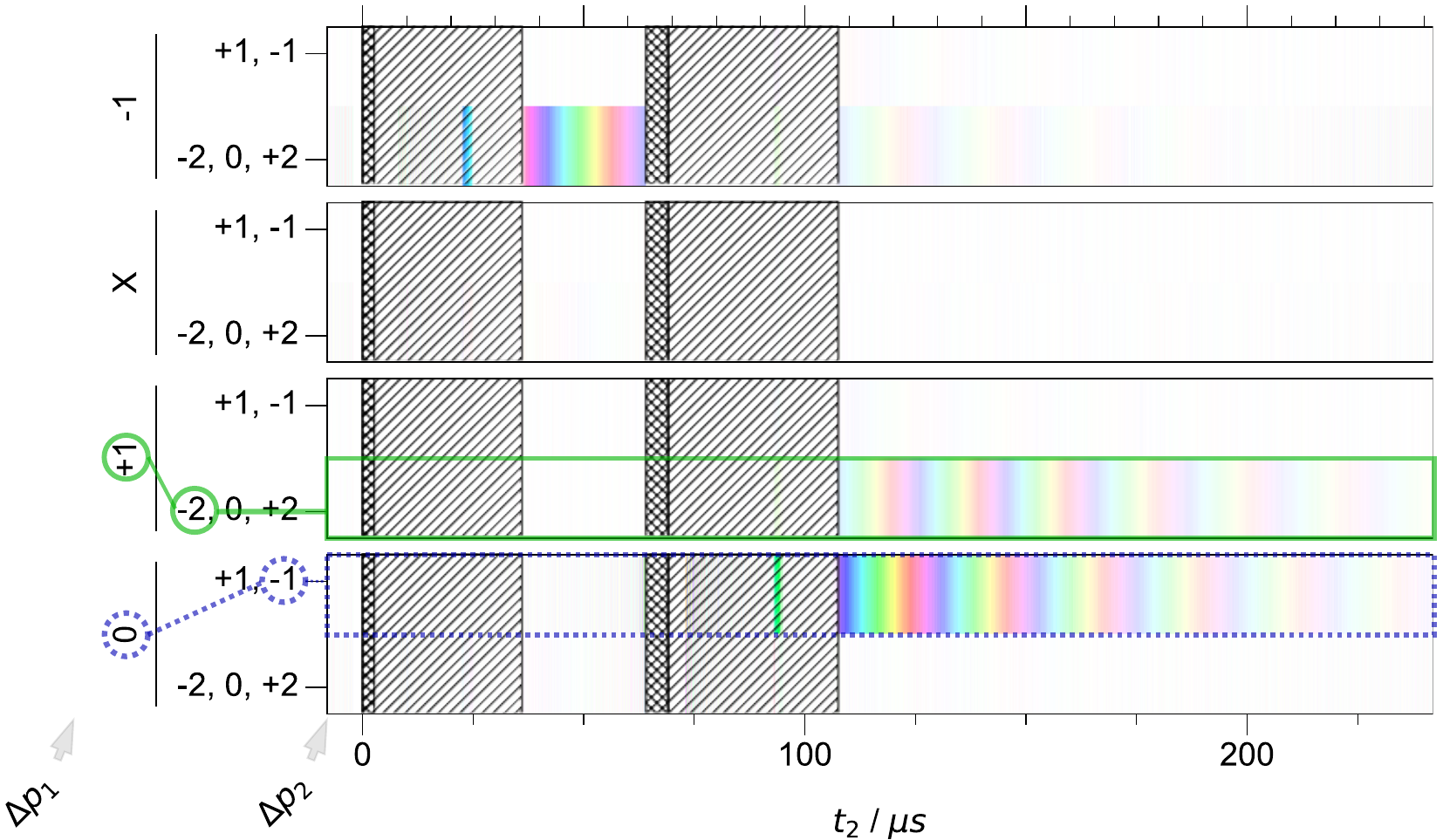}
        }
        \\
        \reffixed
        & \hspace*{-2em}\raisebox{-\height+1ex}{ 
            \includegraphics[width=\linewidth-2em]{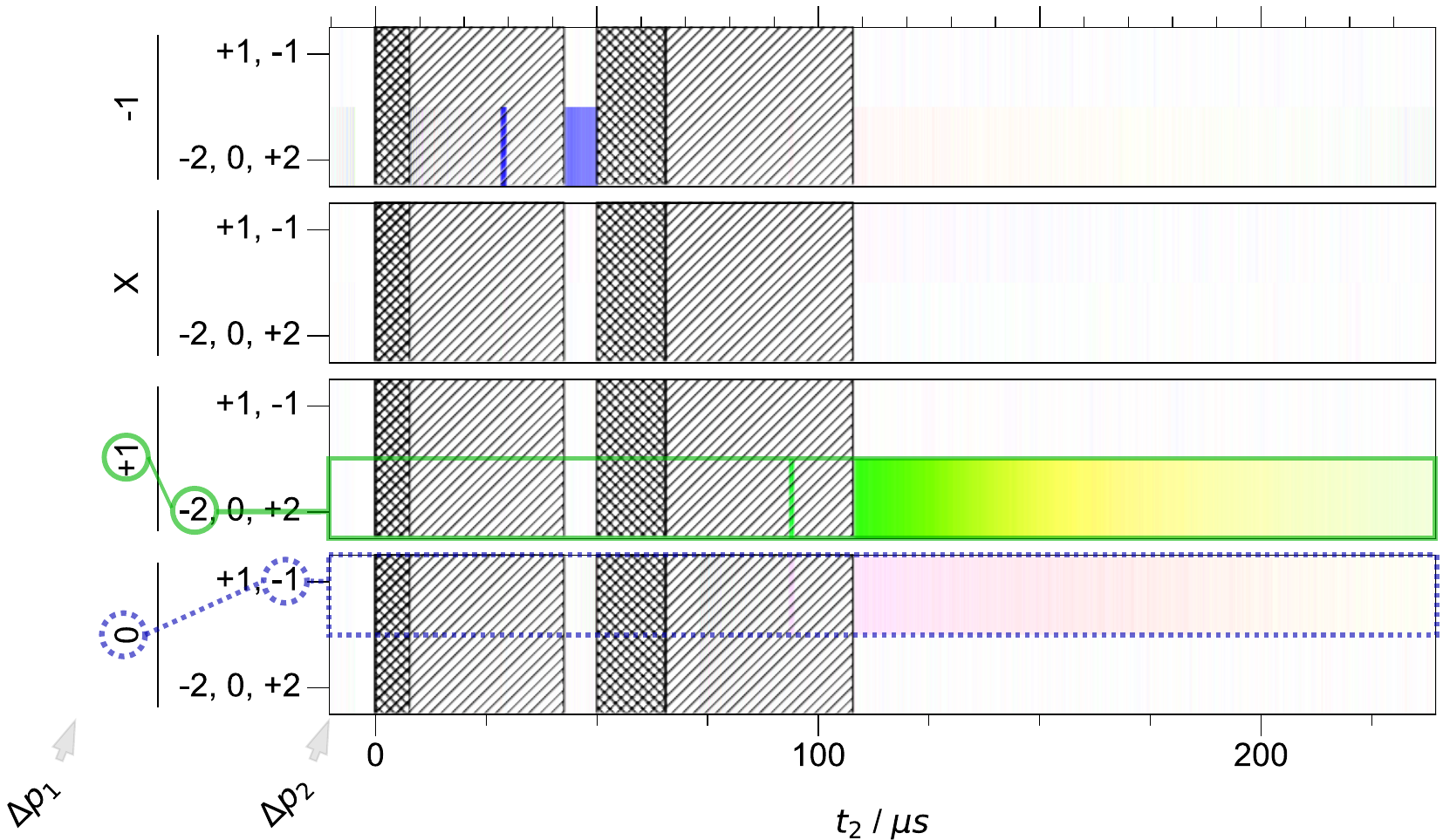}
        }
        \\
        \refcoh
        & \hspace*{-2em}\raisebox{-\height+1ex}{ 
            \includegraphics[width=\linewidth-2em]{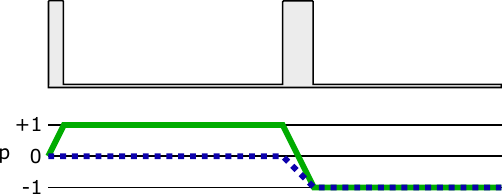}
        }
    \end{tabular}
    \caption{Spin echo acquired
        by an NMR spectrometer
        constructed from off-the-shelf test and
        measurement equipment
        for 13~mM NiSO$_{4}$-doped water
        \refmisset\ under suboptimal experimental
        parameters and \reffixed\ optimal
        parameters.
        The data distinctly peaks 
        around 120 \textmu s at the expected
        location in the coherence transfer domain --
        specifically, the $\Delta p_1=+1$, $\Delta p_2=-2$ pathway
        where $\Delta p_n$ is the Fourier conjugate of the pulse phase
        ($\varphi_n$) dimension.
        The receiver deadtime extends to approximately
        35~μs after the center of the 180° pulse.
        Low-$B_1$ regions of the sample
        contribute to
        the subtle signal visible at
        $\Delta p_1=0$
        $\Delta p_2=-1$,
        which is an \gls{fid} arising from the second pulse.
    }
    \label{fig:ScopeData}
\end{figure}}
\newcommand{\figModCoil}{\begin{figure}[tbp]
    \centering
    \subfig{fig:ModCoilAttachedTime}{attachedTime}
    \subfig{fig:ModCoilDetachedTime}{detachedTime}
    \subfig{fig:ModCoilAttachedFreq}{attachedFreq}
    \subfig{fig:ModCoilDetachedFreq}{detachedFreq}
    \begin{tabular}{ll}
        \quad\quad Mod Coil Attached: & \quad\quad Mod Coil Detached:
        \\
        \refattachedTime &
        \refdetachedTime
        \\
        \raisebox{-\height+4ex}{
            \includegraphics[width=0.5\linewidth-2em]{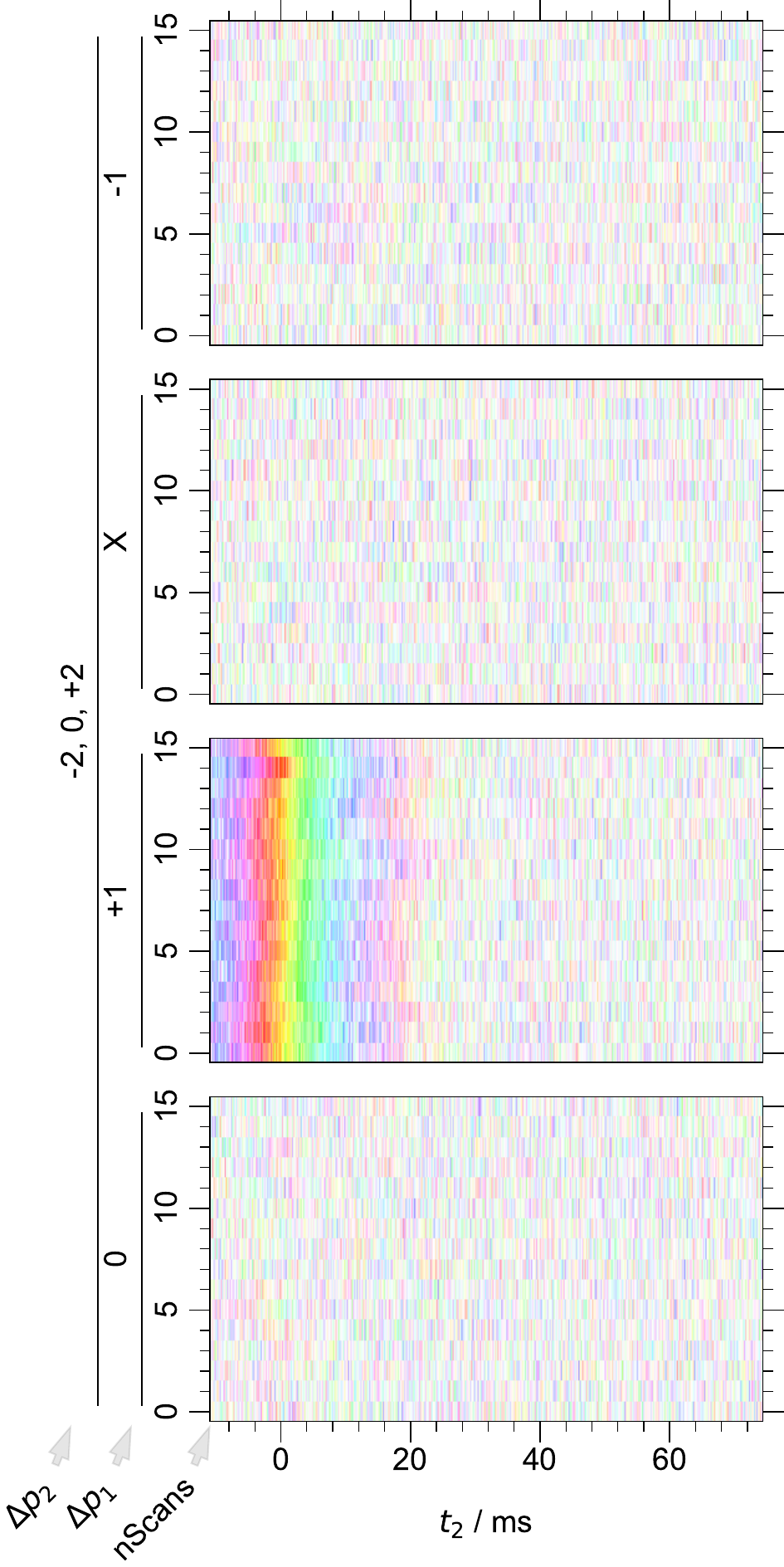}
        }
        & \raisebox{-\height+4ex}{
            \includegraphics[width=0.5\linewidth-2em]{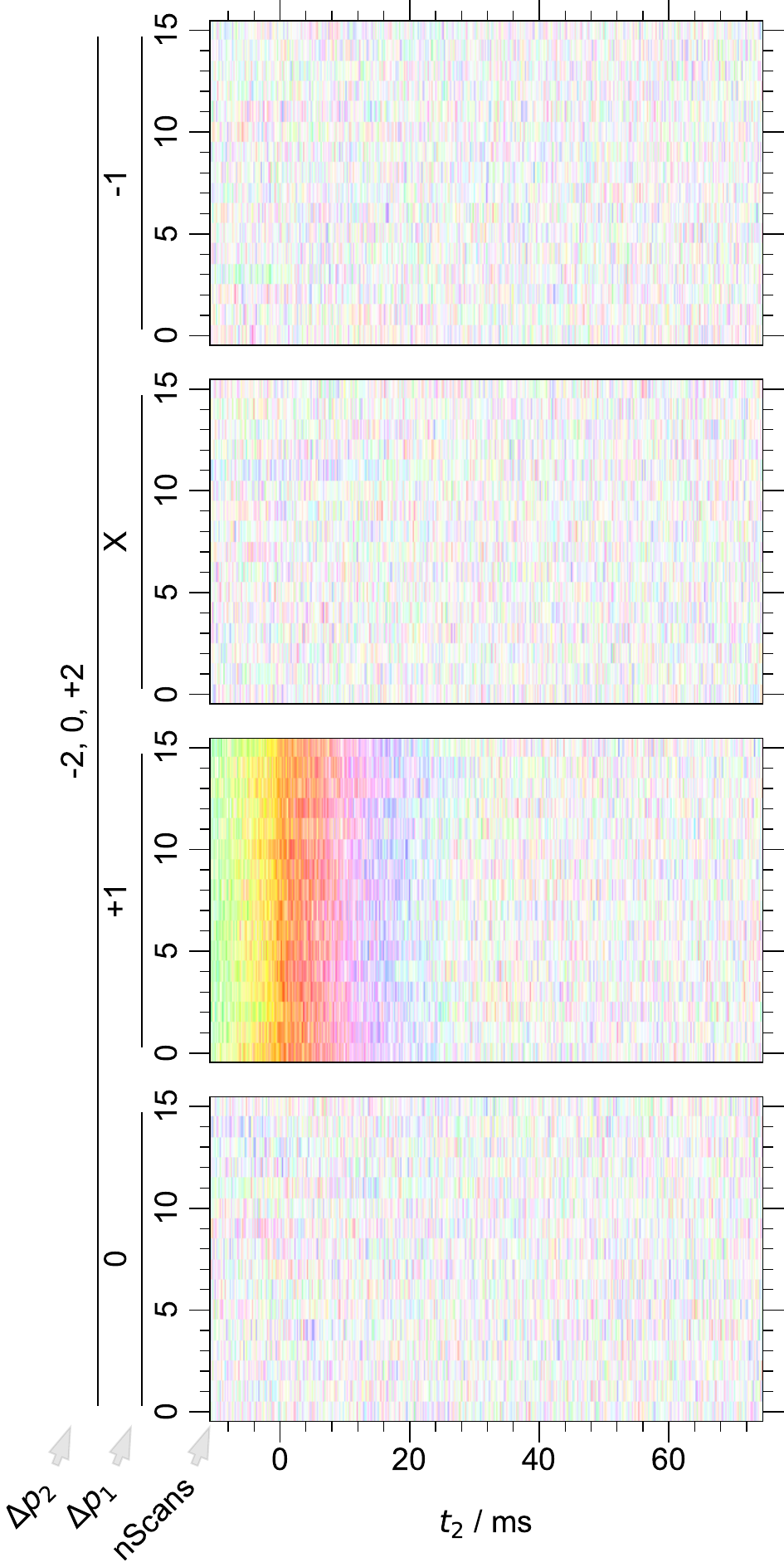}
        }\\
        \refattachedFreq
        &
        \refdetachedFreq
        \\
        \raisebox{-\height+4ex}{
            \includegraphics[width=0.5\linewidth-2em]{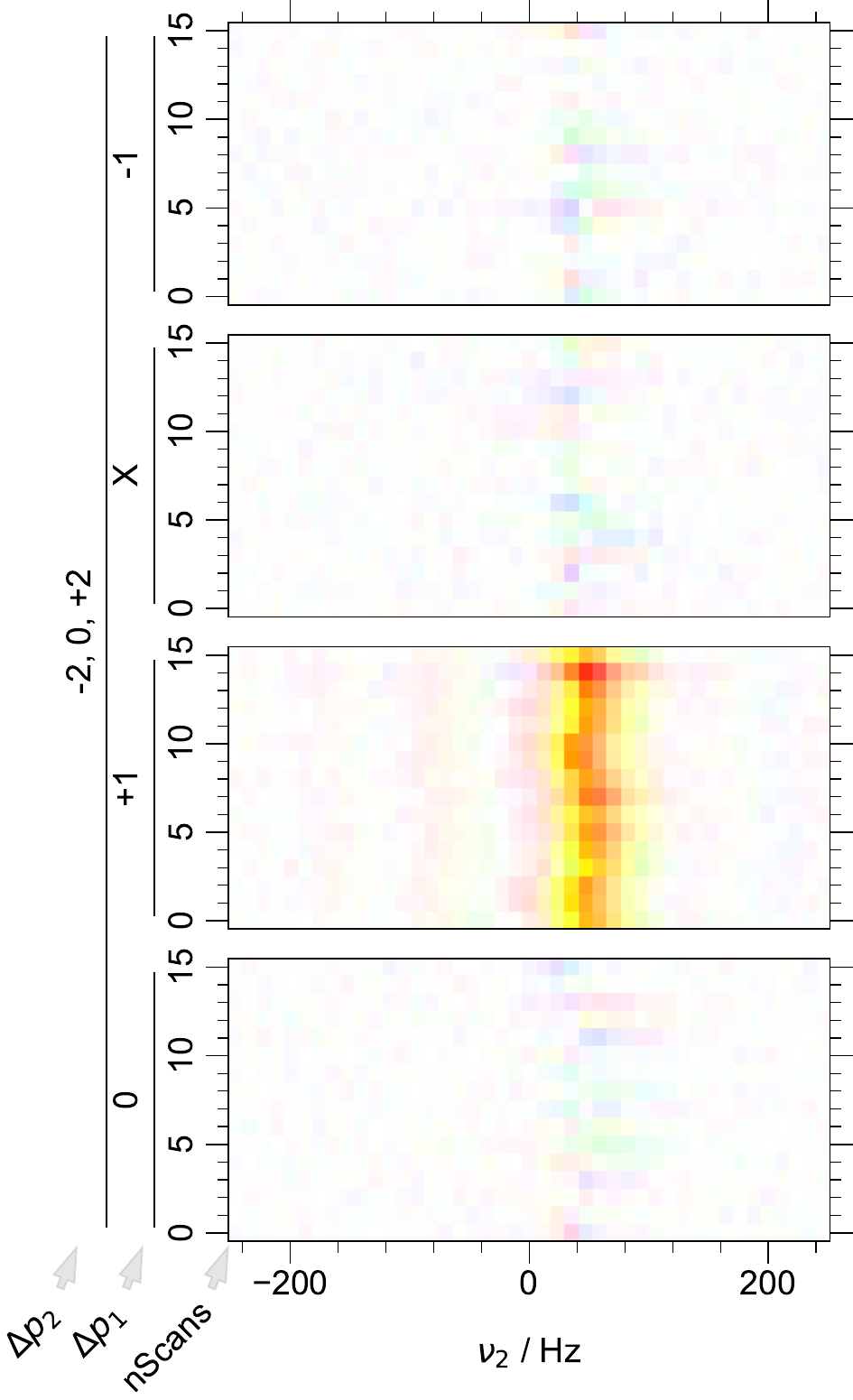}
        }
        & \raisebox{-\height+4ex}{
            \includegraphics[width=0.5\linewidth-2em]{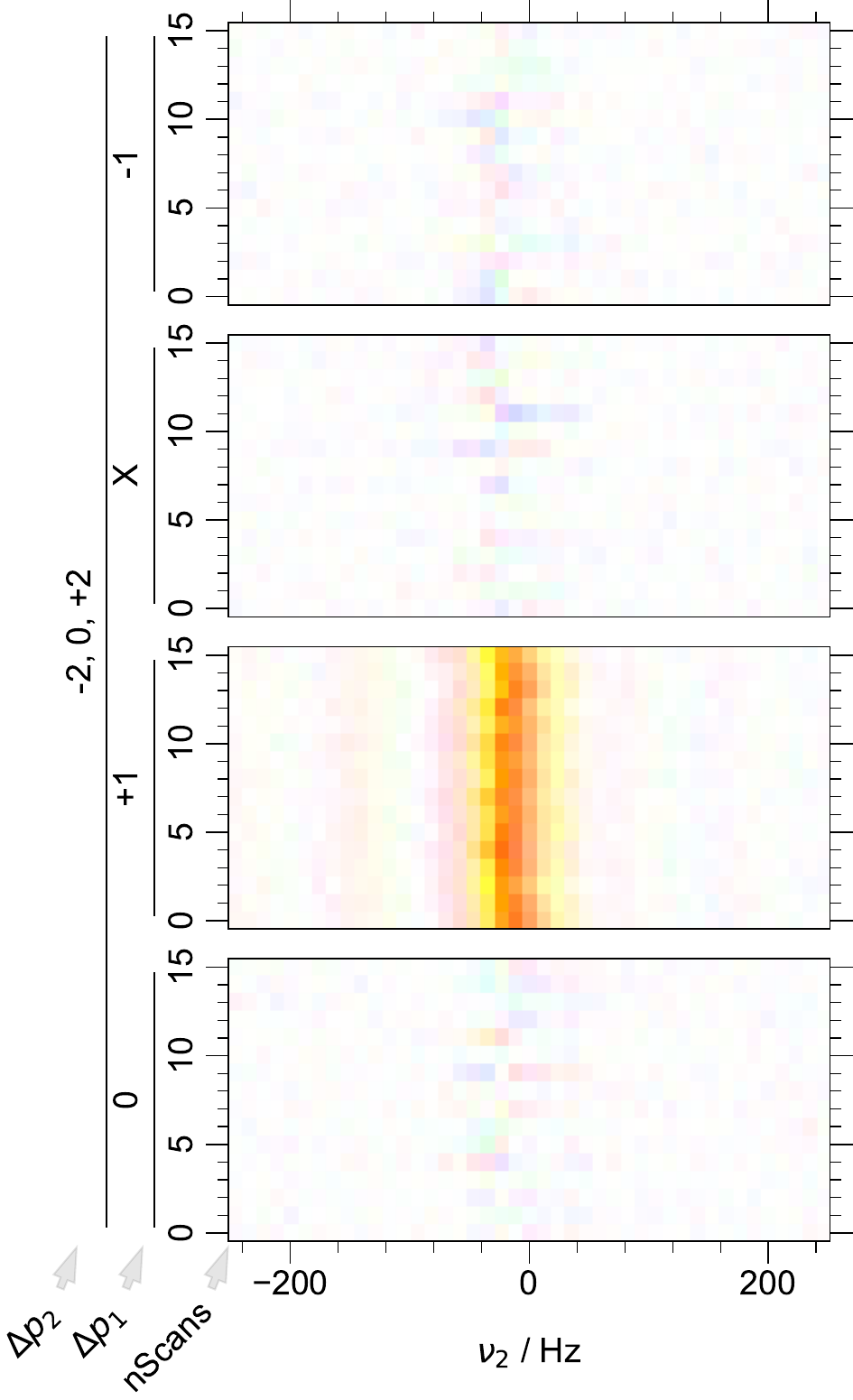}
        }
    \end{tabular}
    \caption{Attaching vs detaching the \gls{esr}
        modulation coil allows for a controlled
        test of 2 situations with different field
        stability, as shown for both the time domain
        (\refattachedTime\ and \refdetachedTime)
        and frequency domain
        (\refattachedFreq\ and \refdetachedFreq),
        respectively.
    Three notable effects include the variation of the average frequency,
    increased amplitude of the signal in inactive
    coherence transfer pathways
    elements of the
    coherence dimension,
    and variation of the echo center.
    With the mod coil detached \refdetachedFreq,
    the average signal value in the desired coherence
    pathway is $1.04\times$ that with the mod coil
    attached \refattachedFreq,
    while the root mean squared amplitude of the noise
    in all other (\ie, inactive) coherence pathways decreases by a
    factor of $0.86$.
    Therefore,
    while the signal amplitude remains roughly
    equivalent,
    detaching the mod coil leads to a noticeable
    decrease in artefacts linked to improper phase
    cycling
    (variation of static field's magnitude or time
    dependence from one step of the phase cycle to
    the next).
    }
    %
    \label{fig:ModCoil}
\end{figure}}
\newcommand{\figModCoilHermitian}{\begin{figure}[tbp]
    \centering
    \includegraphics[width=\linewidth]{figures/mod_coil_transients_lineplot.pdf}
    \sepcomment\pdfcommentJF{I re-ran this, and the
    figure looks quite different vs. before -- can you
    confirm that everything is OK?
    In fact, I realize that it now looks *worse* with the
    mod coil disconnected, which is opposite what we are
    saying in the text (and what you see in the \gls{dcct} plot)}
    \caption{
        When rapid field instabilities are present
        (here a 100~kHz modulation field),
        the center of the echo position may vary
        significantly by as much as 1.5~ms (red line);
        however, it remains otherwise reliably
        stable to within about $\pm 500\;\text{μs}$
        (blue line).
    }
    \label{fig:ModCoilHermitian}
\end{figure}}
\newcommand{\figGaussianApo}{\begin{figure}[tbp]
        \centering
        \begin{tabular}{cc}
            \includegraphics[width=0.5\linewidth]{auto_figures/signal_before_apodization_aposimul200622.pdf}
            &
            \includegraphics[width=0.5\linewidth]{auto_figures/signal_after_apodization_aposimul200622.pdf}
        \end{tabular}
        \caption{Simulated signal before (left) and after
            (right) apodization with the filter
            given by~\cref{eq:simplesignal}.
        }
        \label{fig:GaussianApodization}
\end{figure}}
\newcommand{\figGaussianApoReal}{\begin{figure}[tbp]
        \centering
        \subfig{fig:GaussianApodizationRealBefore}{Before}
        \subfig{fig:GaussianApodizationRealAfter}{After}
        \subfig{fig:GaussianApodizationRealEnvelope}{Envelope}
        \subfig{fig:GaussianApodizationRealOneDRe}{OneDRe}
        \setlength{\templen}{\dimexpr(0.5\linewidth-0.25em)\relax}%
        \setlength{\templen}{\dimexpr(2.5\templen)\relax}%
        \begin{tabular}{cc}
            \refBefore
            &
            \refAfter
            \\
            \setlength{\templen}{\dimexpr(0.5\linewidth-0.25em)\relax}%
            \includegraphics[width=\templen]{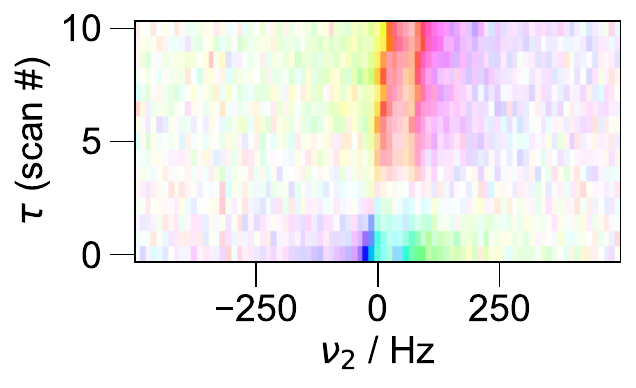}
            &
            \setlength{\templen}{\dimexpr(0.5\linewidth-0.25em)\relax}%
            \includegraphics[width=\templen]{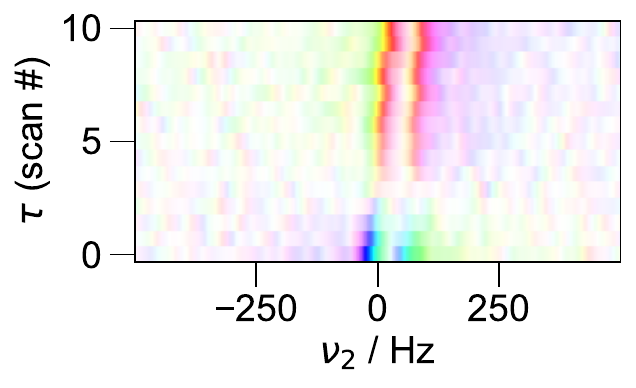}
            \\
            \multicolumn{2}{c}{\refEnvelope}\\%
            \multicolumn{2}{c}{%
                \includegraphics[width=\linewidth]{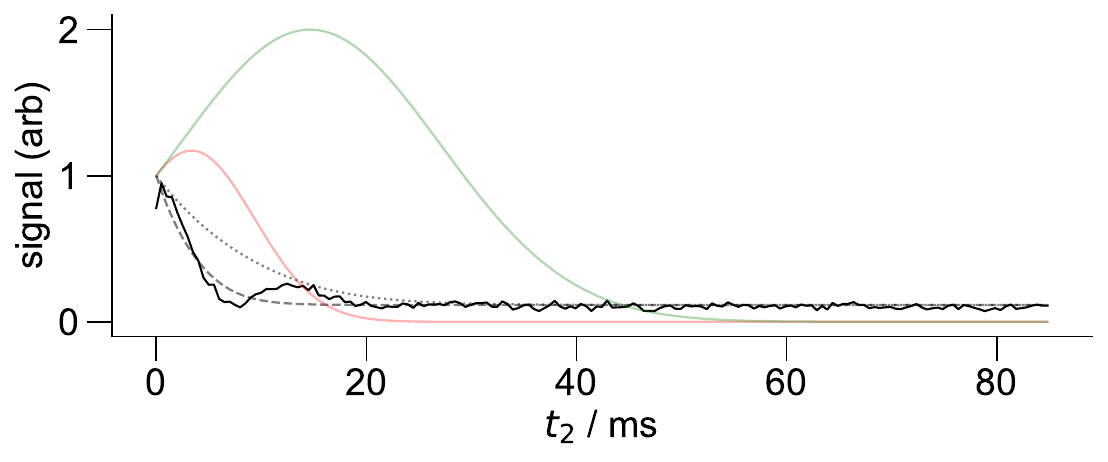}
            }
            \\
            \multicolumn{2}{c}{\refOneDRe} \\%
            \multicolumn{2}{c}{%
            \includegraphics[width=\linewidth]{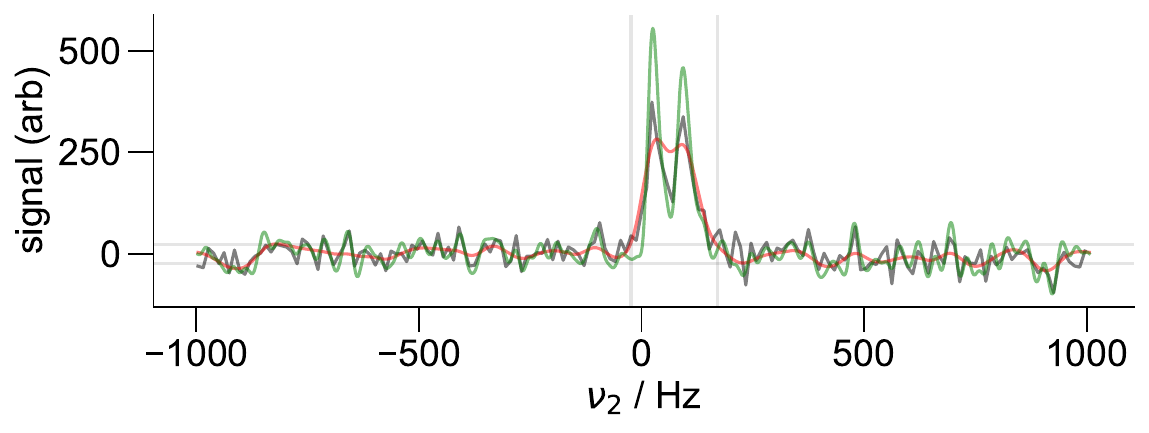}
            }
        \end{tabular}
        \caption{NMR signal for an inversion recovery dataset
            of a 4.8~μL toluene sample, showing
            the subset of the \gls{dcct} map
            \refBefore: before and \refAfter:
            after equal linewidth \gls{l2g}
            apodization, which shows an
            improvement in resolution with no
            noticeable change in \gls{snr}.
            \refEnvelope:
            The \gls{fid} signal envelope (solid line)
            is fit to a model (dashed
            line) and then expanded to trace the
            envelope edge (dotted line),
            along with the resulting apodization
            functions with equal energy \gls{l2g}
            shown in red
            and equal linewidth \gls{l2g} shown in green.
            \refOneDRe: The real 1D spectra 
            demonstrates the application of
            filters from \refEnvelope,
            where red corresponds again to equal
            energy and green to equal
            linewidth,
            compared against the original 1D spectra
            in black.
            The grey horizontal lines superimposed on the spectra
            indicate the noise levels
            (from the fit in \refEnvelope)
            and the vertical lines suggest a choice of
            integration bounds.
        }
        \label{fig:GaussianApodizationReal}
\end{figure}}
\newcommand{\figAlignIR}{\begin{figure}[tbp]
    \centering
    \includegraphics[width=\linewidth]{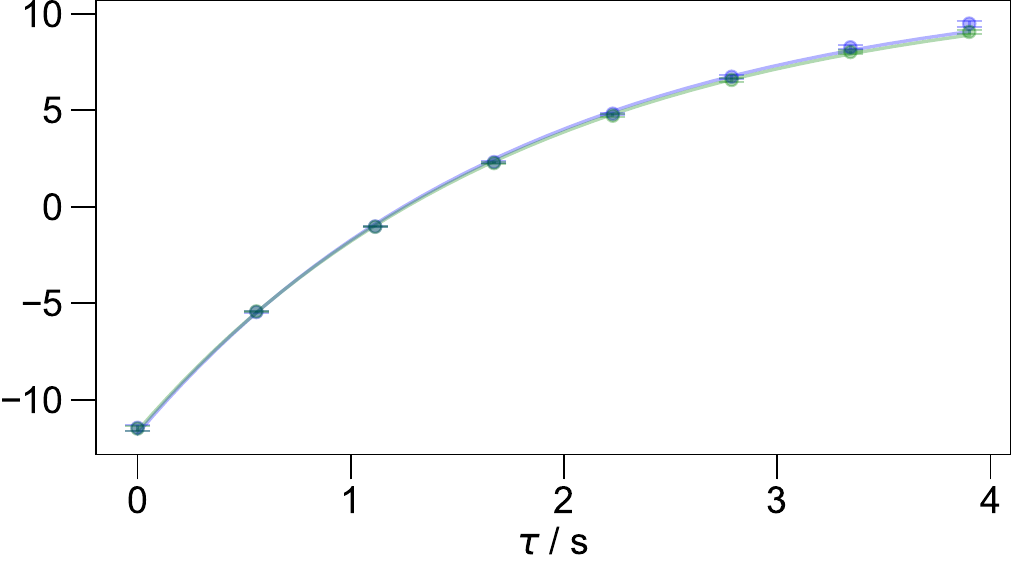}
    %
    \caption{
        Integrated inversion recovery signal (with
        interpulse recovery delay $\tau$)
        before correlation alignment in
        blue and after alignment in green.
        The integrals before and after alignment
        show essentially no difference, apart from
        tighter error bounds signifying that the
        alignment improves experimental error. 
    }
    \label{fig:AlignIR}
\end{figure}}

\newcommand{\figAlignEp}{\begin{figure}[tbp]
    \subfig{fig:AlignEpOverlay}{Overlay}
    \subfig{fig:AlignEpComparison}{Comparison}
    \centering
    \begin{tabular}{cc}
        \refOverlay
        &
        \setlength{\templen}{\dimexpr(\linewidth-2em)\relax}%
        \hspace*{-2em}%
        \raisebox{-\height}{ 
            \includegraphics[width=\templen]{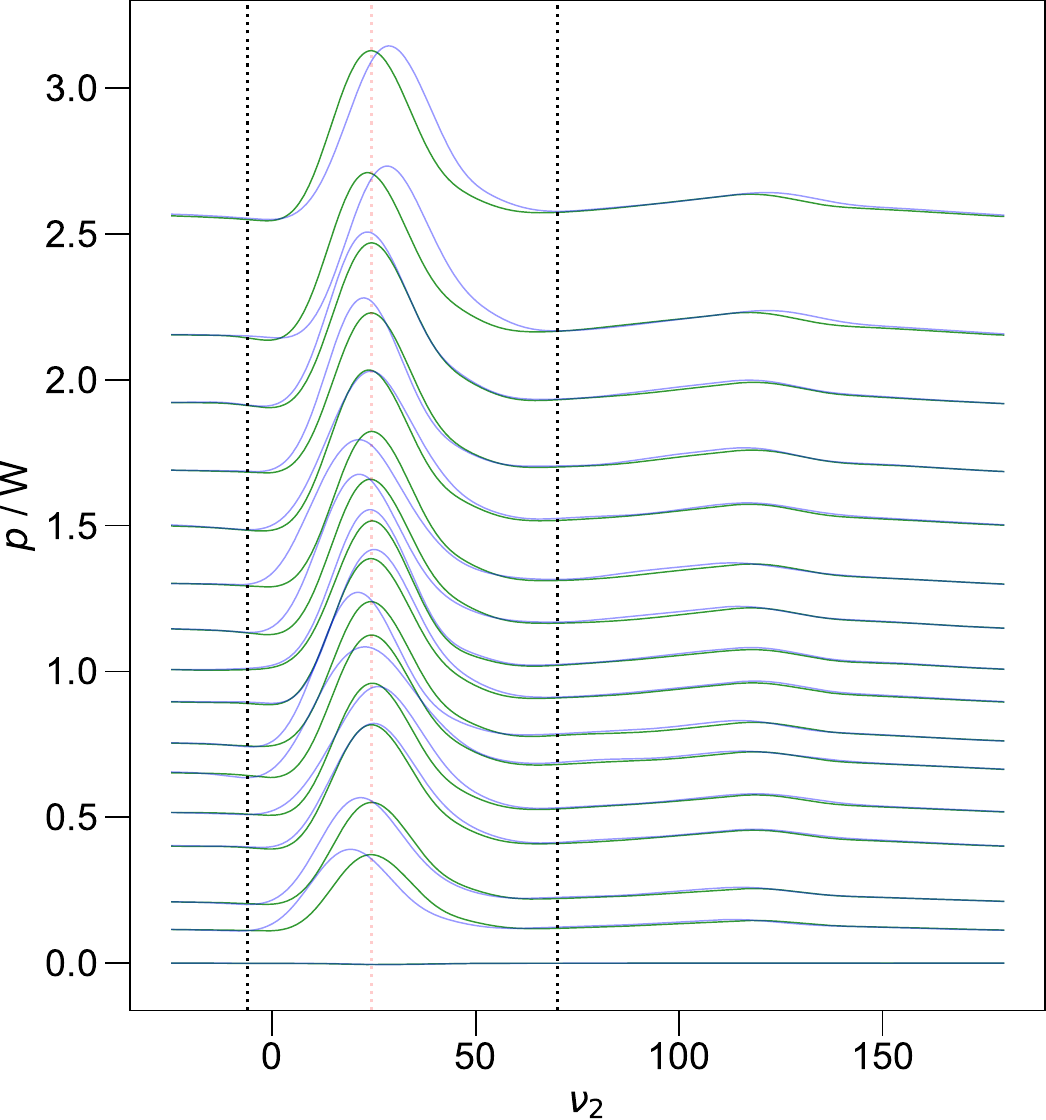}
        }
        \\
        \refComparison
        &
        \setlength{\templen}{\dimexpr(\linewidth-2em)\relax}%
        \hspace*{-2em}%
        \raisebox{-\height}{ 
            \includegraphics[width=\templen]{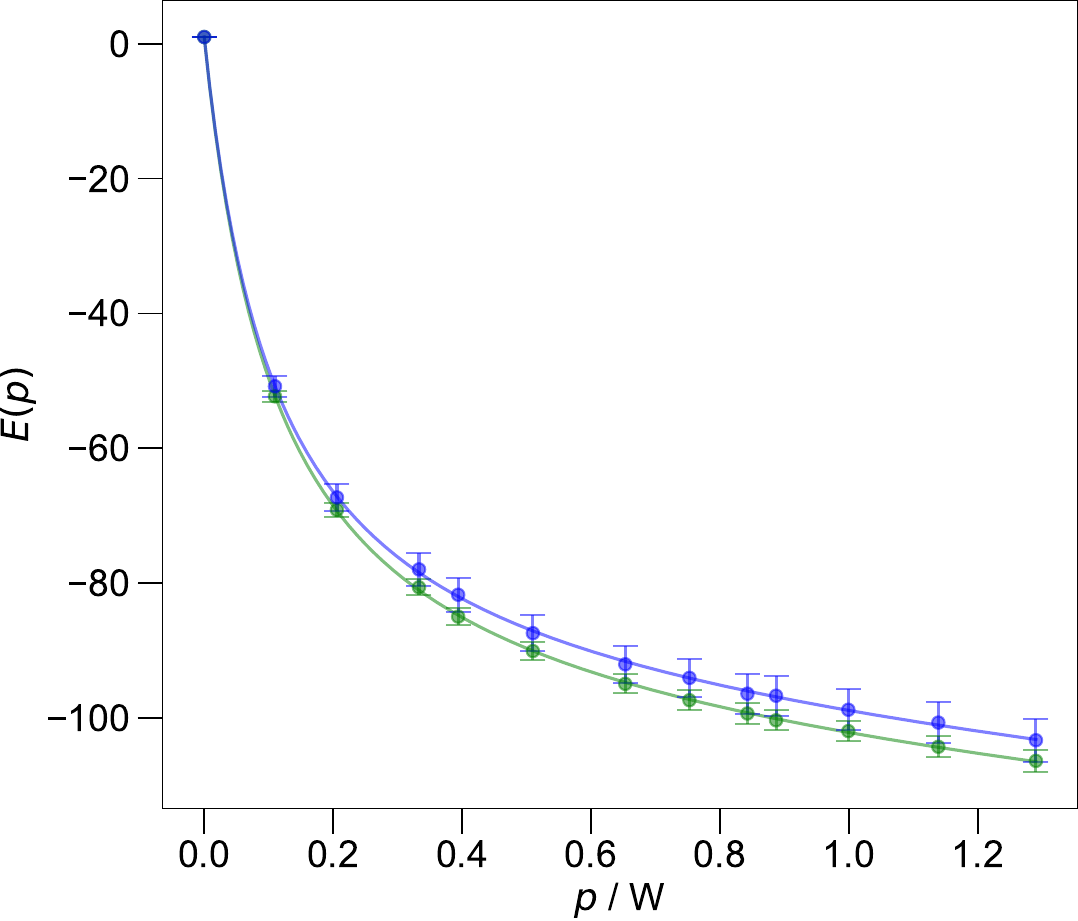}
        }
    \end{tabular}
    \caption{
        The NMR signal intensity
        is determined for
        a progressive
        enhancement experiment on
        6~mM aqueous TEMPOL
        (same data as \cref{fig:StepByStepEp}),
        both without (blue) and with (green)
        alignment.
        \refOverlay: The real part of the spectrum after slicing out the
        \gls{fid} and applying equal-energy
        \gls{l2g} apodization
        with integration limits
        indicated, before alignement in blue and
        after alignment in green.
        The dashed vertical lines provide the integration
        bounds,
        while the red line provides a guide to the eye.
        All signal has been inverted
        and two scans have been omitted
        for the sake
        of presentation clarity,
        \refComparison:
        Normalized integral intensities are
        denoted by circles with
        accompanying error bars,
        while the smooth line gives the least squares
        fit to \cref{eq:epsilon_p}.
        Both sets of data were phase corrected. 
        In both cases, error propagation of the noise
        in inactive coherence channels give the noise
        of the integrals,
        which standard error propagation formulas then
        modify upon normalization.
        Correlation alignment refines integration bounds,
        overall improving \gls{snr}.
    }
    \label{fig:AlignEp}
\end{figure}}

\newcommand{\figNutationResults}{
\begin{figure}[tbp]
    \centering
    \subfig{fig:NutationResultsTdom}{tdom}
    \subfig{fig:NutationResultsFdom}{fdom}
    \begin{tabular}{ll}
        \reftdom
        & \raisebox{-\height+1ex}{ 
            \includegraphics[width=\linewidth-2em]{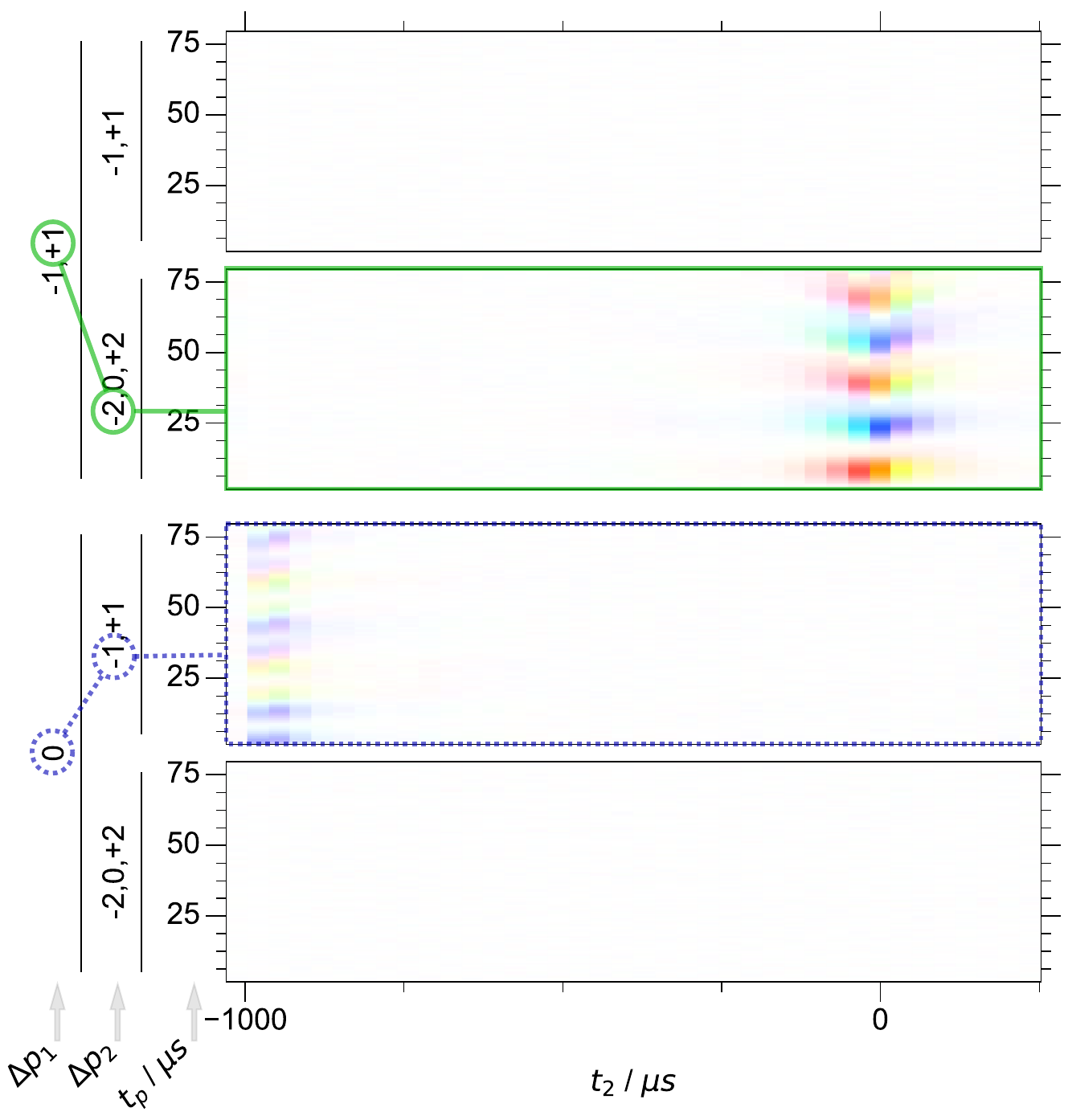}
        }\\
        \reffdom
        & \raisebox{-\height+1ex}{ 
            \includegraphics[width=\linewidth-2em]{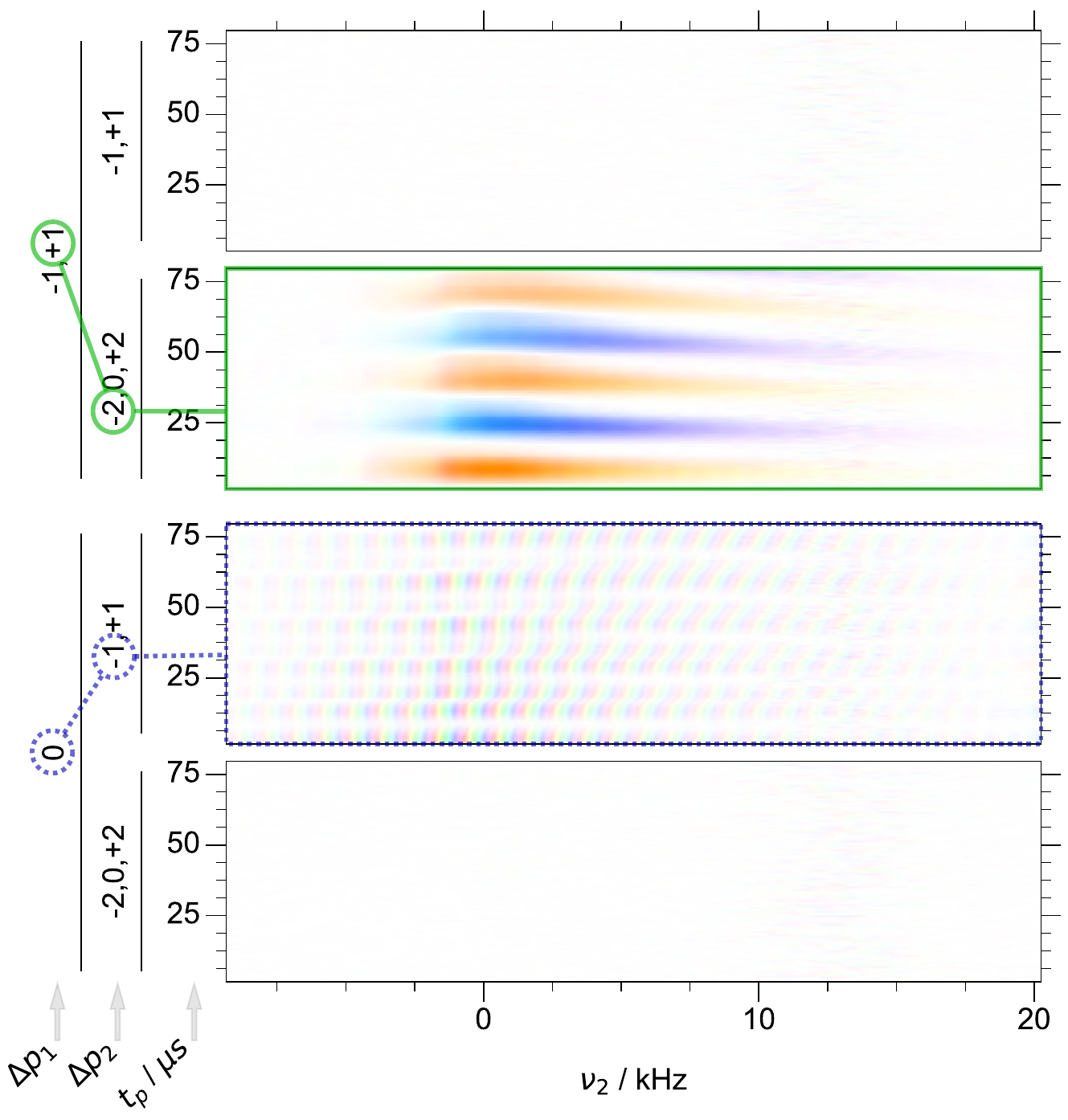}
        }
    \end{tabular}
    \caption{
        Full \gls{dcct} map 
        from a
        ($t_p$ pulse)-($\tau$ delay)-($2 t_p$
        pulse)
        sequence,
        where $t_p=2-80\;\text{μs}$
        and $\tau=1\;\text{ms}$ in both the
        \reftdom: time domain and \reffdom:
        frequency domain .
        The green
        boxes outline the
        echo \gls{ct} pathway (presented exclusively in
        \cref{fig:DomainColorIntro}),
        while the dotted blue boxes outline the
        \gls{fid}-like \gls{ct} pathway. These pathways follow the coherence 
        diagram outlined in 
        \cref{fig:ScopeDataCoh}
        alongside the pulse sequence.
        The only adjustments made to the raw data are
        (1) slicing to a frequency bandwidth where
        signal is observed and (2) setting $t_2=0$ to
        the maximum of the echo coherence pathway.
    }
    \label{fig:NutationResults}
\end{figure}
}
\newcommand{\figSolenoidRelaxometry}{
    \begin{figure}[tbp]
        \centering
        \href{https://jmfrancklab.slack.com/archives/CLMMYDD98/p1607623552020000}{Alex
        is working on this}
        \caption{A solenoid probe and standard \gls{odnp}
        probe yield indistinguishable $T_1$ relaxation
        rates,
        but the solenoid probe provides a higher sample
        volume,
        a better filling factor,
        and permits a design with better shielding from
        electromagnetic interference,
        yielding much better \gls{snr} and more accurate
        determination of the $T_1$ relaxation constant.
        }
        \label{fig:SolenoidRelaxometry}
    \end{figure}
}
\newcommand{\figScopeOneD}{\begin{figure}[tbp]
    \centering
    \subfig{fig:scopeOneDDeadtime}{scopeDataDeadtime}
    \subfig{fig:scopeOneDDeadtimeAlt}{scopeDataDeadtimeAlt}
    \begin{tabular}{cc}
        \refscopeDataDeadtime
        & \hspace*{-2em}\raisebox{-\height+1ex}{ 
            \includegraphics[width=\linewidth-2em]{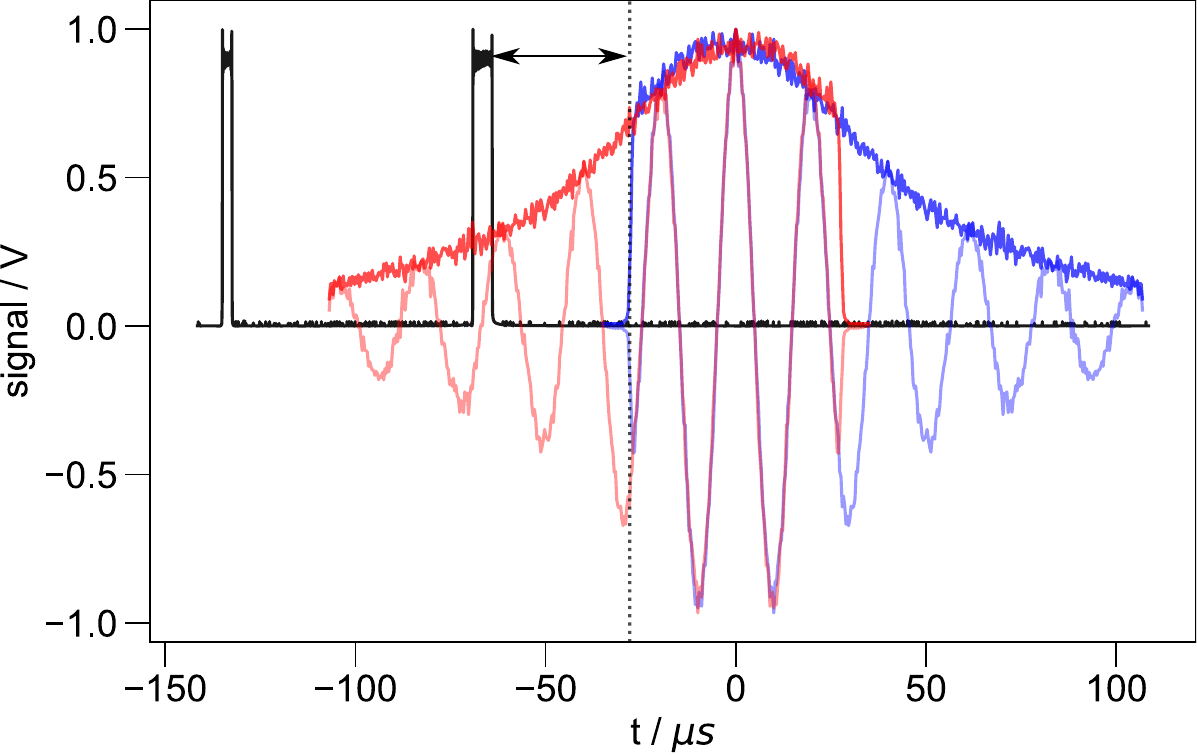}
        }
        \\
        \refscopeDataDeadtimeAlt
        & \hspace*{-2em}\raisebox{-\height+1ex}{
            \includegraphics[width=\linewidth-2em]{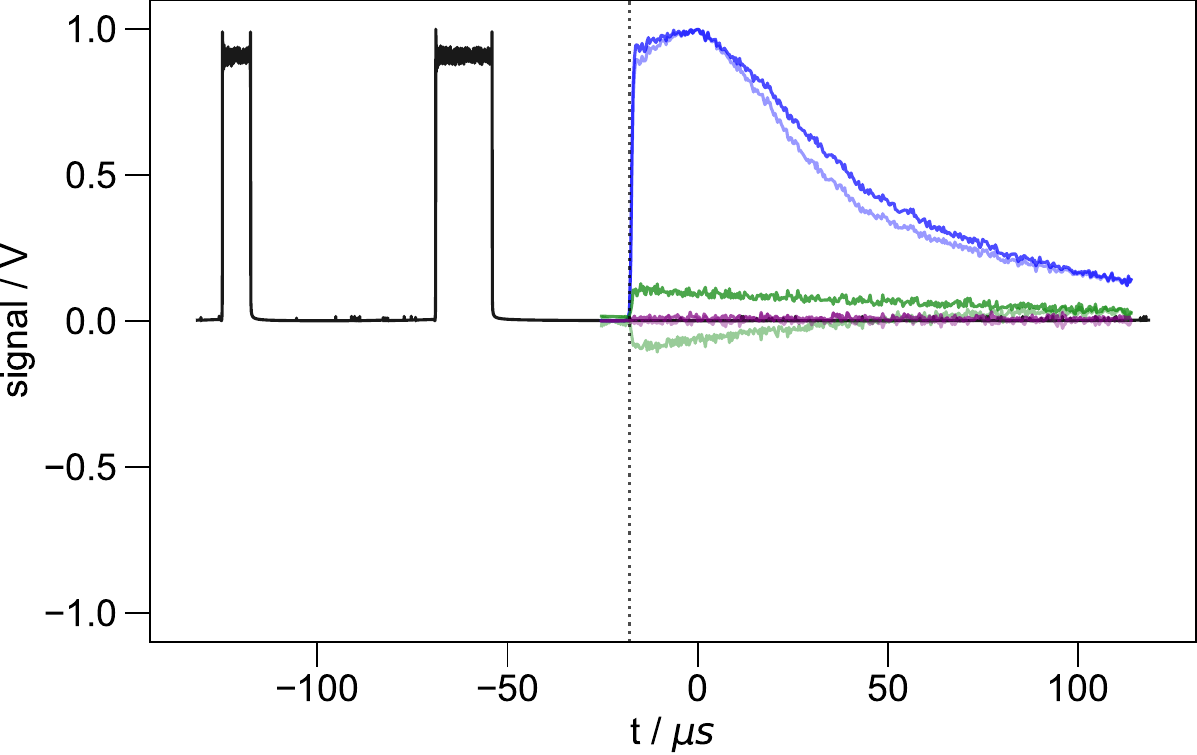}
        }
        \\
    \end{tabular}
    \centering
        \caption{
            \refscopeDataDeadtime: 
            The subset of signal from \cref{fig:ScopeDataMisset}
            that follows the echo coherence pathway of
            \cref{fig:ScopeDataCoh}
            is shown, with the complex magnitude of the pulse waveform
            (black) and signal (blue, with the real part in
            fainter blue)
            captured by the oscilloscope
            shown as 1D line plots
            and
            with $t=0$ set to the echo center.
            A comparison to the Hermitian conjugate of the signal
            (red) clearly illustrates the time at which
            the signals diverge. 
            Due to $T_2^*$ (inhomogeneous) decay
            over the timescale of the deadtime
            (dashed black vertical line, 35.7 μs),
            the signal loses some 25\% of its
            amplitude. 
            In \refscopeDataDeadtimeAlt,
            the pulse
            length has been optimized and the carrier
            frequency set on resonance, showing
            the echo pathway (blue), the residual
            \gls{fid} pathway (green), and the pathway which
            should not contain any signal
            (purple).
            Here, the echo pathway is much
            more intense than the unwanted \gls{fid}
            pathway.
        }
        \label{fig:scopeOneD}
    \end{figure}
}
\newcommand{\figRMdomainOverview}{
    \begin{figure}[tbp]
        \centering
        \includegraphics[width=\linewidth]{figures/RM_domain_phase.pdf}
        \caption{AOT \nts{XXXX}
        \nts{the averaged signal is broad},
        while the \nts{individual transients are
        essentially invisible via traditional plotting methods}}
        \label{fig:RMdomainOverview}
    \end{figure}
}

\newcommand{\figDomainColorIntro}{\begin{figure}[tbp]
        \centering
        \subfig{fig:DomainColorIntroWheel}{wheel}
        \subfig{fig:DomainColorIntroNutation}{nutation}
        \subfig{fig:DomainColorIntroNutationcorr}{nutationcorr}
        \begin{tabular}{rc}
            \refwheel
            & \hspace*{-2em}\raisebox{-\height+1ex}{ 
                \includegraphics[scale=0.4]{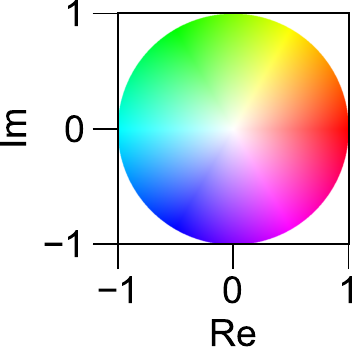}
            }
            \\
            \refnutation 
            & \hspace*{-2em}\raisebox{-\height+1ex}{ 
                \includegraphics[width=\linewidth-2em]{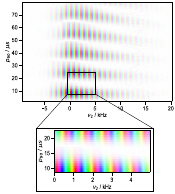}
            }
            \\
            \refnutationcorr 
            & \hspace*{-2em}\raisebox{-\height+1ex}{ 
                \includegraphics[width=\linewidth-2em]{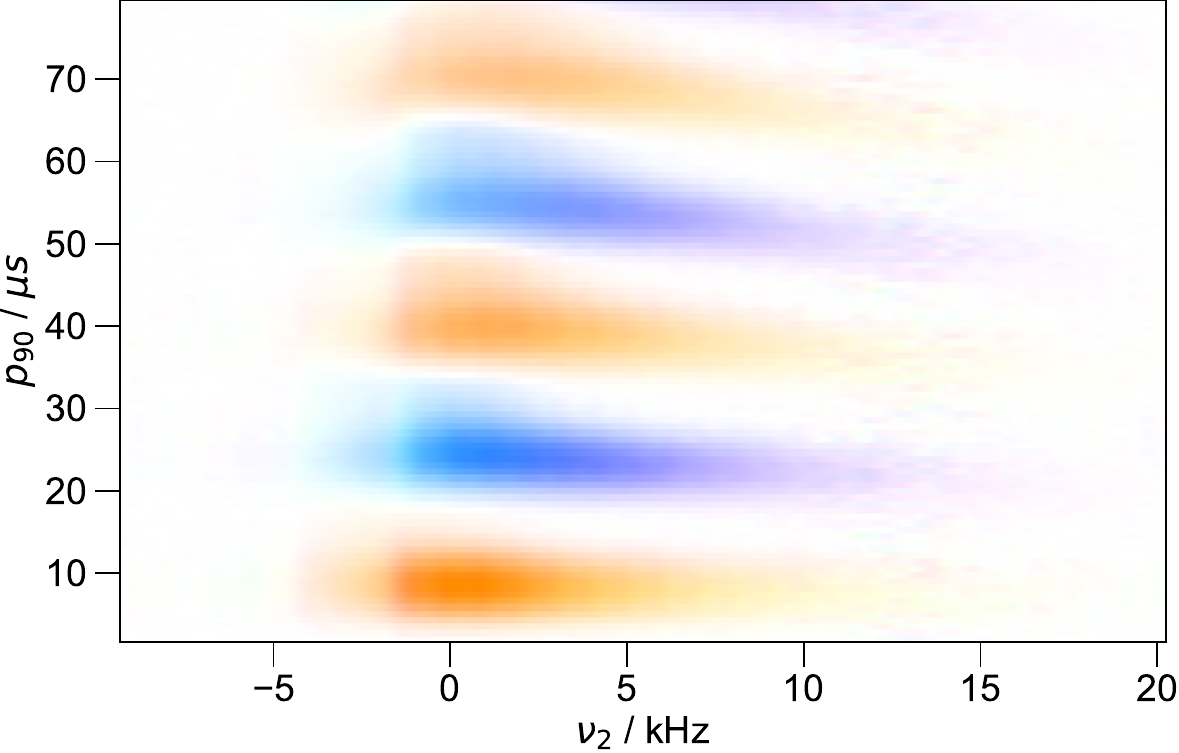}
            }
        \end{tabular}
        \caption{
            \refwheel: The coloring of
            data in the complex plane used in the
            \gls{dcct} map,
            where intensity indicates magnitude,
            and color indicates phase.
            \refnutation\
            The raw signal from an echo-based
            nutation curve
            after selection of the appropriate
            coherence pathway.
            Frequency-dependent phase shifts manifest
            in the domain coloring display as a rainbow
            banding (inset), here with a
            periodicity of 5 cyc/5 kHz = 1 ms.
            When zoomed out,
            the image may appear grey,
            which indicates a rapid variation in colors
            (phases) in the grey area.
            The sudden change in color
            from bottom to top along any
            particular column as the signal passes
            through a faint patch (amplitude near zero)
            indicates the inversion of the signal along 
            the indirect dimension.
            \refnutationcorr: Application of the 1 ms time shift
            yields coherently phased signal.
        }
        \label{fig:DomainColorIntro}
    \end{figure}
}

\newcommand{\figRevMicBefore}{\begin{figure}[tbp]
    \centering
    \begin{tabular}{cc}
    \includegraphics[width=\linewidth]{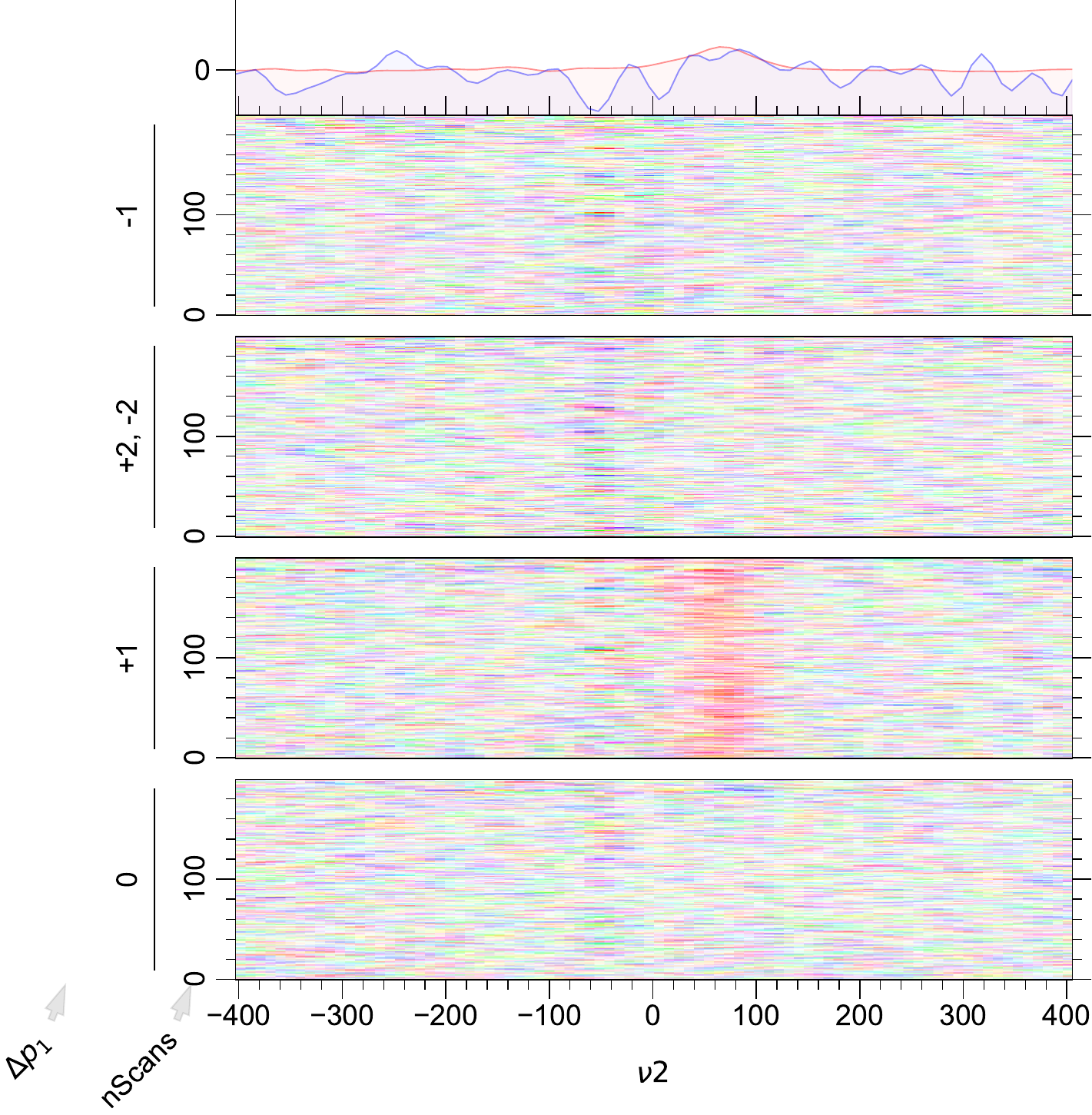}
    \end{tabular}
    \caption{A portion of the \gls{dcct} map for the signal from a reverse
        micelle sample, before undergoing the
        correlation alignment.
        The \gls{dcct} map allows the separate
        display of scans,
        while also emphasizing the phase-coherent signal.
        This map only shows
        the $\Delta p_2 = -2$ portion of the
        coherence domain for the second pulse,
        as the other \gls{ct} pathways 
        include only noise similar to the
        noise-only pathways that are shown here. The signal 
        appears in red in the expected
        pathway but is very faint. The 1D plots 
        illustrate that signal is not observable 
        from a single phase cycled scan of 
        the 1600 scans (blue) and is faintly 
        detectable upon averaging over all 
        of the scans (red).
    }
    \label{fig:RevMicBefore}
\end{figure}}

\newcommand{\figRevMicAfter}{\begin{figure}[tbp]
    \centering
    \begin{tabular}{cc}
    \includegraphics[width=\linewidth]{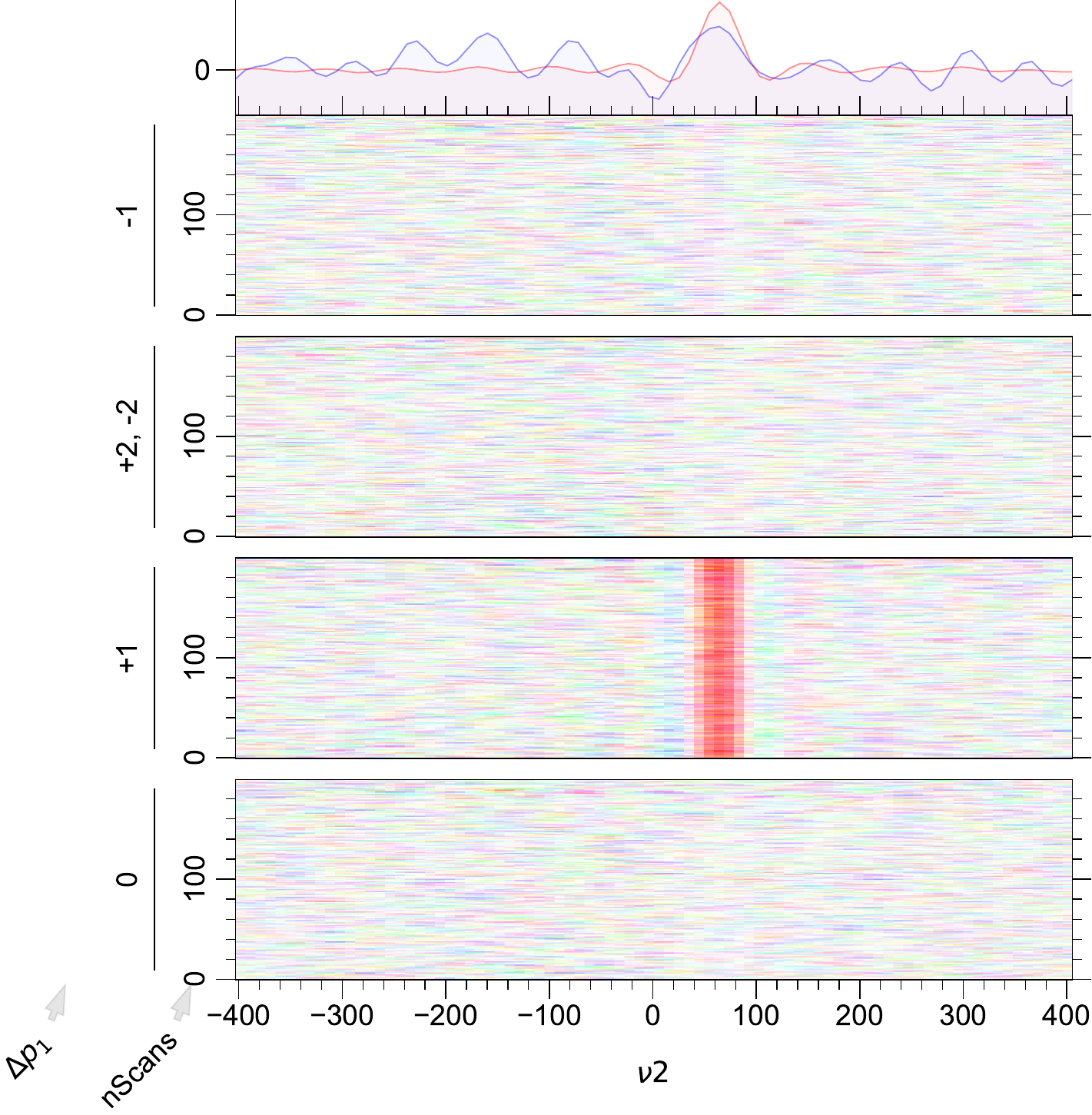}
    \end{tabular}
    \\
    \caption{
        The same portion of the \gls{dcct} map for the
        data in \ref{fig:RevMicBefore} after undergoing the
        correlation alignment in
        \cref{eq:2DcorrelationWithMask}.
        Signal concentrates
        into a bright red band in the appropriate
        coherence pathway.
        The 1D plots illustrate that
        signal is marginally observable from a single
        scan out of the 1600 scans (blue);
        this arises from the alignment of the various
        transients in the phase cycle.
        An average taken over all scans significantly
        improves the signal amplitude (red)
        and does so with much greater efficiency when
        the signal has been previously aligned than in
        the absence of alignment.
    }
    \label{fig:RevMicAfter}
\end{figure}}

\newcommand{\figMask}{\begin{figure}[htbp]
    \centering
    \subfig{fig:Mask}{mask}
    \begin{tabular}{cc}
    \includegraphics[width=\linewidth]{figures/RM_alignment_mask.pdf}
    &
    \end{tabular}
    \caption{The mask applied to the alignment
        data for a spin echo pulse sequence.
    }
    \label{fig:Mask}
\end{figure}}

\newcommand{\figRevMicZoom}{\begin{figure}[tbp]
    \subfig{fig:RevMicZoomBefore}{before}
    \subfig{fig:RevMicZoomAfter}{after}
    \centering
    \begin{tabular}{rc}
        \\
        \refbefore
        & \raisebox{-\height+1ex}{ 
            \includegraphics[width=\linewidth-2em]{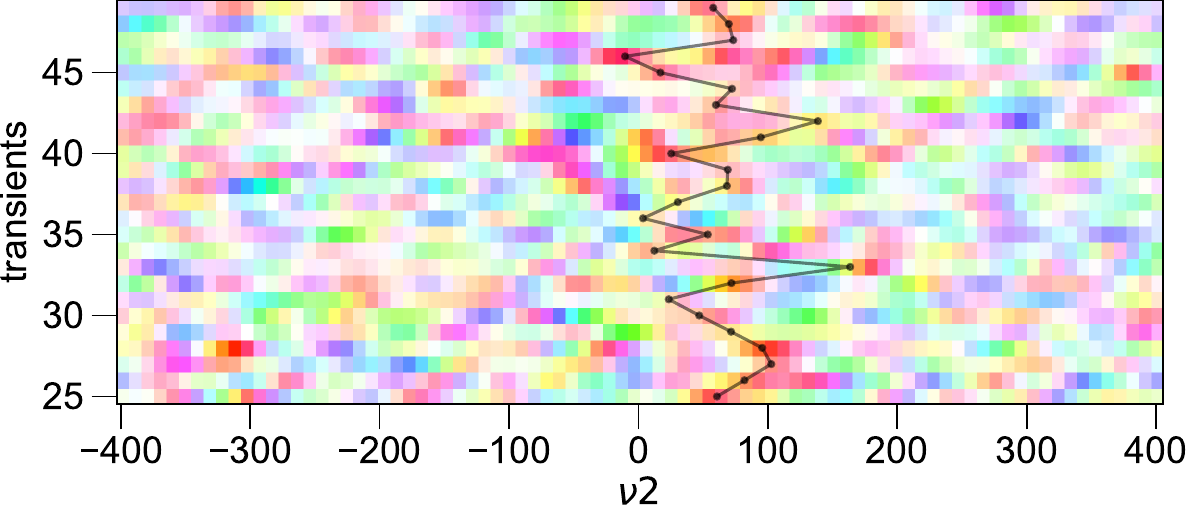}
        }
        \\
        \refafter
        & \raisebox{-\height+1ex}{ 
            \includegraphics[width=\linewidth-2em]{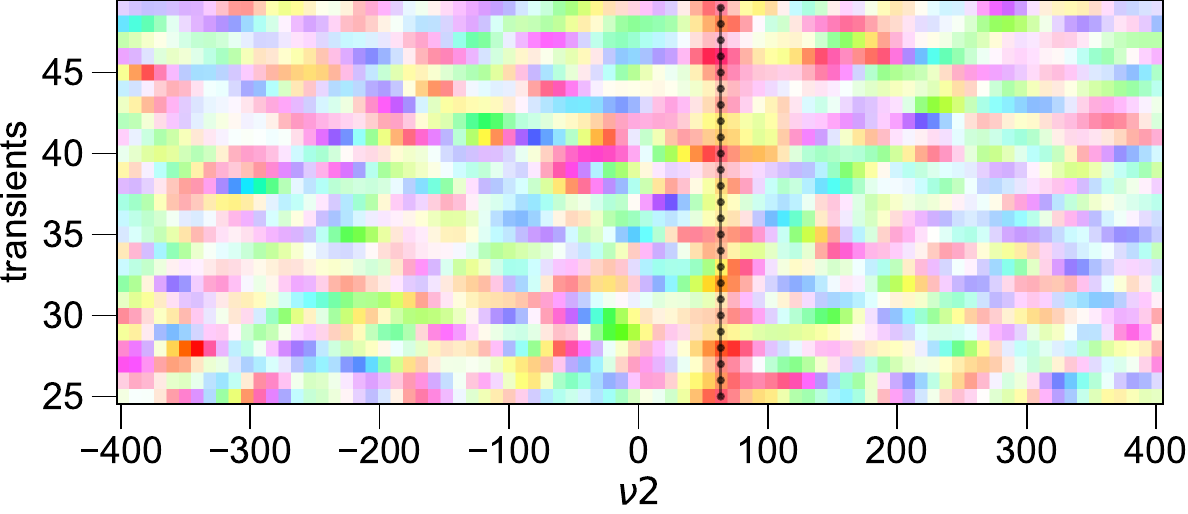}
        }
    \end{tabular}
    \caption{
        \refbefore\
        The \gls{dcct} map shows
        only 50 transients of the coherence channel of
        interest, for the reverse micelle sample
        in \cref{sec:exampleRM}
        before undergoing the correlation
        alignment. 
        The transients axis represents a direct
        product of the pulse phase dimensions
        and the nScans dimension of
        \cref{fig:RevMicBefore,fig:RevMicAfter}.
        A black line at 50 Hz
        indicates the expected signal offset.
        \refafter\
        After undergoing the correlation
        alignment, signal is clearly discernible from
        noise as a red streak across all scans
        centered at 50 Hz (indicated by the black line).
        Note that the alignment routine chooses the
        center of the signal to be not only typically in-phase
        red (0°) signal,
        but also that the signal $\sim
        40\;\text{Hz}$ to the left (smaller
        frequencies) of the chosen point is more
        frequently purple or blue ($-60°$ or $-120°$)
        when compared to frequency with random
        noise.
    }
    \label{fig:RevMicZoom}
\end{figure}}

\newcommand{\figPhCycleDemo}{\begin{figure}[tbp]
    \centering
    \subfig{fig:alignedPhCycleDemo}{aligned}
    \subfig{fig:misalignedPhCycleDemo}{misaligned}
    \begin{tabular}{cc}
        \refaligned 
        & \raisebox{-\height+1ex}{ 
            \includegraphics[width=\linewidth-1.5em]{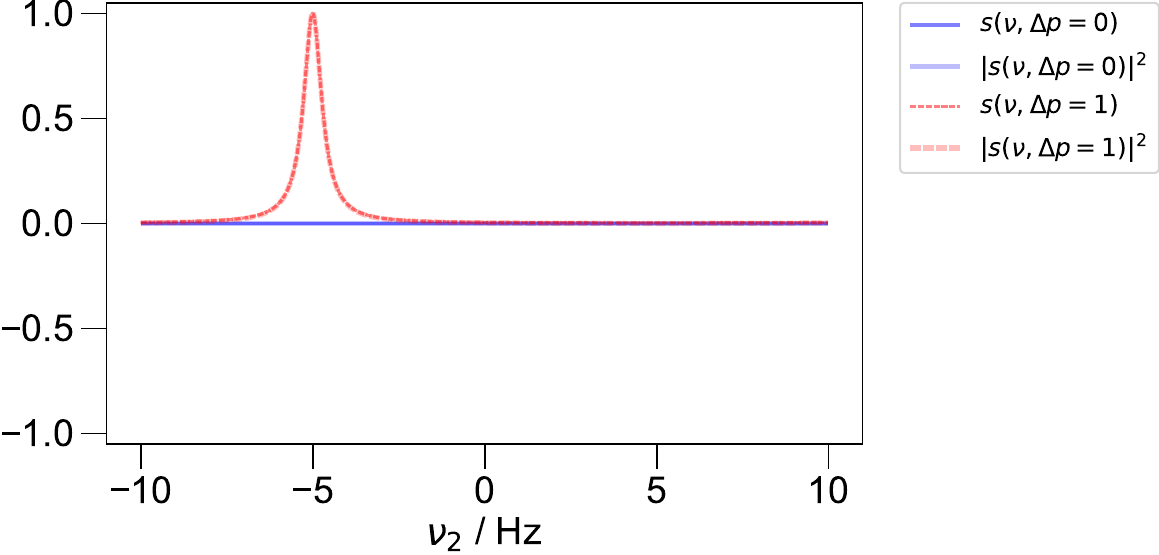}
        }
        \\
        \refmisaligned 
        & \raisebox{-\height+1ex}{
            \includegraphics[width=\linewidth-1.5em]{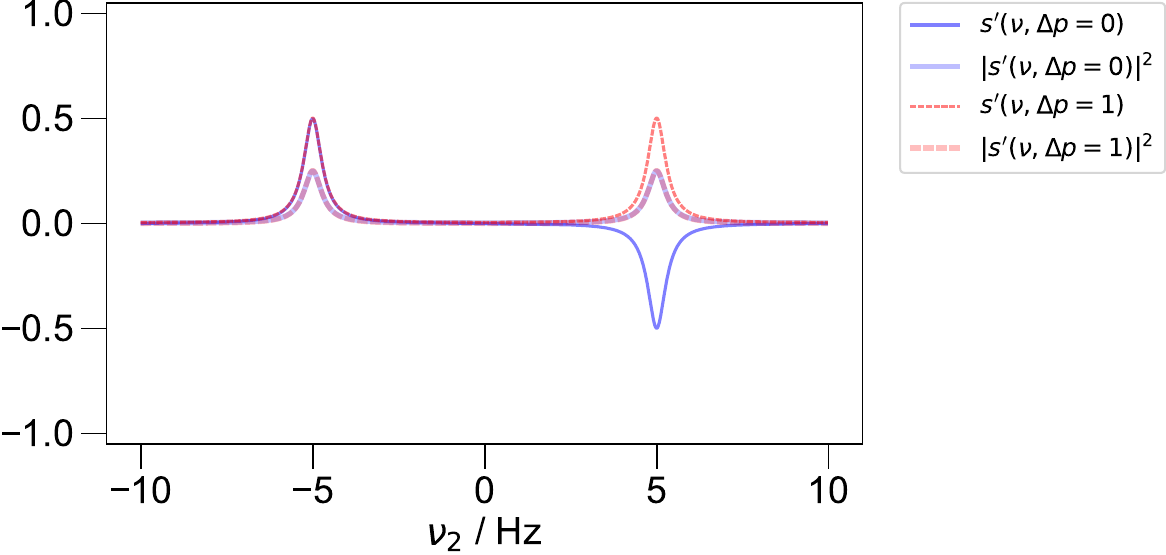}
        }
        \\
    \end{tabular}
    \centering
    \caption{
        A Lorentzian peak,
        following \cref{eq:simpleSignal} with
        $\lambda_L=2/\pi$,
        illustrates the simplest
        example of how field fluctuations
        (typically present in low-field data)
        impact phase cycled data
        and offers insight into a method
        of alignment.
        The resonance frequency remains
        fixed at $-5\;\text{Hz}$ for both steps of the
        phase cycle in \refaligned,
        while the resonance frequency
        for the two steps of the
        phase cycle is set to
        $-5\;\text{Hz}$
        and
        $+5\;\text{Hz}$, respectively,
        in \refmisaligned,
        simulating misaligned data.
        Dashed red represents the desired coherence pathway
        ($\Delta p = +1$) and solid blue represents
        the undesired pathway ($\Delta p = 0$).
        The bolder lines show the data's magnitude
        squared,
        and,
        for the choice of $\lambda_L=2/\pi$,
        the energy of the peak
        (the area under the magnitude squared)
        is equal to the peak height.
        The lineshapes plotted here emphasize that the
        overall signal energy of the
        data remains preserved, whether the data is
        aligned or not, even though misalignment
        spreads signal intensity over a wider
        bandwidth in both the frequency ($\nu$) and
        coherence ($\Delta p$) domains.
    }
    \label{fig:PhCycleDemo}
\end{figure}}

\newcommand{\figPhCycleNoise}{\begin{figure}[tbp]
    \centering
    \subfig{fig:PhCycleNoiseBefore}{PhCycleNoiseBefore}
    \subfig{fig:PhCycleNoiseAfter}{PhCycleNoiseAfter}
    \begin{tabular}{cc}
        \refPhCycleNoiseBefore 
        & \hspace*{-2em}\raisebox{-\height+1ex}{ 
            \includegraphics[width=\linewidth-2em]{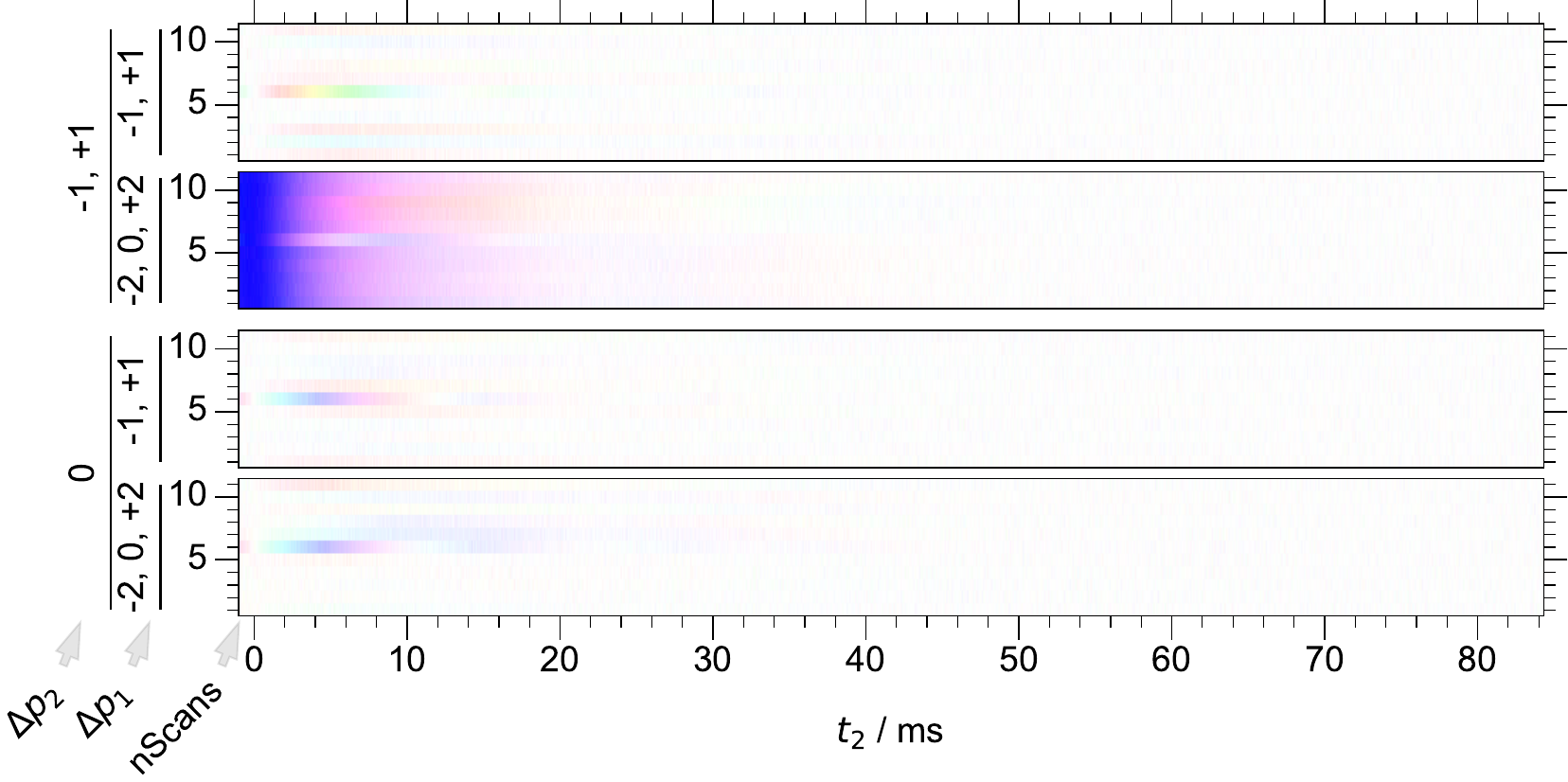}
        }
        \\
        \refPhCycleNoiseAfter
        & \hspace*{-2em}\raisebox{-\height+1ex}{
            \includegraphics[width=\linewidth-2em]{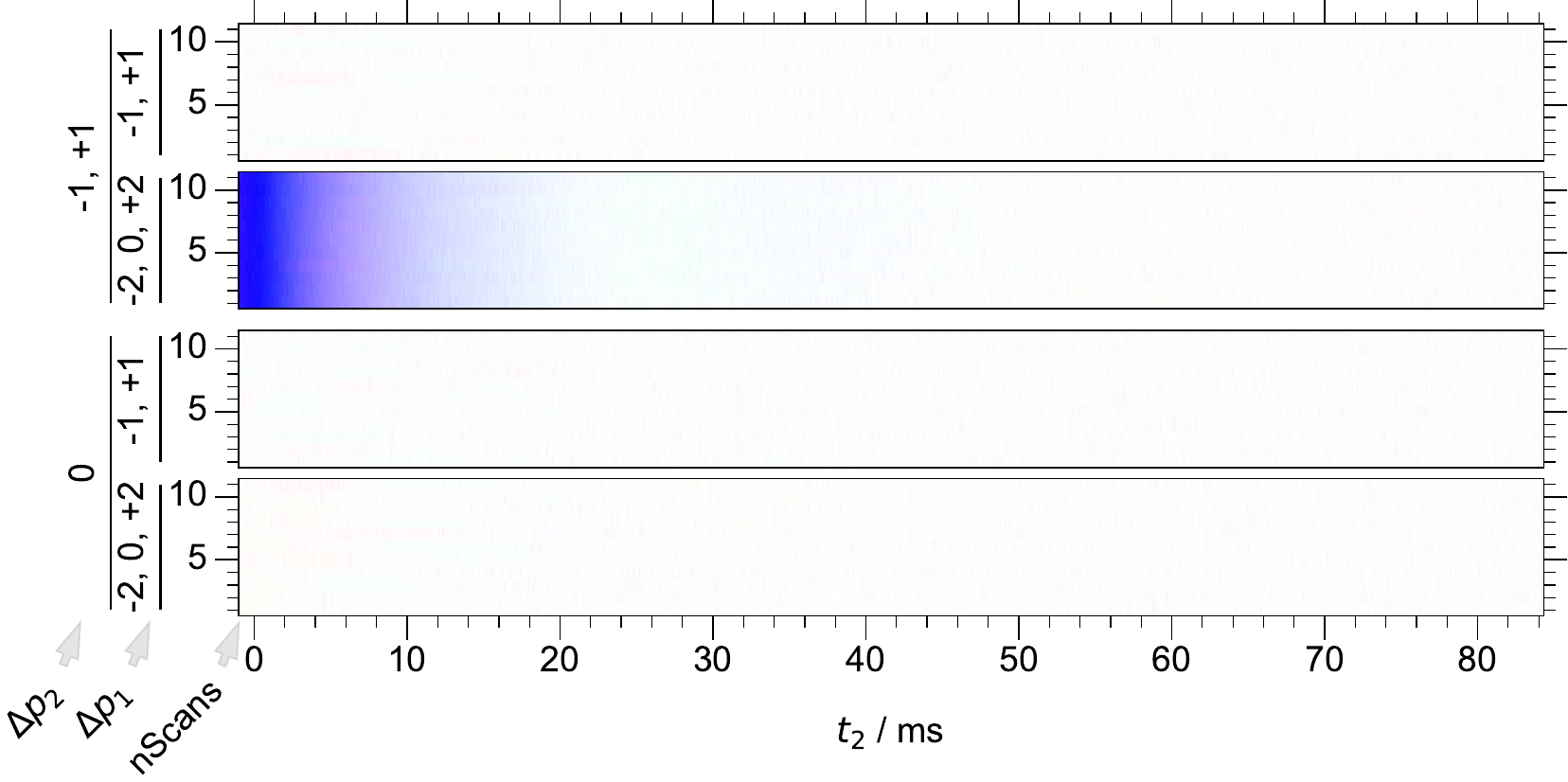}
        }
        \\
    \end{tabular}
    \centering
    \caption{The \gls{dcct} map (time domain) of a spin echo for
        10 complete phase cycles
        with first-order phase correction.
        The undesired coherence pathways contain a
        significant amount of phase cycling noise
        (\refPhCycleNoiseBefore: note especially the
        $6^{th}$ scan),
        which grows in amplitude moving away from
        $t=0$ before falling off as the signal phase
        off.
        The alignment procedure 
        almost completely mitigates this phase cycling
        noise \refPhCycleNoiseAfter, as discussed in
        the text.
    }
    \label{fig:PhCycleNoise}
\end{figure}}

\newcommand{\figModCoilE}{\begin{figure}[tbp]
    \centering
    \subfig{fig:ModCoilEnergyAttached}{attachedEnergy}
    \subfig{fig:ModCoilEnergyDetached}{detachedEnergy}
    \begin{tabular}{cc}
        \refattachedEnergy
        & \raisebox{-\height+1ex}{ 
            \includegraphics[width=0.8\linewidth]{figures/TEMPOL_capillary_probe_16Scans_ModCoil_energy.pdf}
        }
        \\
        \refdetachedEnergy
        & \raisebox{-\height+1ex}{
            \includegraphics[width=0.8\linewidth]{figures/TEMPOL_capillary_probe_16Scans_noModCoil_energy.pdf}
        }
        \\
    \end{tabular}
    \centering
    \caption{Inactive coherence pathways with
        modulation coil attached (\refattachedEnergy)
        and detatched (\refdetachedEnergy).
    }
    \label{fig:ModCoilEnergy}
\end{figure}}

\maketitle
\section{Introduction}
\glsresetall
\paragraph{new way of looking at coherence pathways -- drill down echoes}
This manuscript focuses on a comprehensive, non-standard approach to
    processing and
    presenting data from the various coherence pathways
    accessed by a \gls{mr} experiment.
This formalized approach--\ie, ``schema''--significantly
    improves the speed with which
    spectroscopists can develop
    new types of experiments.
When confronted with adverse experimental
    circumstances,
    it also guides the
    identification of data and the
    development of routines that improve the
    quality of that data.
As the authors' lab has developed an \gls{odnp} system,
    these methods serve here to improve the quality of
    \gls{odnp} data,
    with particular focus on mitigating the
    effects of inhomogeneity and time instability of the fields
    offered by conventional room temperature
    electromagnets.
\paragraph{what is ODNP + its challenges}
Despite the rapid,
    sub-nanosecond exchange of hydration water
    with bulk water,
    \gls{odnp} has the capability to measure the
    variation in
    the dynamics of hydration
    water molecules at specific sites,
    selected with nanometer-scale
    specificity~\cite{Armstrong_jacs,FranckPNMRS},
    as well as
    the accessibility of those
    sites to water~\cite{Doll2012,Segawa2016}.
The spectroscopist can choose sites
    based on the location of a small molecule within a
    heterogeneous mixture,
    by linking labels to the surface of a macromolecule,
    or by linking them within the core of a macromolecule
    (or macromolecular assembly)~\cite{Uberruck2018,Ortony_njp}.
Previous literature has advocated \gls{odnp}
    as a tool for analyzing
    the hydration layer for more than 12
    years~\cite{Armstrong_portable,Franck2013}.
Furthermore, the technique employs
    relatively low magnetic fields
    (typically 0.35~T),
    permitting dissemination in concert with cw~\gls{esr}
    spectroscopy~\cite{Armstrong_jacs,FranckPNMRS} and extension to permanent magnet
    systems~\cite{Armstrong_portable,neudertOverhauser2012,Uberruck2020},
    making it an eminently customizable technique.
Nonetheless,
    \gls{odnp} has not yet been widely adopted for
    studying the hydration layer,
    which we ascribe in part to the practical barriers
    imposed by the differences between \gls{odnp} and more
    traditional \gls{nmr} methods.
\paragraph{take a new perspective on NMR}
\gls{odnp} studies conducted at
    higher resonance
    frequencies~\cite{denysenkov_liquid_2012,neugebauerHighfield2014} tend to
    sample intramolecular motions
    of the spin label~\cite{Sezer_xi},
    or even correlated motions~\cite{Concilio2021}
    that obfuscate the translational
    dynamics that many studies seek to
    recover~\cite{FranckMethEnz2018}.
Thus,
    the study of dynamics via \gls{odnp}
    demands relatively low fields
    that provide low resonance frequencies
    tuned to
    the translational dynamics of solvent
    molecules.
Lower fields, of course, give rise to challenges
	with respect to spin polarization and signal
    amplitude in reference experiments.
The integration of
    the \gls{nmr} spectrometer
    with the \gls{esr} spectrometer and high power microwave amplifier
    required for \gls{odnp} also gives rise to a host of
    practical challenges.
Furthermore, a full analysis of the dynamics
    available from \gls{odnp} involves collection of both a series of
    measurements of spin
    polarization at different levels of microwave (\gls{esr})
    saturation (progressive enhancement),
    as well as several inversion recovery experiments
    acquired under different conditions to account
    for the slight heating,
    and subsequent change in relaxation rates,
    introduced by microwave irradiation.
Careful manual analysis of the resulting data generally demands a
    level of fastidiousness that is
    unreasonable given the quantity of data under
    consideration.
\paragraph{what we do vs. hardware}
For particular sample
    configurations,
    one can achieve reasonable \gls{odnp} results with
    mostly commercial
    hardware~\cite{krahn_shuttle_2010,valentine_reverse_2014,dey_unusual_2019,FranckMethEnz2018,FranckPNMRS}.
Many \gls{odnp} studies follow a strategy that employs
    standard approaches to phase cycling and where
    90°-pulse \gls{fid} experiments
    or, occasionally,
    standard CPMG (echo train) experiments quantify the signal
    intensity.
Recent studies have demonstrated spectral resolution
    without the need for added shim coils~\cite{Uberruck2020},
    as well as examples that implement a practicable shim
    stack~\cite{Keller2020}.
Importantly, thus far, all such advances detail
    engineering of the spectrometer device itself,
    rather than offering strategies for finer control
    of data visualization and processing,
    which
    could apply to all instruments.
This manuscript addresses the latter
    concern.
\paragraph{introduce DCCT map}
Overall, the development of
    \gls{odnp} spectroscopy and other emerging
    \gls{mr} techniques mandates
    a new approach to conventional \gls{nmr}.
In order to attend to all the previously mentioned
    problems,
    this manuscript
    carefully draws fundamental concepts of
    coherence transfer from the original
    literature~\cite{stejskalComparisons1974,stejskalData1974,houltCritical1975,bodenhausenSuppression1977,bainCoherence1984,Bodenhausen1984},
    and motivation from more recent
    efforts~\cite{ivchenko_multiplex_2003,schlagnitweitMQD2012,baltisbergerCommunication2012},
    to formalize a schema
    in which standard code
    can most effectively
    and quickly acquire, organize, and present
    signal for a wide range of contexts.
In particular,
    this schema offers a means for
    maximizing the information
    available from phase-cycled \gls{mr}
    experiments,
    while still presenting (plotting)
    and organizing (deploying object-oriented data structures)
    the data in a convenient fashion.
The approach to a formalized procedure for cycling the 
    phases of pulses
    goes back
    decades~\cite{stejskalComparisons1974,stejskalData1974,houltCritical1975,bodenhausenSuppression1977,bainCoherence1984,Bodenhausen1984}, 
    when \gls{ct} maps
    (also known as \gls{ct} pathway diagrams)
    were introduced
    as a means to 
    understand the multiple \gls{ct} pathways
    existing in a phase cycled pulse sequence.
However,
    traditional experiments have been primarily concerned with how
    to select the desired \gls{ct} pathway of interest
    quickly and efficiently while discarding all other
    information.
As previously
    noted~\cite{ivchenko_multiplex_2003,schlagnitweitMQD2012,baltisbergerCommunication2012},
    this choice was originally motivated by now-antiquated hardware.
Contributions toward utilizing the theory of quantum
    coherence
    transfer~\cite{wokaunSelective1977} and phase cycling~\cite{Bodenhausen1984}
    have been revisited in the past two decades on various
    occasions~\cite{ivchenko_multiplex_2003,gan_enhancing_2004,schlagnitweitMQD2012,baltisbergerCommunication2012,waudbyTwodimensional2020},
    and the recommendation of forgoing cycling of the receiver
    phase and separately storing each acquired transient
    has repeatedly arisen.
Such methods rely on post-acquisition processing to select
    the \gls{ct} pathway of interest.
This work builds on these contributions by
    also highlighting the value in
    assessing ``undesired'' \gls{ct} pathways by
    inspecting the other ``routes'' the signal follows
    throughout a phase cycled experiment.
\paragraph{}
In this contribution, the abbreviation \gls{dcct}
    refers to two key elements of the strategy:
    domain coloring for plotting the complex-valued data,
    and the formalization of \gls{ct}
    dimensions.
While the ``\gls{dcct} map'' refers to the colored
    map of
    the signal flowing through all coherence pathways,
    the ``\gls{dcct} schema'' refers to the overall strategy:
    not only plotting of the signal map,
    but also data organization and coding.
The \gls{dcct} schema is posed as an alternative to the
    more traditional
    schema that moves quickly toward discarding all
    but one coherence pathway.
\paragraph{outline -- structure of paper}
While the \gls{dcct} schema may be expected to aid in
    diagnosing instrumental shortcomings,
    in the applications outlined here,
    it has also proven useful for elucidating signal in
    systems and experiments that previously precluded
    standard \gls{odnp} analysis,
    thus widening the
    expanse of applications for \gls{odnp}.
More broadly, the \gls{dcct} schema should benefit not only \gls{odnp},
    but
    other \gls{mr} methods and, more
    generally, other forms of spectroscopy capable
    of accessing multiple different coherence
    pathways~\cite{linHighResolution2000,Raymer2013,shimFemtosecond2006a,deanCoherence2017}.
\paragraph{outline}
The paper is organized as follows:
    \cref{sec:theory}
    covers the theoretical basis of this paper,
    including some review,
    with emphasis on the mathematical basis of phase
    cycling (\cref{sec:phaseCycling}).
\cref{sec:dataProc} introduces the approach to data processing
    and rationalizes echo detection as yielding the
    benefits of the
    EXORCYCLE~\cite{bodenhausenSuppression1977},
    while permitting routine baseline-free spectral
    acquisition with accurate
    integrals (\cref{sec:derivingIntegrals})
    and phase corrections
    (\cref{sec:zerothOrder,sec:firstorderEcho}).
Next, \cref{sec:dataProc}
    presents the mathematics underlying a
    ``signal averaged''
    alignment routine (\cref{sec:crossCorrel}),
    the specific apodization techniques
    (\cref{sec:GaussianApodization}),
    and concludes with a review of \gls{odnp} theory
    (\cref{sec:ODNPtheory}).
Since compact scripting comprises part of the
    results presented here, equations are
    referenced against an appendix
    (\cref{sec:codeSnippets}) that offers a
    glimpse at the corresponding code.
\cref{sec:experimental} contains all relevant
    experimental details,
    such as sample preparation
    (\cref{sec:experimentalSample}).
It supplies details regarding the modular low field equipment (\cref{sec:minimalExp}),
    including a variation that relies only on a
    few hardware components of
    relatively low sophistication (\cref{sec:NMRwithoutMW}).
It briefly discusses the custom Python software which serves
    as a cornerstone of this work (\cref{sec:softwareExp}).
As novel data visualization/plotting techniques form a
    central component of the new approach,
    the results section (\cref{sec:results})
    begins with a demonstration
    (\cref{sec:show_all_data,sec:treatmentPhase}),
    along with subsequent demonstrations of
    specific \gls{nmr} experiments presented in
    \cref{sec:treatmentPhaseDCCT}.
This plotting technique proves particularly useful for
    experiments that rely on
    unsophisticated hardware,
    such as a ``bare-bones'' spectrometer
    made from non-specialized equipment
    (\cref{sec:data_on_scope})
    and a modular system relying on a
    SpinCore PCI board transceiver
    (\cref{sec:NutationCurve});
    time-variable magnetic fields present in many
    low-field and portable instruments
    also provide an opportunity to showcase the
    applicability of this visualization technique (\cref{sec:fieldInstab}).
However, it can also (\cref{sec:CPMG})
    demonstrate the cooperative effect
    of several pulses and can be deployed on standard
    commercial instrumentation.
The \gls{dcct} schema for organizing and plotting the data
    then motivates and facilitates the algorithms
    in \cref{sec:algorithmsDCCT},
    which presents
    phasing (\cref{sec:phasingDCCT}),
    correlation-based frequency alignment in the
    presence of phase cycling at low \gls{snr} (\cref{sec:alignDCCT}),
    and apodization (\cref{sec:lineshapeDCCT}).
Collectively, these processing procedures prove
    extremely flexible and adaptable to the
    inversion recovery (\cref{sec:exampleIR})
    and to the progressive enhancement (\cref{sec:exampleEp})
    experiments that are essential to recording data relevant to
    hydration dynamics.
The schema proves particularly
    useful in a uniquely low \gls{snr} case scenario
    (\cref{sec:exampleRM}),
    which highlights the utility of this means of data
    handling and presentation.
\cref{sec:discussion,sec:conclusion}
    place the schema
    in the broader context of existing \gls{nmr} literature
    and forecast future applications.
In particular,
    these techniques enable synergy between data
    acquisition and processing, which relies less
    on starting infrastructure and allows for
    greater
    sophistication in the processed results.
Looking forward, the \gls{dcct} map and the resulting signal
    optimization techniques facilitate not only
    improvements in processing methodologies such as those presented
    here but also provide the framework for many
    future advances.

\section{Theory}\label{sec:theory}
\subsection{Data Processing}\label{sec:dataProc}
Several sections below utilize the notation
\begin{equation*}
    c(\Delta x) = f(x)\star g(x)
\end{equation*}
for the correlation function, such that:
\begin{equation}
    c(\Delta x)
    =
    \int_{-\infty}^{\infty} f^*(x)
    g(x+\Delta x) dx
    \label{eq:CorrelSymbDef}
\end{equation}
Aside from providing compactness,
    this notation emphasizes the fact that
    a Fourier domain multiplication
    ($\tilde{c}(\nu) = \tilde{f}^*(\nu)\tilde{g}(\nu)$)
    significantly outperforms the numerical integration
    of \cref{eq:CorrelSymbDef}.
\subsubsection{Phase Cycling}\label{sec:phaseCycling}
\pdfcommentG{added refcheck -- see
https://mirrors.rit.edu/CTAN/macros/latex/contrib/refcheck/refdemo.pdf} 
Throughout,
    the software utilizes the standard
    relationship
    that~\cite{Bodenhausen1984}:
    \begin{equation}
        s(\Delta p,t)
        =
        \frac{1}{\sqrt{n_{\varphi}}}
        \sum_{j=0}^{n_\varphi-1} e^{-i 2 \pi \Delta p_j \varphi_j}
        s(\varphi_j,t)
        \label{eq:basicPhaseCycling}
    \end{equation}
    where
    $n_\varphi$ gives the number of phase cycle steps
    and,
    following standard notation,
    $\Delta p_j$ indicates the coherence change
    during pulse $j$.
The phase angle $\varphi_{j}$ has units of
    $[\text{cyc}]=[\text{rad}]/2\pi$, such that an $x$
    pulse has $\varphi=0\;\text{cyc}$ a $y$ pulse
    has $\varphi=0.25\;\text{cyc}$, \etc; the
    resulting $\Delta p_1$ are then unitless.
This manuscript employs the phrasing that
    \cref{eq:basicPhaseCycling} relates the ``phase
    cycling domain'' ($\varphi_{j}$) to the Fourier
    conjugate ``coherence transfer domain'' ($\Delta
    p_{j}$).
\paragraph{multiple pulses}
When the pulse sequence cycles the phase of multiple pulses,
    the signal in the coherence transfer domain,
    $s(\Delta p_1,\Delta p_2, \cdots,\Delta p_N, t)$,
    derived from an $N$-dimensional Fourier
    transform of the phase cycling domain signal,
    gives the component of the signal that changes
    by $\Delta p_1$ during the first pulse, $\Delta
    p_2$ during the second pulse, \etc
Thus, $s(\Delta p_1,\Delta p_2, \cdots,\Delta p_N, t)$
    includes the signal for all
    distinguishable coherence transfer pathways.
The $\Delta p$ dimensions are subject to standard
    Fourier aliasing, resulting from the Nyquist
    theorem.
When a single coherence transfer pathway
    is selected, this aliasing collapses to the
    well-known rules laid
    out in the seminal phase cycling
    contribution from Bodenhausen, Kogler, and
    Ernst~\cite{Bodenhausen1984,ernst_principles_1987}.
\subsubsection{Deriving Integrals from Echoes}\label{sec:derivingIntegrals}
\gls{odnp} relies on high-quality quantitative \gls{nmr} that
    requires properly phased signal without a baseline.
The inhomogeneous fields and low \gls{snr} conditions
    commonly encountered in \gls{odnp}
    pose some obstacles toward routinely acquiring
    data with correct phasing that avoids baseline
    artefacts.
Fortunately,
    the echo-based signal acquisition advocated in this work
    not only refocuses inhomogeneities
    but also circumvents baseline problems by
    recovering signal typically lost to the pulse
    deadtime,
    thus
    yielding distortion-free early time points of
    the \gls{fid}~\cite{rance_obtaining_1983}.
Echo-based signal acquisition also
    reaps the benefits provided by the EXORCYCLE~\cite{bodenhausenSuppression1977}
    and permits
    straightforward automated approaches for signal
    phase correction.
Typical 90° pulse and subsequent 180° pulse lengths are
    advantageously short ($\le 10\;\text{μs}$) in most
    \gls{odnp} systems,
    while $T_{2}$ times are long
    (hundreds of ms to s),
    garnering these benefits virtually free of
    cost.
\paragraph{slice asymmetric echo to get FID -- perhaps include earlier}
Phase correction routines
    identify the origin of the time axis
    ($t=0$), defined as the time point at which all the
    isochromats in the signal present the same
    phase~\cite{rance_obtaining_1983,callaghanPrinciples1994echoes}.
For an idealized (noise-free) echo,
    $t=0$ corresponds to the peak of the echo.
Echoes symmetric about $t=0$ yield purely real
    Fourier transforms,
but echoes generated in response to short echo times
    (which refocus shortly after the 180° pulse)
    can also be converted
    to \gls{fid}'s \via multiplication by an appropriate
    Heaviside function, $h(t)$:
\begin{equation}
    s_{FID}(t) = s(t) h(t) =
    \begin{cases}
        0 & t<0
        \\
        \frac{1}{2} s(t) & t=0
        \\
        s(t) & t>0
        .
    \end{cases}
    \label{eq:s_FID}
\end{equation}
Before the code implements \cref{eq:s_FID}
    it must 
    adjust
    the origin of the time axis,
    possibly by a non-integral multiple of the dwell
    time (the time-domain sampling interval).
The pySpecData Python library
    (developed in-house and used here throughout)
    facilitates these manipulations 
    by maintaining the time coordinates of the signal
    and providing a compact notation that selects
    and manipulates the signal based on its time
    coordinates (\cref{lst:FID}).
Upon Fourier transformation,
    the pySpecData library automatically
    generates an appropriate axis of frequency
    coordinates and,
    for time axes that do not begin at $t=0$,
    automatically multiplies in the frequency domain by the
    appropriate first-order phase shift.
\paragraph{}
The relatively simple treatment of short-time echoes
    contrasts with the behavior of FIDs arising in
    response to an isolated 90° pulse.
Assuming the frequency-domain signal comprises a
    superposition of Lorentzians,
    beginning acquisition on a time axis ($t'$)
    at some time point $t_d$ after
    the nominal center of the echo
    (\textit{s.t.} $t=t'+t_d$)
    results in
\begin{equation}
    \begin{array}{L}
        \sum_j
        e^{ i 2 \pi \nu_j t - R_j t }
        h\left( t-t_d \right)
        =
        \\
        \quad
        \quad
        \sum_j
        e^{i 2 \pi \nu_j t_d}
        e^{-R_j t_d}
        e^{i 2 \pi \nu_j t'-Rt'}
        h(t')
    \end{array}
    \label{eq:FIDdistortion}
\end{equation}
    -- \ie, each resonance changes slightly in
    amplitude and phase.
Of course, in mild cases,
    a uniform first-order phase shift can correct for
    the change in phases.
However, it is also known that issues arise from
    either distortion of the initial \gls{fid}
    datapoints~\cite{rance_obtaining_1983}
    as well as from the interference of
    nearby peaks
    (especially their dispersive components).
These issues are particularly noticeable
    when
    choosing a value of $t_d$ that is a non-integer
    multiple of the dwell
    time~\cite{baxRemoval1991,zhuDiscrete1993}.
While \gls{lp} and other techniques can mitigate these
    effects in post-processing,
    echoes offer an alternative solution to signal
    detection where no part of the time
    domain must be interpolated or predicted.
\subsubsection{Zeroth-Order Phase Correction}\label{sec:zerothOrder}
Very simple methods for calculating zeroth-order
    (frequency-independent) phase correction behave
    well when all peaks are positive;
    however,
    both for inversion recovery and for \gls{odnp}
    enhancement curves,
    the sign of any given peak is unknown.
Therefore,
    given a collection of complex datapoints ($s_k$) assumed to
    be distributed primarily along the real axis
    of the complex plane
    and then rotated by some
    arbitrary constant (zeroth-order) phase,
    the principle axis of
    the matrix
    \begin{equation}
        \begin{array}{RL}
            I_{ij}
            =&
            \displaystyle
            \sum_{k=1}^N
            \begin{bmatrix}
                \left|s_k \right|^2- \Re\left[ s_k \right]^2
                &
                - \Re\left[ s_k \right]\Im\left[ s_k \right]
                \\
                - \Re\left[ s_k \right]\Im\left[ s_k \right]
                &
                \left|s_k \right|^2- \Im\left[ s_k \right]^2
            \end{bmatrix}
            \\
            =&
            \sum_{k=1}^N
            \begin{bmatrix}
                \Im\left[ s_k \right]^2
                &
                - \Re\left[ s_k \right]\Im\left[ s_k \right]
                \\
                - \Re\left[ s_k \right]\Im\left[ s_k \right]
                &
                \Re\left[ s_k \right]^2
            \end{bmatrix}
        \end{array}
    \end{equation}
    (motivated by the formula for the inertia tensor)
    would provide the vector in the complex plane
    that the real axis had been rotated to.
The zeroth order phase correction
    that corresponds to
    rotating this axis to align with the real axis
    performs well even when the datapoints have a
    variable sign.
\subsubsection{First-Order Phase Correction}\label{sec:firstorderEcho}
\paragraph{first order phasing from echo}
The expression:
\begin{equation}
    \tau_{echo} \approx \tau + 2 t_{90}/\pi
\end{equation}
    provides a reasonable approximation
    to the actual observed $\tau_{echo}$
    based on the ninety time $t_{90}$
    and inter-pulse delay $\tau$ in a spin echo
    experiment~\cite{rance_obtaining_1983,baxImproved1988}.
But, in practice, the existence of hardware trigger delays,
    group delays due to line transmission, \etc, 
    means that the exact position of the echo center
    must be determined experimentally.
\paragraph{barriers to determination}
If the timescale of the inhomogeneous decay
    ($T_2^*$)
    exceeds the required timing correction,
    the determination of the center maximum of the
    echo frequently proves less trivial than expected.
In particular,
    the presence of noise complicates
    attempts to choose between the amplitude of
    time points near the peak of the echo,
    where the magnitude is relatively flat as a
    function of time.
Furthermore,
    echoes resulting from experiments that invert some
    isochromats but not others (or from antiphase signal)
    will not necessarily present maximum amplitude at
    the center of the echo.
\paragraph{introduce two strategies}
Two strategies yield more robust procedures for
    finding the echo center:
\paragraph{correlation format}
The first strategy exploits the fact that
    the echo signal in the time domain
    must have Hermitian symmetry
    ($s(t)=s^*(-t)$).
Therefore, the cross-correlation of the echo waveform with
    its Hermitian conjugate
    can identify the timing discrepancy between the
    instrumentally assigned time axis and the ``true''
    time axis, where $t=0$ corresponds to the echo center.
\paragraph{}
Invoking established procedure for analogous calculations in the
    molecular dynamics literature~\cite{Futrelle1971},
    this strategy begins by zero-filling the signal, $s(t)$,
    to twice its length.
For a discrete signal with $N$ datapoints,
    the resulting zero-filled signal $s_{zf}(t)$,
    has length $2 N t_{dw}$,
    where $t_{dw}$ (the dwell time)
    gives the separation between
    datapoints in the time domain.
The following equations will utilize the fact that
    $s_{zf}$
    (or any signal treated by a discrete Fourier Transform)
    is mathematically modeled as infinitely repeating,
    with 
    a periodicity equal to its length ($2 N t_{dw}$).
For any real inhomogeneous broadening,
    the cost function
    \begin{equation}
        \begin{array}{RL}
            c(\Delta t) =&
            \frac{t_{dw}}{\Delta t + t_{dw}}
            \int_{-\Delta t}^{0}
            \Big|
            e^{i\varphi_0} s_{zf}^*(-t)
            \\
            & 
            \phantom{\frac{1}{t_{len}-\Delta t}\int_0^{t_{len}-\Delta t} \Big|}
            -e^{-i\varphi_0} s_{zf}(t+\Delta t)
            \Big|^2 dt
        \end{array}
        \label{eq:HermitianCost}
    \end{equation}
    should drop to a minimum
    for a value of $\Delta t$,
    named $\Delta t_{min}$,
    that gives the time shift needed to
    align the echo center in $s_{zf}(t)$
    with the echo center in the Hermitian conjugate
    $s_{zf}^{*}(-t)$
    (bearing in mind the periodicity of both functions).
Therefore $\Delta t_{min}/2$
    corresponds to the difference between the original
    (instrumental) $t=0$ and the center of the echo.
In \cref{eq:HermitianCost},
    the integral limits run only over times where the
    integrand is non-zero
    and multiplication with the term outside the integral
    normalizes by the number of integrated
    datapoints;
    $\varphi_0$ signifies that the zeroth-order phasing
    of the signal remains unknown until the determination
    of the echo center.
Expansion of the modulus squared yields
    \begin{equation}
        \begin{array}{RL}
            c(\Delta t) =&
            \frac{t_{dw}}{\Delta t + t_{dw}}
            \Bigg[
            \int_{-\Delta t}^{0}
            \left|s_{zf}^*(-t)\right|^2
            +
            \left|s_{zf}(t+\Delta t)\right|^2
            dt
            \\ &
            - 2
            \int_0^{2N t_{dw}}
            \Re\left[ 
                e^{-i2\varphi_0} s_{zf}(t+\Delta t)s_{zf}(-t)
            \right]
            dt
            \Bigg]
        \end{array}
        ,
    \end{equation}
    where the limits of the third term have been expanded
    into regions where the integrand is zero.
Substitution of integration variables
and utilization of the definition of the correlation
symbol then  yields
    \begin{equation}
        \begin{array}{RL}
            c(\Delta t) =&
            \frac{2 t_{dw}}{\Delta t + t_{dw}}
            \Bigg[
            \int_0^{\Delta t}
            \left|s_{zf}(t)\right|^2
            dt
            \\ &
            - \Re\left[ 
                e^{-i2\varphi_0} s_{zf}^*(-t) \star s_{zf}(t)
            \right]
            \Bigg]
        \end{array}
        \label{eq:fastHermitianWithPhase}
    \end{equation}
Finally, note that the choice of $\varphi_0$
    that will minimize the previous expression
    is simply the phase of $s_{zf}(t)$ at the echo
    center.
Therefore, the expression
    \begin{equation}
        \begin{array}{RL}
            c'(\Delta t) =&
            \frac{2t_{dw}}{\Delta t + t_{dw}}
            \Bigg[
            \int_0^{\Delta t}
            \left|s_{zf}(t)\right|^2
             dt
            \\ &
            - \left| 
                s_{zf}^*(-t) \star s_{zf}(t)
            \right|
            \Bigg]
        \end{array}
        \label{eq:fastHermitian}
    \end{equation}
has a minimum at the same $\Delta t$ as
    \cref{eq:fastHermitianWithPhase}.
As the Fourier transform of the second term in
    \cref{eq:fastHermitian} is the square of the
    Fourier transform of $s_{zf}(t)$, the
    Hermitian conjugate shown in the second term is
    never explicitly calculated.
\paragraph{}
The second phasing strategy relies on the fact that the
    integral of the absolute
    value of the absorptive component of a Lorentzian peak
    is smaller than the absolute value of the
    dispersive component.
Previous literature has extensively employed
    this principle for first order
    phase correction of FIDs~\cite{DeBrouwer2009}.
In the present context, this amounts to finding the
    minimum of the 
    cost function
    \begin{equation}
        C(t_d)
        = 
        \int \left| \Re \left[
            e^{-i \varphi_0} h(t-t_d) s(t)
        \right] \right| dt
        .
        \label{eq:realAbsCostUnmod}
    \end{equation}
Of course, this cost will artificially fall off as
    $t_d$ is pushed to later portions of the signal
    decay,
    so we instead optimize: 
    \begin{equation}
        C(t_d)
        =
        \frac{
            \int \left| \Re \left[
                e^{-i \varphi_0} h(t-t_d) s(t)
            \right] \right| dt
        }{
            \int \left| \Im \left[
                e^{-i \varphi_0} h(t-t_d) s(t)
            \right] \right| dt
        }
        .
        \label{eq:realAbsCost}
    \end{equation}
While this cost function might be expected to fluctuate
    significantly at low signal energies,
    it can still provide useful context,
    especially for sequences that employ standard 90°
    pulses or very short echo times.
\subsubsection{Linewidth and Apodization}\label{sec:GaussianApodization}
\paragraph{our linewidth transformation}
For various purposes, the spectroscopist may desire to determine a
    generic measure of linewidth, defined even for
    complicated lineshapes.
For low-field data,
    standard apodization techniques
    (such as Lorentzian-to-Gaussian transformations)
    prove useful on several fronts.
Therefore,
    a consistent definition of standard lineshapes
    and quantification of
    their signal energy
    ($\propto\int |s(t)|^2 dt=\int \left|s(\nu) \right|^2 d\nu$)
    are advantageous.
\paragraph{gaussian def}
The Fourier transform (with limits $\pm \infty$) of
\begin{equation}
    s_G(t,\lambda_G)
    =
    \exp\left( 
        - \frac{\pi^2 \lambda_G^2 t^2}{4 \ln 2}
    \right)
    \label{eq:FWHMdefGauss}
\end{equation}
is a Gaussian with \gls{fwhm} of $\lambda_G$,
integral 1,
and signal energy 
\begin{equation}
    E_G(\lambda_G)
    =
    \sqrt{
        \frac{2 \ln 2}{\pi}
    }
    \frac{1}{\lambda_G}
    \label{eq:EnergyG}
    ,
\end{equation}
while the Fourier transform of
\begin{equation}
    s_L(t,\lambda_L)
    =
    \exp\left( 
        - \pi \lambda_L |t|
    \right)
    \label{eq:FWHMdefLorentz}
\end{equation}
is a Lorentzian with \gls{fwhm} of
$\lambda_L$,
integral 1,
and signal energy
\begin{equation}
    E_L(\lambda_L) = 
    \frac{1}{\pi \lambda_L}
    \label{eq:EnergyL}.
\end{equation}
The preceding statements are true for spin echo signal
    that includes both a rising refocusing period as
    well as a decay.
Signal comprising exclusively a perfect
    \gls{fid} (\cref{eq:s_FID})
    would yield an integral with half the size and
    an energy of half the magnitude.
\paragraph{new strategy}
Apodization requires some knowledge of the
    original experimental
    linewidth for optimal results.
However, the presence of inhomogeneities can
    complicate the definition of such a linewidth.
A fairly robust strategy for determining the
    generalized linewidth can be adopted from
    a strategy of fitting the envelope
    (the sum of the absolute value of the time domain
    waveform along any indirect
    dimensions)
    of the
    time domain waveform to either a Gaussian or
    decaying exponential function.
The folded normal distribution~\cite{Leone1961} gives the
    functional form of the
    time domain signal envelope
    as $y'(t)$:
\begin{equation}
    y'(t) = \sigma_n \sqrt{\frac{2}{\pi}} \exp\left( 
        -\frac{A^2 y(t)^2}{2 \sigma_n^2}
    \right)
    + A y(t) \mathrm{erf}\left( 
        \frac{A y(t)}{\sqrt{2 \sigma_n^2}}
    \right)
    \label{eq:foldedNormal}
\end{equation}
    where 
    $y(t)$ is either \cref{eq:FWHMdefGauss} or \cref{eq:FWHMdefLorentz},
    $\sigma_n$ is the standard deviation of the
    instrumental noise,
    and $A$ is the signal amplitude.
\paragraph{L2G}
Finally, when considering
    the Lorentzian-to-Gaussian transformation,
    a range of final linewidths could be chosen,
    with narrower linewidths reducing the \gls{snr}.
This manuscript will consider two cases. 
First, when \gls{snr} considerations are significant,
    it is sensible to
    choose the $\lambda_G$ of the resulting (output) Gaussian
    to be of equal integral and energy
    to the Lorentzian of intrinsic linewidth
    $\lambda_L$.
Specifically, this corresponds to
    $\lambda_G = \lambda_L \sqrt{2 \pi \ln 2}$,
    for which $E_G(\lambda_G)=E_L(\lambda_L)$,
    and, overall, an apodization operation of:
    \begin{equation}
        s'(t)
        = 
        s(t)
        \exp\left( 
            \left(- \frac{\pi^{2} \lambda_{L} t^{2}}{2} + \left|{t}\right|\right)
            \pi \lambda_{L}
        \right)
        \label{eq:L2GequalE}
    \end{equation}
    where $\lambda_L$ comes from the generic linewidth
    determined from the signal envelope.
In the second case,
    \begin{equation}
        s'(t)
        = 
        s(t)
        \exp\left( 
            \left(- \frac{\pi
                \lambda_{L} t^{2}}{4 \ln{\left(2
            \right)}} + \left|{t}\right|\right)
            \pi \lambda_{L}
        \right)
        \label{eq:L2GequalWidth}
    \end{equation}
    implements an equal linewidth transformation,
    replacing a Lorentzian of $\lambda_L$ with a Gaussian
    of the same \gls{fwhm}.
\subsubsection{Alignment by Cross-Correlation}\label{sec:crossCorrel}
Field fluctuations of permanent magnets and
    room-temperature electromagnets lead to slight
    shifts in the resonance frequencies of subsequent
    transients.
Some previous methods for spectral alignment have
    relied on iterative Bayesian~\cite{kimBayesian2010}
    or other statistical
    means to align signal~\cite{Ha2014} while others invoke
    correlation of spectral
    fragments~\cite{veselkovRecursive2009,savoraniIcoshift2010a}.
These methods typically operate on the
    absolute value of the signal or well-phased
    signal, have not been implemented for low-field
    signal, or do not allow for independent shifting
    of phase cycled transients.
More recent studies have demonstrated the promise of
    cross-correlation of transients
    as a simple and potent technique to
    align \gls{nmr} transients in the
    presence of a variable field in low-field
    systems~\cite{Keller2020}.
In fact,
    cross-correlation can explicitly deal with
    complex signals
    undergoing phase cycling
    and can be clearly mathematically justified.
Motivated by this, we introduce
    a specific variant of cross-correlation
    (signal-averaged mean-field cross-correlation)
    that can function even
    under circumstances where individual pairs
    of transients do not offer sufficient \gls{snr} for
    alignment.
\paragraph{}
Consider inspecting the signal from two
    transients in the frequency domain,
    $S_j(\nu)$ and $S_m(\nu)$,
    shifting $S_j$ to the left by
    $\Delta\nu_j$ to maximize the norm of the resulting
    signal $\left|S_m(\nu)+S_j(\nu+\Delta\nu_j)
    \right|^2$ --
    \ie, consider maximizing the expression:
    \begin{equation}
        \begin{array}{RL}
            \int \left|S_m(\nu)+S_j(\nu+\Delta\nu_j) \right|^2 d\nu
            \hspace*{-5em}
            &
            \\
            = &
            \int \left|S_m(\nu) \right|^2 d\nu
            \\ &
            + \int \left|S_j(\nu + \Delta\nu_j) \right|^2 d\nu
            \\
            &
            + 2 \int \Re\left[
                S_m^*(\nu) S_j(\nu + \Delta \nu)
                \right] d\nu
        \end{array}
        \label{eq:CorrelOptFull}
    \end{equation}
$S_j$ and $S_k$ are periodic
    (because
    Fourier transformation of
    discretely sampled time domain signals
    generates periodic frequency domain signals).
Therefore, not only the first term in \cref{eq:CorrelOptFull},
    but also the second term,
    remains constant for all values of $\Delta \nu_j$.
Consequently, the problem of aligning the signals in the
    frequency domain to give maximum overlap mathematically
    corresponds \textit{exactly} to
    the much simpler problem of
    finding the maximum of the real part of the
    correlation function ($C(\nu_j)$)
    in the third term,
    where
    \begin{equation}
        \begin{array}{RL}
            C(\Delta \nu_j)
            = &
            \int S_m^*(\nu)
                S_j(\nu+\Delta\nu_j) d\nu
            \\
            = &
            S_m(\nu) \star S_j(\nu)
        \end{array}
        \label{eq:CorrelOptDef}
    \end{equation}
    and where the second equality makes use of the
    $\star$ symbol
    (\cref{eq:CorrelSymbDef}),
    and the FFT makes calculation trivial and
    fast.
Difficulties in applying this mathematical truism to
    real data appear when considering noise
    and phase cycling.
\paragraph{}
As implicitly recognized in previous
    literature~\cite{savoraniIcoshift2010a}, note that
    \cref{eq:CorrelOptDef}
    integrates over regions of $\nu$ that potentially
    contain exclusively noise.
Noisy transients
    will lead to noisy correlation functions,
    meaning
    the position of the maximum of \cref{eq:CorrelOptFull}
    will be influenced by the maximum of the signal and
    the presence of noise.
This effect will degrade \gls{snr} of the correlation
    function even for values of $\Delta \nu$ that
    correspond to well-aligned signal,
    making it important to leave out frequencies that
    contain only noise.
\paragraph{}
Even then,
    a single transient may not provide sufficient
    \gls{snr} to even identify the presence or
    absence of signal.
Importantly,
    the generalization of
    \cref{eq:CorrelOptDef} to the case of more than
    two transients involves the consideration of not only
    the overlap of adjacent transients,
    but the total overlap of all transients,
    and requires
    finding the maximum of the real part of:
    \begin{equation}
        \begin{array}{RL}
            C(\Delta\nu_1,\cdots,\Delta
            \nu_J)
            =&
            \sum_{\substack{m,j\\  m \neq j}}
            \int S_m^*(\nu+\Delta\nu_{m})
            \\[-3ex]
            & \phantom{\sum_{\substack{m,j\\  m \neq j}} \int}
            \times S_j(\nu+\Delta\nu_{j}) d\nu
            \\[-3ex]
            =
            &
            \sum_{\substack{m,j\\  m \neq j}}
            S_m(\nu + \Delta \nu_m)
            \star
            S_j(\nu)
        \end{array}
        \label{eq:CorrelOptMultiNoMF}
    \end{equation}
    to determine the corrective shifts
    in an experiment with $J$ transients.
Note that \cref{eq:CorrelOptMultiNoMF}
    depends on the shifts for all transients,
    simultaneously;
    thus, the ellipsis indicates the presence of
    $J$ arguments to the highly multi-dimensional
    function $C$,
    while the sum varies $m$ and $j$ across all
    possible combinations of the $J$ transients.
The approach to optimizing $C$
    adopted here
    involves
    finding a
    ``mean field''-type solution
    that fixes all but one of the $\Delta\nu_{j}$
\begin{equation}
    \begin{array}{RL}
        C_{m.f.}(\Delta\nu_{j}) &=
        \sum_{\substack{m\\  m \neq j}}
        \Re\Big[ 
            S_m(\nu)
            \star
            S_j(\nu)
        \Big]
        ,
    \end{array}
    \label{eq:CorrelOptMulti}
\end{equation}
where the correlation resulting from $\star$ is a function of
    $\Delta \nu_j$ (\cref{eq:CorrelSymbDef}),
    with the position of all
    other transients
    ($\Delta\nu_m$ in \cref{eq:CorrelOptMultiNoMF})
    held constant.
This expression generates a single one dimensional curve for each
    transient (of subscript $j$), each of which yields a
    clear optimal $\Delta \nu_j$.
This solution requires iterating
    until the list of $\Delta\nu_{j} \rightarrow \Delta\nu_{m}$ values
    remains consistent from one iteration to the next:
    typically 3-10 iterations.
While requiring a more laborious computation
    by demanding calculation of all possible
    correlation functions between all possible transients
    (rather than merely, \eg, adjacent transients in a
    time series),
    the sum in this expression actually involves a
    signal averaging of the correlation function
    and proves particularly important in the case
    where individual transients may have
    particularly low \gls{snr}.
We, therefore, refer to \cref{eq:CorrelOptMulti} and its
    generalization to phase-cycled signal, below, as
    a ``signal-averaged correlation function.''
\paragraph{}
Aligning in the presence of a phase cycle comprises the
    second main consideration here.
Following the \gls{dcct} schema, one can
    treat phase cycling as an added dimension,
    then Fourier transform into
    the coherence domain,
    and then seek
    to optimize the portion of the (Frobenius) norm of the signal
    that varies with the frequency
    shift.
The norm squared is:
\begin{equation}
    \begin{array}{L}
        C(\Delta \nu_{1,1},
        \cdots,\Delta
        \nu_{J,K})
        \\
        \quad
        =
        \sum_l
        \int 
        \Big| \sum_{j,k} e^{-i2\pi\varphi_k\Delta
            p_{l}}
        s_{j,k}(\nu+\Delta\nu_{j,k},
        \varphi_k) \Big|^{2} d\nu
    \end{array}
    \label{eq:2DFrobNorm}
\end{equation}
    for an experiment
    with $J$ repeated scans cycled over $K$ phases.
Here, the sum over $j$ spans all signal
    averaged repeated scans,
    while the sum over $k$ spans all pulse phases
    and the sum over $l$ spans all
    coherence transfer values.
Again (as indicated by the ellipsis)
    $C$ is highly multidimensional.
As in the 1D case, one could consider optimizing $C$ by
    iteratively optimizing the individual mean field
    correlation functions
    arising from the cross-terms
    (between $j$, $k$ \vs $m$, $n$ terms)
    of \cref{eq:2DFrobNorm}
    and iterating
    $\Delta \nu_{j,k} \rightarrow \Delta \nu_{m,n}$
    to convergence.
Isolating the cross terms,
    analogously to 
    \cref{eq:CorrelOptFull}
    $\rightarrow$
    \cref{eq:CorrelOptDef},
    and rearranging slightly yields:
\begin{equation}
    \begin{array}{RL}
        C_{m.f.}(\Delta\nu_{j,k};\Delta p_l)
        =
        \Re\Bigg\{&
            e^{-i2\pi\varphi_k\Delta p_l}
            \sum_n 
            e^{+i2\pi\varphi_n\Delta p_l}
            \\
            & \times
            \sum_{\substack{m\\  m \neq j}}
            \int 
            \big[
                s_{m,n}^{*}(\nu,\varphi_n)
                \\
                & \times
                s_{j,k}(\nu+\Delta\nu_{j,k},\varphi_k)
            \big]
        d\nu \Bigg\} \\
    \end{array}
    \label{eq:2Dcorrelation}
\end{equation}
The cross-terms of \cref{eq:2DFrobNorm}
    involve summing $C_{m.f.}$ over all values of
    $\Delta p_l$;
    however,
    as rationalized later,
    the code will frequently only calculate
    $C_{m.f.}(\Delta \nu_{j,k};\Delta p_l)$
    for a subset of $\Delta p_l$ values,
    so \cref{eq:2Dcorrelation} will prove more convenient.
After defining
    $\Delta\varphi_{n} = \varphi_{k} - \varphi_{n}$,
    rearranging the summations,
    and taking advantage of the $\star$ symbol
    (now redefined to generate a function of $\Delta
    \nu_{j,k}$),
    this becomes
\begin{equation}
    \begin{array}{L}
        C_{m.f.}(\Delta\nu_{j,k};\Delta p_l)
        =
        \sum_n \Re\Bigg\{
            \\
            \phantom{C_{m.f.}}
            \sum_{\substack{m\\  m \neq j}}
            \Big[
                \sum_n
                \big[
                    e^{-i2\pi\Delta\varphi_n\Delta p_l}
                    s_{m,n}(\nu,\varphi_k+\Delta \varphi_n)
                \big]
                \\
                \phantom{C_{m.f.}}
                        \star
                        s_{j,k}(\nu,\varphi_k)
                    \Big]
                \Bigg\}
            \end{array}
    \label{eq:2Dcorrelation3}
\end{equation}
Note that the code that implements the innermost
    sum first introduces a new dimension
    ($\Delta \varphi_n$)
    with the same size and coordinates as both
    $\varphi_n$ and $\varphi_k$,
    and along which the elements of
    the signal are duplicated and then cyclically
    permuted (rolled)
    along the $\varphi_k$ dimension
    by the index $n$.
It then FFTs along the new $\Delta \varphi_n$
    dimension.
For example, if data arises from a pulse sequence
    repeated over eight scans,
    where each scan
    involves cycling the phase of the pulse $\varphi_k$
    in four steps
    and $C_{m.f.}$ is only calculated for a single
    value of $\Delta p_l$,
    then \cref{eq:2Dcorrelation3} yields a set of
    $8\times 4=32$
    functions
    $C_{m.f.}(\Delta \nu_{j,k};\Delta p_l)$
    for the 32 transients
    that not only cross-correlate different transients
    repeated with the same pulse phases
    (subscripted by $j$) but also transients that
    differ only in the choice of pulse phase
    (subscripted by $k$).
The maximum of the real part of each of the 32 functions indicates the
    frequency shift to be applied to the 32 transients.
The $\Delta p_n$ that appears in
    \cref{eq:2Dcorrelation3} represents the difference
    in phase between the two functions that are being
    correlated;
    the choice of $\Delta p_l$ uses $\Delta p_n$ to determine
    whether the phase term in \cref{eq:2Dcorrelation3} 
    leads to addition, subtraction, or quadrature
    addition of the different contributions that sum to
    yield the correlation function.
\paragraph{}
Importantly,
    since the signals are periodic, Plancherel's
    theorem implies that different frequency
    shifts applied to different spectra along
    an indirect dimension cannot change the norm
    of the data.
Therefore,
    unlike the case of frequency shifts
    among transients that are signal
    averaged (the $j$ or $m$ indices in
    \cref{eq:CorrelOptMulti} and
    \cref{eq:2Dcorrelation}), differences between
    the frequency shifts along the phase cycling
    dimension (the $k$ or $n$ indices) cannot lead to
    optimization of~\cref{eq:2DFrobNorm}.
\paragraph{}
Both the specifics of this effect
    and a workaround can be considered with the
    simplest possible example:
    signal acquired with two transients under a
    two-step phase cycle,
    as exemplified in \cref{fig:PhCycleDemo}.
    Specifically, consider an idealized resonance,
    arising from an \gls{fid},
    of \gls{fwhm} $\lambda_L$:
\begin{equation}
    s(\nu,\varphi) =
    \left( 
        \frac{
            e^{i\varphi}
        }{\sqrt{2}}
    \right)
    \frac{\pi\lambda_L}{i2\pi(\nu-\nu_0)+\pi \lambda_L}
    \label{eq:simpleSignal}
\end{equation}
acquired with $\varphi=0\;\text{rad}$ and $\varphi=\pi\;\text{rad}$
    (\ie, using a two-step phase cycle resulting
    in two separate transients).
Upon discrete Fourier transformation
    (dimension of length 2)
    along $\varphi$
    into the conjugate domain $\Delta p$,
    the signal
    ($\Delta p=-1$, aliased to $\Delta p=+1$)
    appears
    centered about $\nu_0$ at $\Delta p=+1$
    (dashed red lines in \cref{fig:alignedPhCycleDemo}).
Such a signal, with $\lambda_L=2/\pi$
    has a signal energy (norm squared) of 1,
    both before and after
    unitary Fourier transformation
    along the $\varphi$ dimension.
For this idealized signal,
    no signal appears in the other coherence pathway
    ($\Delta p=0$; blue lines in
    \cref{fig:alignedPhCycleDemo}).
In the case where field fluctuations
    shift one of the transients by $\Delta\nu\gg \lambda_L$
    (\cref{fig:misalignedPhCycleDemo}),
    one transient presents signal
    centered at $\nu_0$ and another at
    $\nu_{0}+\Delta\nu$.
The signal must still have a signal energy of 1,
    so that
    after Fourier transformation from $\varphi$ to
    $\Delta p$, when the signal energy from both peaks
    ($\nu_{0}$ \vs  $\nu_{0} + \Delta\nu$) spreads
    equally across $\Delta p=0$ and $\Delta p=1$, and breaks into 4
    peaks, each has signal energy of 0.25, as shown in
    \cref{fig:misalignedPhCycleDemo}.
Notably,
    misalignment spreads the signal energy
    both along the
    frequency domain ($\nu$) and the coherence domain
    ($\Delta p$).
This effect will also appear,
    to a less dramatic extent,
    in experimental data (\eg, \cref{fig:StepByStepEp,fig:StepByStepInv}).
\paragraph{}
A different signal metric
    -- specifically a ``masked norm'' --
    provides an analog of
    \cref{eq:2DFrobNorm}:
\begin{equation}
    \begin{array}{LL}
        N'(\Delta \nu_{1,1},
        \cdots,\Delta
        \nu_{J,K})
        \\
        \quad =
        \sum_l
        \int 
        f_{mask}(\nu,\Delta p_l)
        \Big| \sum_{j,k} & e^{-i2\pi\varphi_k\Delta
            p_{l}}
        \\
        & \times s_{j,k}(\nu+\Delta\nu_{j,k}, \varphi_k) \Big|^{2} d\nu
    \end{array}
    \label{eq:maskedCorrel}
\end{equation}
where the new function 
    $f_{mask}(\nu,\Delta p_l)$
    is the (real and positive) ``mask'' function.
\figPhCycleDemo
With an appropriate choice of the mask function,
    the masked norm $N'$
    will indeed only rise to a maximum for
    the choice of $\Delta\nu_{j,k}$ that
    aligns the transients.
\paragraph{refer to example}
For instance, consider
    a mask function that is uniform along $\Delta p$ and
    significantly exceeds 0 along $\nu$ only in
    a bandwidth of similar size to the linewidth.
In the example of
    \cref{eq:maskedCorrel}, only 2
    of the 4 peaks observed in
    \cref{fig:misalignedPhCycleDemo}, each peak with
    an energy of $0.25$, would contribute to the
    calculated masked norm.
Thus, the
    masked norm for properly aligned signal ($\sim1$)
    would exhibit $\sim 2\times$ greater energy than
    that of the masked norm for the unaligned signal
    ($\sim 0.5$).
Attempting to optimize $N'(\Delta \nu_{j,k})$
    for this choice of $f_{mask}$ therefore
    would result in aligned signal.
Considering $f_{mask}$ of opposite construction --
    this time uniform along $\nu$ and only 
    nonzero 
    for $\Delta p_l=+1$ (the expected coherence pathway of
    \cref{eq:simpleSignal}) would lead to a
    similar optimization.
Meanwhile, a $f_{mask}$ selective along both $\Delta p$
    and $\nu$ would lead to a $4\times$ greater
    energy of the masked norm for the aligned signal
    relative to the energy of the unaligned signal
    -- \ie, a 4-fold preference for aligned signal over
    unaligned signal.
\paragraph{}
Thus, to align signal in the presence of phase cycling,
    one should construct a
    $f_{mask}(\nu,\Delta p_l)$
    that is nonzero along $\nu$ only over a bandwidth 
    similar to that of the signal and
    nonzero along $\Delta p_l$ 
    only for values of $\Delta p_l$ where the signal or significant
    artefacts appear.
Then, one can iteratively optimize the masked version of
    the 2D mean-field correlation function
    shown in \cref{eq:2Dcorrelation3}:
\begin{equation}
    \begin{array}{L}
        C'_{m.f.}(\Delta\nu_{j,k};\Delta p_l)
        =
        \sum_n
        \sum_{\substack{m\\  m \neq j}}
        \Re\Bigg\{
                s_{m,n}(\nu,\varphi_k+\Delta \varphi_n)
        \\
        \quad
                \star
                \sum_l 
                \big[
                    e^{-i2\pi\Delta\varphi_n\Delta p_l}
                f_{mask}(\nu,\Delta p_l)
                    s_{j,k}(\nu,\varphi_k)
                \big]
        \Bigg\}
    \end{array}
    \label{eq:2DcorrelationWithMask}
\end{equation}
    where, as before, the $\star$ operation operates along the
    $\nu$ domain to yield a function of $\Delta \nu_{j,k}$
    (and not $\nu$),
    and where $s_{j,k}$ is a function of 3 variables:
    $\nu$, $\varphi_k$, and $\Delta \varphi_n$.
In comparing to other alignment methods,
    note that while the mask can and does serve as a type of
    ``reference spectrum'' for the alignment,
    the signal-averaged correlation function still
    drives alignment of the spectra;
    in fact, the two operate cooperatively,
    with the mask helping to improve the signal to
    noise (\textit{via} filtering) of the correlation
    function,
    and the signal-averaged correlation function
    driving alignment of the sharper features of the
    spectrum.
\subsection{ODNP}\label{sec:ODNPtheory}
\paragraph{introduce two different relaxivities}
An \gls{odnp} experiment simultaneously excites \gls{esr} with \gls{mw}
radiation and detects \gls{nmr} with rf.
It analyzes the mobility of water
    and discriminates motion at
    the timescale of the \gls{nmr} resonance (here $\sim 15\;\text{MHz}$)
    from motion at the timescale of the \gls{esr}
    resonance ($\sim 9.8\;\text{GHz}$).
Two different relaxivities, $k_{low}$ and
    $k_{\sigma}$ $[\text{M}^{-1}\text{s}^{-1}]$,
    sample the nuclear spin single-flips,
    and the cross relaxation 
    between the unpaired electron and the nuclei, 
    respectively~\cite{FranckMethEnz2018}.
\paragraph{$\varepsilon$ to determine $k_{\sigma}$ (⚠ rewrite with this focus)}
Some of the experiments reported here measure the
    signal intensity, $I(p)$,
    as a function of microwave power, $p$,
    where $I(0)$ gives the thermally polarized
    (\ie, Boltzmann, non-hyperpolarized)
    signal intensity.
The results follow
    the established convention~\cite{Doll2012} of defining
    the transferred polarization as $\varepsilon(p) =
    (I(0)-I(p))/I(0)$, so that 
\begin{equation}
    \varepsilon(p) = \left( \frac{s(p)}{R_{1}(p)} \right)
    k_{\sigma} C_{SL}
    \left|\frac{\omega_{e}}{\omega_{H}}\right|
    \label{eq:epsilon_p}
\end{equation}
where $s(p)$ is the electron spin saturation factor
    (averaged across all hyperfine transitions)
    as a function of microwave power, 
    $R_{1}(p)\;[\text{s}^{-1}]$ is the longitudinal relaxation rate
    of the $^{1}$H nuclei as a function of microwave power,
    $\omega_{e}$ $[\text{rad}/\text{s}]$ is
    the Larmor frequency of the electron, $\omega_{H}$ $[\text{rad}/\text{s}]$
    is the Larmor frequency of the $^{1}$H nucleus,
    with, \eg,
    $|\omega_e/\omega_H|=659.33\pm 0.05$
    for TEMPOL in aqueous
    solution~\cite{FranckPNMRS}.
\paragraph{temperature variation of $R_1$}
Importantly, minuscule
    variations in temperature can significantly change
    $R_1(p)$, thus affecting the overall
    $\varepsilon(p)$~\cite{FranckPNMRS}.
For this reason,
    \gls{odnp} measurements of dynamics
    depend on efficient $R_1(p)$ measurements to
    enable accurate quantification
    of the product of the
    cross-relaxivity ($k_{\sigma}$) and the
    saturation factor:
\begin{equation}
    k_{\sigma} s(p) = \frac{\varepsilon(p) R_{1}(p)}{C_{SL}}
    \left|\frac{\omega_{H}}{\omega_{e}}\right|
    .
    \label{eq:ksp}
\end{equation}
\cref{eq:ksp} typically follows an asymptotic form,
    even when $\varepsilon(p)$ does
    not~\cite{FranckPNMRS,Franck_crowding}.
$R_1(p)$ is found from inversion recovery
    experiments at different powers
    by fitting each set of data
    to the equation
\begin{equation}
    M(\tau) = M_\infty (1 - (2-e^{-W R_1}) e^{-\tau R_{1}}
    \label{eq:invRec}
\end{equation}
    where $\tau$ represents the variable delays used
    in each inversion recovery experiment and $W$
    represents the magnetization recovery time
    \cite{weiss_choice_1980}.
\pdfcommentJF{let's follow up on this thread:
    https://jmfrancklab.slack.com/archives/CLMMYDD98/p1654460737886009
    }
\pdfcommentAB{so it seems like we don't have
    this, but next time someone is on magnet I
    could acquire quickly}
\pdfcommentJF{let's not let this hold us up, but
    sounds good}
Therefore, the typical procedure for \gls{odnp} entails collecting
    several individual \gls{nmr} experiments: a
    series of 1D \gls{nmr} spectra, recorded at different microwave powers to
    obtain a progressive enhancement ($E(p)=1-\varepsilon(p)$) curve,
    and also several inversion recovery
    experiments,
    recorded at different microwave powers to
    obtain $R_{1}(p)$.
This contribution does not focus on the
    determination of $k_{\sigma}$ for particular
    samples
    but on the acquisition of progressive enhancement
    and relaxation curves themselves.
\section{Experimental}\label{sec:experimental}
\subsection{Sample Preparation}\label{sec:experimentalSample}
\paragraph{description of samples}
The spin probe TEMPOL (4-hydroxy-2,2,6,6-tetramethylpiperidin-1-oxyl, Sigma-Aldrich)
    provides the unpaired electron for
    all \gls{odnp} measurements shown here,
    including several studies that utilize pure
    solutions of the spin probe in water and
    in toluene (Sigma-Aldrich). 
\paragraph{}
For reverse micelle measurements,
    CTAB (hexadecyltrimethylammonium bromide, 0.186~g, 57~mM, Sigma-Aldrich)
    was dissolved in 8.39~mL of CCl$_{4}$,
    into which hexanol co-surfactant (0.460~g, 502~mM)
    was added.
After addition of water (68~mg, 424~mM), a total
    H$_{2}$O:CTAB:hexanol of 7.45:1:8.82 was
    obtained.
The sample was then vortexed $2\times
    30\;\text{s}$, allowed to rest for 5~min at r.t.,
    and loaded into a $0.6\times 0.8\;\text{mm}$
    capillary tube
    (Fiber Optic Center, New Bedford, MA, USA)
    that was flame-sealed at both ends.

For measurements of water samples not requiring \gls{odnp},
    a 13~mM
    NiSO$_{4}$ (Fisher) solution was prepared.
\subsection{Spectrometer with Minimalistic, Modular Design}\label{sec:minimalExp}
Most data was acquired on a modular \gls{nmr}
    spectrometer operating in tandem with a Bridge12
    microwave power source and the magnet of a Bruker
    E500 cw EPR with SuperX bridge.
The modular design of the system makes several
    different configurations possible.
\subsubsection{ODNP Configuration and Probe}
\paragraph{standard commercial components}
    \pdfcommentLater{note for later -- cite arxiv for
    hardware paper}
The most frequently used configuration employs
    a home-built probe (4.8~μL sample volume, 17~mm length)
    that integrates specifically
    with the Bruker Super High Sensitivity
    Probehead X-Band resonator (ER 4122 SHQE).
A SpinCore RadioProcessor-G
    transceiver
    (a TTL and rf waveform generator and rf digitizer),
    packaged as a PCI board,
    interfaces with the probe
    by way of a homebuilt passive duplexer
    and standard LNA receiving chain,
    as well as a SpinCore rf amplifier powered by generic power
    supply electronics.
\subsubsection{Large Sample 15~MHz Probe}
Some experiments that do not require high-power
    microwaves instead utilize a probe with a
    solenoid coil, rather than the typical hairpin
    loop employed in the \gls{odnp} experiments.
The solenoid probe, enclosed in a shielding box,
    accommodates
    a sample size of 390~μL
    (approximate sample height of 20~mm in a 5~mm o.d. \gls{nmr} tube),
    allowing for better signal to noise.
Notably, however,
    the larger sample size also leads to significantly
    larger field inhomogeneities
    when placed in the same location
    (within the EPR magnet gap) as the \gls{odnp} probe.
\subsubsection{NMR spectrometer independent of microwave electronics}\label{sec:NMRwithoutMW}
A simple off-the-shelf
    oscilloscope and
    arbitrary waveform generator
    can function
    as an alternative to the SpinCore transceiver
    board.
The GW-Instek AFG-2225 waveform generator
    generates the pulse waveforms that are amplified by
    an ENI 3100L RF
    amplifier (does not require de-blanking),
    while
    the GW-Instek GDS-3254 oscilloscope digitizes the
    signal after passing through an analog
    low-pass filter (MiniCircuits SLP-21.4+).
The bandwidth (250~MHz) and sampling rate (5~GSPS) of
    the oscilloscope both exceed the bandwidth of the
    low-pass filter.
\paragraph{compare to spincore}
In practice,
    this configuration was used before the SpinCore
    for setup and diagnostic purposes.
\subsection{Commercial High-Field NMR}\label{sec:highFieldExp}
A high-field Bruker AVANCE III HD
    spectrometer, equipped with a broadband room
    temperature SMART probe with Z-gradient, acquired
    the high-field \gls{nmr} data.
By default, phase cycling on a Bruker spectrometer
    results in the selection of only a single coherence
    pathway.
To implement the \gls{dcct} schema,
    we present a robust
    template for saving all transients of a phase
    cycle in \cref{lst:bruker}.
\subsection{Software Strategy}\label{sec:softwareExp}
\paragraph{introduce pyspecdata}
The pySpecData library,
    developed in part for this work,
    plays a crucial role in the software strategy
    deployed here.
Aside from storing data in an object-oriented
    format (with a structure not
    dissimilar to the xarray library~\cite{hoyer2017xarray}),
    pySpecData offers key advantages for
    spectroscopic data.
\paragraph{tracking multiple dimensions is a pain}
As a specifically relevant example,
    most of the methods presented here
    require treatment of the phase cycle as an
    additional dimension of the data,
    beyond the standard direct and indirect dimensions.
To avoid confusion tracking the meaning of the multiple
    dimensions 
    (\eg, whether the $1^{st}$, $2^{nd}$, or 
    $5^{th}$ dimension encodes phase cycling at a 
    particular point in the code),
    pySpecData utilizes
    modern object-oriented capabilities.
Specifically,
    the class of objects designed for storage of data
    includes a ``label,''
    along with optional units and axis coordinates,
    associated with each dimension.
It also incorporates
    methods for addressing and
    selecting the data with compact notation
    and for performing common operations such as Fourier
    transformation.
As a result, relabeling a time axis
    (such as centering an echo about $t=0$)
    automatically leads to
    an appropriate
    frequency-dependent phase shift upon
    Fourier transformation.
It also enables automatic dimensional
    alignment and creation
    that facilitates, \eg,
    vectorized computation
    of cost functions by easily introducing new
    dimensions corresponding to different choices of
    optimization parameters.
Other benefits include
    (1) uniform storage and processing of data both from
    proprietary file formats and directly
    acquired from instruments,
    (2) automatic interpretation of symbolic functions
    supplied for fitting
    (including the identification of fitting parameters
    \vs data coordinates),
    and (3) automatic propagation of errors.
An important observation reported here is that the
    spectroscopist can capitalize not just on
    the object-oriented
    organization of data and associated information,
    but also on
    the more singularly object-oriented
    capabilities of the Python programming language
    such as operator overloading and property
    definitions.
\section{Results}\label{sec:results}
While advancing the \gls{odnp} methodology,
    the authors noted a need for
    a sweeping reevaluation of basic elements
    of \gls{nmr} acquisition and data processing
    along several fronts.
Although various ad-hoc solutions have been developed over
    the years,
    this work focuses on addressing the lack of 
    a consistent, modern,  and well-explained
    schema (approach, plan, and organization)
    for presenting and
    optimally processing all the raw data acquired
    during an \gls{odnp} experiment (and, more generally,
    \gls{nmr} techniques under active development).
Conceptually, this strategy for data manipulation and
    visualization integrates the synergistic benefits
    of three techniques:
    (1) domain coloring for visualizing complex
    data~\cite{Poelke2012},
    which removes the need for phase correction before
    data can be interpreted,
    and
    (2) object-oriented capabilities that
    facilitate the treatment of new dimensions
    introduced to store all the information in a
    phase-cycled experiment,
    as well as Fourier transform operations,
    and (3) open-source libraries that aid in
    visualization.
\paragraph{point out this is an outline}
In the following text,
    \cref{sec:show_all_data} first demonstrates how
    domain coloring can provide a compact
    representation of \gls{nmr} signal phase.
Formal implementation of
    ``phase cycling dimensions''
    as additional dimensions in an \gls{nmr}
    dataset yields a variety of benefits,
    starting with the rapid setup
    and optimization of the \gls{nmr} experiment and
    instrumentation (\cref{sec:data_on_scope}).
A visualization of standard nutation curve data
    provides a
    straight-forward introduction to this non-standard
    plotting technique (\cref{sec:NutationCurve}).
Notably,
    a simultaneous
    presentation of all (distinguishable) coherence
    transfer pathways,
    which typically have different signal phases or
    timings,
    in a domain-colored format
    proves surprisingly useful.
We refer to the resulting image as
    a ``\gls{dcct} map,''
    alluding to the term `\gls{ct} map'
    in the seminal literature~\cite{Bodenhausen1984}.
While traditional techniques involve throwing out
    at least some data (undesired \gls{ct} pathways,
    the imaginary part of data, \etc),
\gls{dcct} maps provide a comprehensive overview of all
    acquired data in one image, visualizing the
    relationship between signal in the
    desired coherence pathway
    and correctly separated artefacts,
    as well as signal improperly sorted into
    undesired \gls{ct} pathways.
This manuscript refers to these three contributions,
    respectively,
    as: ``desired signal'' or ``desired pathways,''
    ``artefacts'' or ``artefactual pathways,''
    (\cref{sec:treatmentPhaseDCCT})
    and ``phase cycling noise'' (\cref{sec:fieldInstab}).
In the results that follow,
    \gls{dcct} maps
    enrich signal
    analysis to inform experimental design and
    the development of data-processing algorithms
    (\cref{sec:algorithmsDCCT}),
    without adding any
    additional time cost to a more
    traditionally-acquired \gls{nmr} experiment.
\subsection{Display of Signal Phase}\label{sec:show_all_data}
\paragraph{talk about domain coloring plot -- separate this}
\figDomainColorIntro
Before considering the visualization of signal in
    the coherence domain,
    this manuscript first tests the
    applicability of the domain coloring concept
    to raw \gls{nmr} data
    (\ie, data that may be Fourier transformed, but
    without any type of phasing/timing corrections
    or other post-processing).
Domain coloring plots appear in other
    fields:
    \eg, in solid state physics studies,
    hue frequently
    represents directionality~\cite{Foster2019,Yu2012a}.
Similarly, several public-domain webpages present example diagrams
    illustrating phase with hue.
Nevertheless, such plots remain
    under-exploited in \gls{mr} data
    visualization~\cite{meriles2003bpm,Franck2009}.
As indicated in~\cref{fig:DomainColorIntroWheel},
    hue corresponds to
    the complex phase and value/intensity corresponds
    to the complex magnitude.
The Matplotlib Python plotting
    library~\cite{Matplotlib} specifically enables the
    development of libraries such as pySpecData
    to deploy a wide variety of such custom plotting styles
    with relative ease.
Armed with this plotting technique,
    the spectroscopist can assess the success of an
    experiment at a glance, without
    requiring any phase or timing corrections.
\paragraph{}
A 390~μL (Ni$^{2+}$-doped water) sample inside a solenoid coil probe provides the
    signal for this section.
As mentioned previously,
    this sample size
    experiences a much greater range of field
    inhomogeneity than a probe with a capillary sample.
\cref{fig:DomainColorIntroNutation} illustrates the raw
    signal (no phasing corrections),
    from the desired \gls{ct} pathway
    of an echo-based nutation curve
    (pulse sequence:
    $\theta$--$\tau$--$2\theta$--\textit{acq.},
    with tip angle $\theta=\gamma B_1 p_{90}$ increasing along
    the indirect dimension).
The inset of \cref{fig:DomainColorIntroNutation}
    shows all frequencies where the pulse excites
    signal while also 
    clearly demonstrating the inversion of the signal
    as $\theta$ passes through 180°.
During inversion,
    the signal passes through zero (white)
    while also changing color to indicate the 180°
    phase change
    (\eg, red to cyan or blue to yellow,
    or \textit{vice versa}).
A quick glance at a
    complex domain coloring plot can ascertain
    information that may otherwise require
    a(n inverse) Fourier transform
    and separate plot in order to be understood.
For example, a time-shift appears in the
    frequency-domain plot as a ``rainbow'' color
    variation,
    and in the case of a spin echo,
    indicates that the origin of the time axis ($t=0$)
    does not properly align with the center of the
    echo signal.
Here, an inspection of
    \cref{fig:DomainColorIntroNutation}~(inset)
    indicates
    5~cycles ($10\pi$~rad) of phase rotation over
    a span of 5~kHz, corresponding to a time shift of
    1~ms.
\paragraph{timing correction}
\cref{fig:DomainColorIntroNutationcorr}
    introduces this 1~ms
    timing/phase correction,
    achieved by shifting the time axis
    in the pySpecData processing code (as in
    \cref{lst:TimeSetting}) or,
    equivalently, applying the appropriate
    first-order phase shift
    in the frequency domain.
As expected from an echo signal properly centered about
    $t=0$,
    \cref{fig:DomainColorIntroNutationcorr}
    yields a uniform phase
    lineshape.
\paragraph{}
\cref{fig:DomainColorIntro} provides just one of many
    possible examples in which domain
    coloring plots both quickly guide simple manual
    data processing of raw data and also allow
    instant access to information.
Note how a real-valued plot of
    \cref{fig:DomainColorIntroNutation}
    would generate ambiguity between the phase
    variation and amplitude variation.
Here, a simple glance reveals phase variation along the
    direct dimension and amplitude variation along the
    indirect dimension.
Also note that information such as the
    time delay of the signal is accessible
    in both domains (time and frequency),
    while other plotting strategies might
    make this information apparent in only one
    domain.
\subsection{Treatment of Phase Cycling}\label{sec:treatmentPhase}
\paragraph{benefits of phase cycling as a separate dimension}
As demonstrated in this and the following subsections,
    the incorporation of many short dimensions
    corresponding to the cycling of individual
    pulses
    clearly illustrates the
    effects of drifting fields,
    rf amplitude misset, resonance frequency
    offset effects, and
    pulse ringdown.
It also yields a simple scheme for quantifying
    the \gls{snr} ratio.
\paragraph{introduce phase cycling domain}
With the increased practicality offered by object
    oriented programming in hand,
    the original conceptual approaches to phase cycling -- pioneered
    by Wokaun, Bodenhausen, Ernst, and others --
    assume new meaning;
    these approaches
    incisively focus on the concept that
    the change in coherence
    order during a pulse ($\Delta p$) 
    is, quite simply, the Fourier transform of the
    dimension along which pulse phase is cycled
    ($\varphi_{j}$, \cref{eq:basicPhaseCycling})~\cite{wokaunSelective1977,drobnyFourier1978,Bodenhausen1984}.
While the traditional schema adds up the
    effect of all pulse phase variations to
    determine the overall phase cycle of the
    desired coherence pathway,
    the \gls{dcct} schema keeps matters simpler by treating each pulse
    involved in the phase cycle separately.
To achieve this,
    the processing code
    organizes data into one or
    more short
    (typically 2-4 elements long) dimensions
    for each pulse involved in the phase cycling,
    in addition to the traditional
    direct ($t_2$) and indirect dimensions
    typical of multidimensional \gls{nmr} experiments.
Each of these phase cycling dimensions can be in the
    ``phase cycling domain'' ($\varphi_i$ for pulse $i$)
    or the conjugate ``coherence transfer domain'' ($\Delta p_i$).
More typical phase cycling procedures
    effectively filter and discard these additional
    dimensions.
However, a central result presented here is
    that modern instrumentation and open-source coding~\cite{Matplotlib,harris2020array,pySpecData}
    standards
    facilitate both this
    treatment of the phase cycle as an additional
    dimension
    and the ability to visualize
    these new dimensions with automatically generated
    \gls{dcct} maps.
Together, these offer a more comprehensive view of the
    underlying spin physics and of the impacts of
    experimental imperfections.
\subsection{Domain Colored Coherence Transfer (DCCT) Schema}\label{sec:treatmentPhaseDCCT}
Developing new methods under adverse circumstances,
    such as new \gls{odnp} instrumentation or
    protocols,
    requires
    unambiguous identification of signal as well as
    quick insight into any potential issues with the
    setup.
The \gls{dcct} map, importantly,
    displays the artefactual signal arising from undesired
    coherence pathways.
Instrumental errors that lead to
    imperfections in the phase cycle itself
    can lead to miscategorization of the desired
    signal as arising from undesired pathways--resulting
    in signal loss and distortion,
    or miscategorization of artefactual signal as
    arising from the desired coherence pathway--leading
    to artefacts in the final result.
By displaying these features,
    \gls{dcct} maps provide
    greater insight into the
    effects of experimental imperfections on the
    signal.
These are most easily demonstrated
    through a series of examples.
\subsubsection{Example: Phase-cycled NMR with Standard Test + Measurement Equipment}\label{sec:data_on_scope}
\paragraph{introduce exp + motivate AFG/Scope setup}
Non-specialized test and measurement equipment
    can acquire a reasonable \gls{nmr}
    signal~\cite{hibinoSimple2018}.
The \gls{dcct} schema enables rapid setup and diagnosis of
    such ``bare bones'' \gls{nmr} instrumentation.
For example, here, an arbitrary function generator
    operates as an rf source and a
    digital oscilloscope
    operates as
    a digital to analog converter
    whose bandwidth and sampling rate exceed
    the bandwidth of an attached analog low-pass filter
    (\cref{sec:NMRwithoutMW}).
Such an inexpensive,
    uncomplicated \gls{nmr}
    spectrometer is invaluable for diagnostic
    purposes -- \eg,
    when assembling a modular \gls{odnp} spectrometer.
\paragraph{coherence display useful for non-specialized setup}
Object-oriented Python code~\cite{pySpecData}
    controls both the function generator and the
    oscilloscope via USB~2.0 communication,
    an instrumental setup similar to that which was described
    previously~\cite{Chonlathep2017}
    (see \cref{lst:oscope}).
Despite the fact that a standard
    oscilloscope has no built-in phase-cycling
    capabilities,
    the software can trivially save separate
    transients acquired with different pulse phases,
    along with the pulse waveform as a phase reference.
The object-oriented code implemented here
    frequency filters the results,
    digitally mixes down and phase references by
    comparison to the captured pulse waveform,
    and saves data in an HDF5
    format.
For a spin echo experiment,
    as in \cref{fig:scopeOneD,fig:ScopeData},
    it stores the data immediately available from the
    acquired transients
    in a 3-dimensional array of data, with shape
    $n_{\varphi_1} \times n_{\varphi_2} \times
    n_{t_2}$,
    representing the function $s(\varphi_1,\varphi_2,t_2)$~\footnote{Note that the
        ordering of the dimensions in the $\times$
        expression and in the function are
        meaningful with respect to the code and
        imply the ordering of the data dimensions
        following a ``C-type'' (as opposed to ``Fortran-type'')
        convention,
        such that datapoints with adjacent indices along the
        right/inner most dimension
        reside in adjacent memory locations,
        while adjacent values in the left/outermost
        dimension reside further apart in memory.}.
Here
    $n_{\varphi_1}$
    and
    $n_{\varphi_2}$
    are the number of phase cycling steps for the
    two pulses,
    and $n_{t_2}$ is the number of time points
    along the direct dimension.
\paragraph{}
After 3-dimensional Fourier transformation (without
    zero filling)
    along the $\varphi_1$, $\varphi_2$, and $t_2$
    dimensions,
    the signal becomes:
\begin{equation}
    \begin{array}{RL}
        \tilde{s}(\Delta p_1,\Delta p_2,\nu_2)
        =
        \iiint &
        e^{-i2\pi 
            \left( \nu_2 t_2 + \Delta p_1 \varphi_1
                + \Delta p_2 \varphi_2
            \right)
        }
        \\
        & \times
        f_{sampling}(\varphi_1,\varphi_2,t_2)
        \\
        & \times
        s(\varphi_1,\varphi_2,t_2)
        \;d\varphi_1 d\varphi_2 dt_2
    \end{array}
    \label{eq:phaseCycling}
\end{equation}
    where
    $\Delta p_j$ indicates the coherence change
    during pulse $j$, $\nu_2$ gives the offset
    frequency (in Hz) along the direct dimension,
    and $f_{sampling}$ gives the function representing
    the instrumental and/or digital filtering and discrete
    sampling of the otherwise continuous signal
    (see \ref{sec:fTnorm} for further notes).
As a reminder, the coherence level change ($\Delta p_i$)  and 
    the cycled pulse phase ($\varphi_i$ ) are Fourier 
    conjugates (\cref{eq:basicPhaseCycling,eq:phaseCycling},
    see example code \cref{lst:chunkExample}).
\paragraph{describe plot}
\cref{fig:ScopeData} presents the \gls{dcct} map that
    results from an initial attempt at finding signal
    with a spin echo pulse sequence
    (\cref{fig:ScopeDataMisset}),
    as well as an optimized attempt
    (\cref{fig:ScopeDataFixed}).
In both \cref{fig:ScopeDataMisset} and
    \cref{fig:ScopeDataFixed}, the \gls{ct} pathways
    $\Delta p_1=+1$ and $\Delta p_2=-2$ contain
    signal unambiguously identified as arising from the spin echo. 
Coherence pathways that are not physically
    meaningful are marked with an X -- \eg, in
    \cref{fig:ScopeData},
    the coherence order change marked by the first pulse,
    $\Delta p_1$ has been
    marked with an X due to the fact
    that the initial pulse acting on polarization
    only generates single quantum coherence here.
On the other hand,
    when several reasonable $\Delta p$ values are
    Fourier aliased together,
    all such values are presented,
    separated by commas.
\paragraph{}
In this example, 
    the \gls{dcct} map clearly demonstrates that,
    despite the non-specialized instrumentation
    employed here,
    the phase cycling of the pulses on the
    function generator
    and phase referencing of the oscilloscope
    operate as expected,
    giving clean isolation of the \gls{ct}
    pathways.
Furthermore, the sign associated with the phases can
    vary on different spectrometers from different
    manufacturers, as previously noted,
    \cite{levitt_signs_1997,levitt_signs_2000},
    and this is observable in the \gls{dcct} map.
In \cref{fig:ScopeData},
    the convention is properly applied,
    but an inversion of the sign of $\Delta p$
    (indicating an inversion of the sign of $\varphi$)
    could be easily observed and corrected.
\paragraph{main obs}
Notably, \cref{fig:ScopeDataMisset}
    identifies that the majority of
    the coherences generated by the pulse sequence
    do not contribute to the spin echo
    (green solid line, \cref{fig:ScopeDataCoh}),
    but rather to undesired/artefactual signal,
    where the second, longer pulse
    (nominally the 180° pulse) excites an \gls{fid} ($p=-1$)
    directly from the polarization ($p=0$)
    (blue dashed line, \cref{fig:ScopeDataCoh}).
The \gls{dcct} map, therefore, offers insight into
    experimental improvements by identifying the
    different pathways where the polarization has been
    utilized.
In this simple example,
    after increasing the length of both pulses,
    the pulse sequence predominantly
    generates coherences that contribute to the spin
    echo signal.
At the same time, the rainbow banding in the time
    domain data of \cref{fig:ScopeDataMisset}, also
    indicates that the resonance frequency differs
    significantly from the carrier frequency. 
\cref{fig:ScopeDataFixed}
    corrects both the 90-pulse time and the carrier
    frequency to yield
    an intense, single-colored band in the
    desired \gls{ct} pathway for the echo signal.
\paragraph{detailed obs}
Note that
    artefacts due to pulse ringdown
    cycle with the same phase as the
    pulses, in the $\Delta p=-1$ pathway for the
    relevant pulse, \ie,
    in $\Delta p_1=0 \rightarrow \Delta p_2=-1$
    and $\Delta p_1=-1 \rightarrow \Delta p_2=0$.
These artefacts overlay with simple (\gls{fid}-like)
    excitation and appear here as regions of alternating
    signal (colored) and no signal (white),
    with the latter arising from times where the
    oscillating high-intensity ringdown saturates the
    duplexer diodes and/or low-noise amplifier.
Thus,
    with a single phase-cycled scan,
    the data can highlight how spin echo acquisition gives
    uniquely unambiguous confirmation that true signal
    has been observed.
\figScopeData
\paragraph{}
\figScopeOneD
The \gls{dcct} schema introduces the \gls{dcct}
    map as an intermediate step between acquisition
    and presentation of the final data in a traditional
    phase cycled NMR experiment.
The \gls{dcct} schema also opens up the possibility
    of other plots 
    that display a subset of the data offered by the
    \gls{dcct} map,
    tailored to particular
    experiments.
For example, after plotting
    \cref{fig:ScopeData},
    extraction of a subset of the data in a customized
    plot yields \cref{fig:scopeOneD},
    which serves to emphasize the strong preference for
    echo-based detection for this large sample,
    due to the rapid $T_2^*$.
Meanwhile, traditional \gls{nmr} acquisition
    employing
    on-board signal averaging through a
    phase-cycled receiver
    would select an even more exclusive subset of the available information,
    detecting and saving exclusively the signal
    from the desired echo \gls{ct} pathway
    $\Delta p_1=+1 \rightarrow \Delta p_2=-2$
    (the blue line
    in \cref{fig:ScopeDataCoh});
    the resulting data would only indicate the
    presence or absence of
    signal.
\subsubsection{Example: Nutation Curve}\label{sec:NutationCurve}
\figNutationResults
\paragraph{off-the shelf result shows what we want from spincore}
In these results,
    a SpinCore RadioProcessorG replaces the
    USB~2.0 function generator and oscilloscope to enable
    faster rates of data transfer to the computer,
    as well as the use of digital
    filtering and downsampling
    that permit the
    distortion-less capture of longer signals~\footnote{%
    The instrument constructed from these non-specialized test
    and measurement equipment suffered from two
    drawbacks: first, a data size limitation, and,
    second,
    slow communication between programmed
    commands and actual execution on the hardware.
For a spectrometer constructed from non-specialized
    components, 
    higher-end test and measurement equipment
    with a larger storage memory
    could address
    the first limitation,
    as could a standard analog quadrature mixer and
    filter scheme, both of which were beyond the
    scope of this study.
The second limitation may be overcome with
    test and measurement instruments providing USB 3.0
    capabilities.}.
\paragraph{concrete example with symbol labels -- nutation}
The same spin echo nutation experiment
    from \cref{fig:DomainColorIntro}
    offers more information when presented as a
    \gls{dcct} map (\cref{fig:NutationResults}).
The full dataset comprises a 4-dimensional function,
    $s(t_p, \varphi_1, \varphi_2, t_2)$,
    with a size of
    $n_{t_p} \times n_{\varphi_1} \times
    n_{\varphi_2} \times 
    n_{t_2}=100 \times 2 \times 2 \times 1024$.
A 3D Fourier transform converts $s$ to
    $\tilde{s}(t_p, \Delta p_1, \Delta p_2, \nu_2)$.
The \gls{dcct} map demonstrates all possible conversions of
    the polarization to coherent signal
    (here both echo-like and \gls{fid}-like),
    at all frequencies where it
    occurs,
    while simultaneously preserving all relative
    phase/sign information.
Here, domain coloring proves crucial,
    given that signal from different coherence
    pathways occurs at different times and with
    different absolute phases.
For example, \cref{fig:NutationResults}
    represents data from an early trial on a modular
    system with new components,
    and it was unclear if the phase cycling was
    functioning optimally.
The \gls{dcct} map highlights the
    contribution of an unwanted signal artefact
    (blue-dashed line in \cref{fig:NutationResults})
    that,
    in the absence of a phase cycle,
    would
    interfere with the desired signal
    (solid green line)
It clarifies that simple hardware and data
    processing are capable of cleanly separating
    this artefact from desired signal.
\cref{fig:NutationResults} demonstrates another
    general advantage of the \gls{dcct} map:
    the same information may be displayed and
    interpreted in
    either the time (\cref{fig:NutationResultsTdom})
    or the frequency (\cref{fig:NutationResultsFdom})
    domain.
Specifically,
    note how the appearance of \gls{fid}-like (blue dashed)
    and echo-like (green solid) signal at two different times in
    the time domain can be deduced directly from the
    rainbow banding of the \gls{fid}-like signal in the
    frequency domain,
    in contrast to the constant color of the echo
    signal.
\subsubsection{Example: Field Instabilities and
the Estimation of Noise}\label{sec:fieldInstab}
\figModCoil
\paragraph{why we care about mod coil attachment}
Another example of relevance to \gls{odnp} arises when considering field
    instabilities.
The Bruker \gls{esr} system comes equipped with a system for
    generating a standard modulation field
    (typically varying with a period of 10~μs),
    and one practical concern of \gls{odnp} spectroscopy
    involves
    what influence this might have on the \gls{nmr} signal
    (when the spectrometer is set to ``standby'' mode).
While a specific practical interest related to \gls{odnp} motivates this
    measurement,
    it also, more importantly,
    provides a controlled demonstration of
    general effects arising from unstable fields.
The \gls{dcct} maps
    in \cref{fig:ModCoil}
    represent
    signal from a spin echo experiment comprising
    an 8-step phase cycle with 16 repeats,
    when
    the modulation cable is left attached to the cavity
    (\cref{fig:ModCoilAttachedTime,fig:ModCoilAttachedFreq})
    \vs when it is detached from the cavity
    (\cref{fig:ModCoilDetachedTime,fig:ModCoilDetachedFreq}).
\paragraph{results}
The timing and phase of the echo in
    \cref{fig:ModCoilAttachedTime,fig:ModCoilDetachedTime}
    vary from scan to scan,
    as evidenced by, respectively,
    the left \vs right translation of the echo position
    relative to the $x$-axis
    and change to the color of the signal at $t=0$.
Notably, domain coloring helps to emphasize these
    concurrent variations.
Again,
    the frequency domain \gls{dcct} maps
    (\cref{fig:ModCoilAttachedFreq,fig:ModCoilDetachedFreq})
    make the same effect evident in a different form:
    as changes in how the color varies from left to
    right across the spectrum
    and in the average (overall) color across the
    spectrum for each scan.
As expected, the signal remains more stable when the connector for the
    modulation coil is disconnected, as in
    \cref{fig:ModCoilDetachedTime,fig:ModCoilDetachedFreq}.
\paragraph{enumerate effects}
The echo signal responds to the increased field variation
    from the modulation coil in three specific ways.
First, as already noted, the phase (color) of the signal at
    $t=0$ (or across all frequencies)
    varies more across consecutive scans.
This inconsistency arises from the residual modulation field
    driving changes in the $B_0$ field
    strength before \vs after the 180° pulse.
This scan-to-scan variation of the signal phase
    would lead to reduced signal amplitude
    if these scans were averaged.
Second, greater frequency variation of the individual
    transients contributing to the signal in
    \cref{fig:ModCoilAttachedFreq}
    (also evident from the differences in the rainbow
    banding of \cref{fig:ModCoilAttachedTime})
    results in the more jagged appearance of the
    signal ``bands'' as
    compared to \cref{fig:ModCoilDetachedFreq}.
If the various scans of \cref{fig:ModCoilAttachedFreq}
    were averaged together,
    a distorted lineshape with increased linewidth
    and reduced signal energy and amplitude
    would result,
    underscoring the usefulness of interrogating the
    impact of each transient.
Third, and perhaps most significantly,
    in both cases, the \gls{dcct} map
    shows noise/artefacts
    in the \gls{ct} pathways
    outside the echo-like ($\Delta p_1=+1$, $\Delta p_2=-2$)
    pathway.
These artefacts are
    concentrated within the
    signal bandwidth.
No explanation based on the transitions between
    different coherence levels of the density matrix
    can rationalize the appearance of significant
    artefactual signal in these ``inactive'' \gls{ct} pathways
    shown in \cref{fig:ModCoil}.
Rather, since the phase cycle must be sorting signal into the
    wrong \gls{ct} pathway,
    we refer to the noise-like artefacts that appear
    in the inactive pathways
    as ``phase cycling noise.''
The difference is subtle,
    but with the modulation coil
    attached,
    more intense spikes in the
    phase cycling noise are observed,
    as can be quantified from the
    mean squared phase cycling noise amplitude
    ($1.3\times$ higher with the modulation coil attached).
\paragraph{analysis of noise}
Clearly, \gls{odnp} should be acquired with the modulation
    coil detached under all cases.
Even then, because the signal energy is diverted into
    energy for phase cycling noise
    (\cref{sec:crossCorrel}),
    the standard deviation across the coherence domain
    provides an appropriate source for error bars on
    datapoints in the spectrum.
However,
    it will be noted that care must be taken when
    applying these errors to integrated signal,
    since they can be correlated.
\figCpmg
\figBrukerCpmg
\subsubsection{Example: CPMG}\label{sec:CPMG}
As noted in other publications,
\gls{odnp} can be acquired with a CPMG
    (Carr-Purcell Meibloom-Gill)
    sequence,
    as is typical of many studies in low-field and
    portable \gls{mr}~\cite{Neudert2014_Sband,Franck2013,Uberruck2020}.
An interesting result of the \gls{dcct} map
    arises when applied to a fully phase cycled
    CPMG -- that is, a four step phase cycle on
    a 90° excitation pulse,
    followed by a train of evenly spaced 180°
    pulses, phase cycled in concert.
Some analysis schemes
    bin the results of the phase cycle into a ``CP''
    component, where the phases of the 90° pulse
    match the phase of the 180° pulses
    or have a 180° phase difference,
    and a ``CPMG'' component,
    with orthogonal phases.
Since the \gls{dcct} map offers a new means
    for visualizing signal,
    it provides the opportunity to revisit the
    decay of CP components
    and the persistence of CPMG
    components~\cite{ahola_multiecho_2006}.
Briefly,
    following the \gls{dcct} treatment,
    signal should
    alternate between
    $\Delta p_1=+1$ and $\Delta p_1=-1$,
    as in the initial echoes of~\cref{fig:CPMGsignal}; however,
    signal starts to ``bleed''
    into the opposite $\Delta p_{1}$ value
    following pathways like those shown in
    \cref{fig:CPMGbleeding},
    ultimately yielding a
    constant signal for both values of $\Delta p_1$.
This corresponds to the previously observed decay of
    the ``CP'' component~\cite{ahola_multiecho_2006},
    viewed from a new perspective.
This example also allows the introduction of phase and
    coherence dimensions that group the effects of more
    than one pulse.
\paragraph{coherence transfer domain display shows absolute spectrometer phase}
\paragraph{separate even and odd echoes}
In more detail,
    naively following the methodology laid out in
    \cref{sec:show_all_data},
    the software can sort the data into a 4D dataset
    (a single-line command in pySpecData),
    with the signal given by the discretized function
    $s(\varphi_1, \varphi_2, \tau_{echo}, t_2)$
    of shape
    $n_{\varphi_1} \times n_{\varphi_2} \times
    n_{\tau_{echo}} \times n_{t_2}$;
    where $t_2$ gives the points within each echo
    and $\tau_{echo}$ gives the center position of
    each echo.
It then Fourier transforms the signal along the
    $\varphi_1$, $\varphi_2$, and $t_2$
    dimensions to permit filtering by resonance
    frequency and coherence pathway.
\paragraph{what should happen}
The initial (nominal) 90° pulse
    changes the
    coherence order of the initial polarization by
    $\Delta p_1 = \pm 1$.
Here, since the 180° pulses are phase cycled
    together, $\Delta p_2$ refers to the net change in
    coherence order due to all 180° pulses.
Therefore, odd-numbered echoes
    harvest signal from the pathway that experiences
    $\Delta p_1 =+1$
    and $\Delta p_2 = -2$.
In contrast,
    even-numbered echoes
    harvest signal from the pathway that experiences
    $\Delta p_1 =-1$
    and $\Delta p_2 = 0$.
\paragraph{}
Thus,
    the CPMG experiment particularly motivates
    an informed choice of phase cycling
    dimensions that simplifies the analysis.
Specifically,
    since only two types of pulses are phase cycled
    (along dimensions $\varphi_1$ and $\varphi_2$),
    a coherence-domain dimension
    $\Delta p_1 + \Delta p_2$ only yields signal
    for $\Delta p_1 + \Delta p_2 = -1$.
The only signal that appears for other values of
    $\Delta p_1 + \Delta p_2$
    arises from instrumental artefacts.
A rearrangement of the expression
    $\Delta p_1 \varphi_1+\Delta p_2 \varphi_2$
    (which is the effect of phase cycling on the
    phase angle of the transients)
    yields
\begin{equation}
    \Delta p_1 \varphi_1+\Delta p_2 \varphi_2
    =
    \Delta p_1 \left( \varphi_1-\varphi_2 \right)
    + \left( \Delta p_1+\Delta p_2 \right) \varphi_2
    ,
    \label{eq:rearrangePhase}
\end{equation}
and motivates rearranging the signal to the form
    $s(\varphi_1-\varphi_2,\varphi_2,n_e,t_2)$
    whose (3D) Fourier transform is given by
    $\tilde{s}(\Delta p_1,\Delta p_1+\Delta p_2,n_e,\nu_2)$
    where $n_e$ represents the echo number.
\paragraph{purpose of Δp₁}
The $\Delta p_1$ dimension disentangles the
    pathways that should give rise to the even \vs odd echoes.
Odd-numbered echoes
    (following 1, 3, \etc, inversion pulses)
    should only present signal for
    $\Delta p_1=+1$
    since the expected/desired coherence pathway
    with $\Delta p_1=+1$
    (green line in \cref{fig:CPMGbleeding})
    only reaches $p=-1$ for odd echoes.
Analogously, even-numbered echoes
    should only present signal for $\Delta p_1=-1$
    (blue line in \cref{fig:CPMGbleeding}).
The observed signal shown in the \gls{dcct} map of
    \cref{fig:CPMGsignal} derives from the alternating
    green and blue lines along the $p = -1$ coherence
    level.
In this scheme, as long as even \vs odd echoes
    remain separated along $\Delta p_1$,
    any phase encoded between the excitation and
    first echo pulse would be properly preserved.
\paragraph{bleed}
However, due to offset or misset effects,
    imperfect 180° pulses can store
    a small fraction of transverse magnetization along
    the $z$-axis for one or more echo periods
    and then re-excite it as observable magnetization
    (the red pathway in \cref{fig:cpmg}).
In \cref{fig:CPMGbleeding},
    the signal
    cleanly separates into the desired coherence
    channels for the first few echoes.
However, near the 7\textsuperscript{th} echo,
    significant amounts of
    $\Delta p_1=+1$ signal begin to appear in
    even-numbered echoes,
    as well as significant amounts of
    $\Delta p_1=-1$ signal in odd-numbered echoes
    -- \ie, the signal ``bleeds'' from the even echo
    pathway into the odd echo pathway, and
    \textit{vice versa}.
\paragraph{}
\cref{fig:CPMGsignal} also demonstrates that
    signal from the two $\Delta p_1$ pathways
    can be combined into a single decay
    with coherent phase by
    subtracting (adding with a phase rotation of 180°)
    the $\Delta p_1 = -1$ signal and
    the $\Delta p_1 = +1$ signal.
Such an operation reduces to:
\begin{equation}
    \begin{array}{L}
        \frac{1}{2}
        \left( 
            \tilde{s}(t_2,+1)
            -
            \tilde{s}(t_2,-1)
        \right)
        \\
        \quad =
        \frac{1}{2}
        \sum_{n=0}^3
        \left( 
        (-i)^{n}
        s\left(t_2,\frac{n}{4}\right)
        -
        (i)^{n}
        s\left(t_2,\frac{n}{4}\right)
        \right)
        \\
        \quad =
        -i s\left( t_2,\frac{1}{4} \right)
        +i s\left( t_2,\frac{3}{4} \right)
    \end{array}
    \label{eq:CPCPMG}
\end{equation}
    where $s$ is a function of $t_2$ and
    $\varphi_1 - \varphi_2$,
    and $\tilde{s}$ a function of $t_2$ and
    $\Delta p_1$,
    where $n$ gives the integral steps of
    $\varphi_1-\varphi_2=\frac{n}{4}$,
    and where the second line uses the standard identity
    $e^{\frac{n\pi}{2}}=i^n$ in conjunction with
    \cref{eq:basicPhaseCycling},
    and where the third line keeps only the surviving
    terms of the sum.
Thus, subtracting signal from the $\Delta p_1=+1$
    and  $-1$ pathways
    is mathematically equivalent to
    adding the two CPMG components
    ($\varphi_1 - \varphi_2 = \frac{1}{4},$
    $\frac{3}{4}\;\text{cyc}$).
Similarly, a sum of the $+1$ and $-1$ pathways
    -- equivalent to
    isolating the CP
    ($\varphi_1 - \varphi_2 = 0$, $\frac{1}{2}\;\text{cyc}$)
    component -- would cancel
    out the long-lived tails of the signal,
    which are 180° out of phase.
Thus, interestingly,
    the ``CP'' \vs ``CPMG'' components of the
    signal arise naturally from this
    analysis
    as the Fourier conjugate of $\Delta p_1$,
    the $\varphi_1-\varphi_2$ dimension.
Specifically,
    transients for which $\varphi_1-\varphi_2 = 0\;\text{cyc}$
    or $\frac{1}{2}\;\text{cyc}$ ($\pi$ rad)
    are the ``CP'' component
    and those for which
    $\varphi_1-\varphi_2= \frac{1}{4}\;\text{cyc}$ or
    $\frac{3}{4}\;\text{cyc}$
    ($\frac{\pi}{2}$, $\frac{3\pi}{2}\;\text{rad}$)
    are the ``CPMG'' component.
\paragraph{}
In the case where field inhomogeneity
    is much smaller than the $B_1$ multiplied by the number of echoes
    ($\Delta \Omega \ll |\gamma B_1 n_{echo}|$),
    both CP and CPMG
    signals present a significant
    amplitude~\cite{ahola_multiecho_2006}.
The ``bleeding'' observed here
    illustrates the oscillation and decay of the
    CP-component~\cite{ahola_multiecho_2006}
    arising from
    unanticipated storage of coherence along $z$
    (the red line in \cref{fig:CPMGbleeding})
    and clearly indicates when phase encoding
    information is not valid.
Because only magnetization orthogonal to the
    direction of the pulse field in the rotating frame
    (\eg, $x$-magnetization subjected to a 
    $y$-pulse)
    is stored along the $z$-axis,
    this affects only the CP transients of the
    phase cycle,
    and not the CPMG transients.
The resulting loss of phase-sensitive
    information, in turn,
    demands the application of an alternate acquisition
    scheme for many applications~\cite{ahola_multiecho_2006}.
However,
    for various reasons,
    one may wish to harvest as much phase information
    as possible from the standard pulse sequence,
    while optimizing the \gls{snr}.
Here,
    the \gls{dcct} map offers guidance on optimally
    filtering the signal.
In the most rudimentary case,
    this involves zeroing
    the even/odd echoes of each pathway that contain
    only noise
    and frequency filtering along $\nu_2$.
This corresponds to preserving some information
    arising from the CP component
\paragraph{Bleeding of signal}
The \gls{dcct} schema may also be applied to a
    CPMG measurement on a high-field spectrometer
    (\cref{fig:BrukerCPMG}).
Here, the bleeding occurs at a later echo
    number due to the greater homogeneity of the
    superconducting magnet,
    pointing to the possibility of utilizing phase
    information as long as the number of echoes is kept
    limited.
\paragraph{}
Finally,
    note that these results represent examples where
    the phase cycling is constrained to
    a 90° step.
Notable examples in the literature,
    such as the PIETA sequence,
    can maximize the use of concurrent phase cycles of
    multiple pulses for cases where finer-level phase
    incrementation is possible
    (and where \gls{dcct} coding and visualization should
    also prove fruitful)~\cite{baltisbergerCommunication2012}.
\subsection{Algorithms Motivated and Analyzed by \gls{dcct}}\label{sec:algorithmsDCCT}
\paragraph{outline}
The remaining results introduce several realistic
    examples of how the \gls{dcct} map
    visualizes the transformation of the data
    from raw signal, through phasing and alignment,
    and ultimately through integration of the real part of
    the signal.
Importantly, the \gls{dcct} schema provides a means for organizing and
    visualizing the data that enables the validation,
    implementation, and
    optimization
    of these algorithms
    for the purpose of analyzing \gls{odnp}
    data.
Here,
    \cref{sec:exampleIR,sec:exampleEp,sec:integration} are specifically
    attuned  to acquiring \gls{odnp} data while
    \cref{sec:phasingDCCT,sec:alignDCCT,sec:exampleRM,sec:lineshapeDCCT}
    should prove of more general interest.
\subsubsection{Phasing of Echo-Detected Signal}\label{sec:phasingDCCT}
\paragraph{why readdress first-order phasing}
In the field of quantitative \gls{nmr},
    various studies have treated the seemingly
    trivial,
    but ultimately pervasive and fundamentally
    linked issues of automated baseline correction
    and first-order phase
    correction~\cite{DeBrouwer2009,bao_robust_2013,sawallMultiobjective2018}.
The presence of shot-to-shot instabilities of
    the magnetic field and the desire for a seamless
    transition between 1D spectroscopy and
    stroboscopic (\eg, CPMG)
    acquisition further complicates such attempts.
As discussed in theory~\cref{sec:firstorderEcho},
    echo-based signals can simplify the timing/phase
    correction of the signal
    by removing the need to account for the distortion
    or loss of the first few points in the \gls{fid}.
Specifically, acquisition of echo-based signals
    replaces the potentially
    iterative phase and baseline correction with the
    problem of locating the
    center of the echo.
\paragraph{how we find the center}
\cref{fig:ModCoilAttachedTime} highlights the fact that
    even when
    the temporal (left to right) variation of the
    echo intensity is subtle,
    the phase (color) of successive echoes varies
    noticeably.
This encourages the use of the full complex signal
    -- \ie, phase as well as magnitude --
    from as much of the echo as possible in order to
    identify the echo center.
\figHermitianPhasing
\paragraph{demonstration of hermitian phasing}
In fact, a simple algorithm that utilizes the phase
    information as well as the amplitude information of the
    echo can generate well-phased baseline-free signal
    from spin echoes.
As a demonstration,
    a standard sample of water and
    TEMPOL
    generated a series of signals as part of a
    progressive enhancement sequence.
\cref{eq:fastHermitianWithPhase}
    calculates a cost whose minimum locates the
    center of an echo signal with Hermitian
    symmetry.
\cref{fig:HermitianPhasingCost} displays this
    cost averaged across 28 indirect power steps
    at microwave powers ranging from 0 to 4 W.
The cost function exhibits a well-defined minimum
    at $\Delta t_{min}/2=10.6 \text{ms}$.
After subtraction of this value from the time coordinates
    and application of a uniform zeroth order phase correction
    to all scans,
    all signals display approximate Hermitian symmetry.
The residual calculated by subtracting the signal from
    its Hermitian transpose ($s^*(-t)$) barely rises
    above the level of noise,
    and the imaginary
    components of all scans cross zero at $t=0$
    (\cref{fig:HermitianPhasingImaginary}).
In other words, the detected echo accurately comprises a
    rising signal ($s_{rising}(t)$) for $t<0$ and a
    mirror image (in the sense of
    Hermitian symmetry, such that $s(t) = s^{*}(-t)$)
    \gls{fid}
    ($s_{FID}(t)$) for $t>0$.
Under standard electromagnet conditions,
    the refocusing of inhomogeneities is dramatic
    with $T_2^* \ll T_2$,
and detection of echoes does not
    lead to a detectable decrease in signal amplitude
    relative to an \gls{fid} acquired after a 90° pulse
    (\cref{fig:FIDvsEchoAmp}).
Finally,
    \gls{fid} slicing (\cref{eq:s_FID})
    followed by Fourier transformation
    gives rise
    to in-phase, baseline-free signals
    (\cref{fig:HermitianPhasingFT}).
\paragraph{how long does $\tau$ need to be for Hermitian phasing?}
Slightly longer echo times are preferred as they
    allow the spectrometer to acquire a
    sufficient amount of rising echo signal between the
    pulse dead-time and the center of the echo.
Echo times of as low as 3~ms
    enable~\cref{eq:fastHermitianWithPhase} to easily
    determine the echo center for 15~MHz
    \gls{nmr}.
While the timing correction needed may be a
    conglomerate of various contributions,
    such as those mentioned in
    \cref{sec:firstorderEcho},
    the system in the authors' lab typically requires a
    timing correction of 100-500\us relative to the
    expected center of the echo
    ($t_{echo}=\tau+2 t_{90}/\pi$~\cite{rance_obtaining_1983}).
Repeated experiments typically reproduce the
    echo location satisfactorily and enable signal averaging.
\subsubsection{Simple NMR Signal Alignment}\label{sec:alignDCCT}
\paragraph{introduce alignment}
Even with a reasonable effort to maintain a stable
    resonance frequency,
    significant phase cycling noise tends to appear in the
    \gls{dcct} map of signal acquired on an
    electromagnet,
    as shown in~\cref{fig:StepByStepInvPhaseCorr,fig:stepTwoEp}.
\Cref{fig:PhCycleNoiseBefore}
    illustrates the \gls{dcct} map of the signal from a
    simple spin echo with a short ($\sim 1\;\text{ms}$) echo time on
    the 15~MHz electromagnet system,
    repeated for 10 scans (\ie, 10 complete phase cycles)
    for an aqueous $100\;\text{mM}$
    TEMPOL.
During this relatively short echo time,
    the field has little opportunity to drift,
    and the signal refocuses at the echo center ($t=0$)
    almost completely
    and with a consistent phase.
The amplitude of the phase cycling noise
    relative to the amplitude of the signal increases
    noticeably for times increasingly further away from the
    center of the echo.
This effect matches the expected effect of a transient
    to transient variation of resonance (field) offset,
    which does not affect signal at the echo center,
    but does affect the evolution of signal phase
    moving away from the
    echo center.
Thus, an initial inspection implies that a significant
    portion of phase cycling noise results from field
    fluctuations.
\paragraph{noticeable drift}
\figPhCycleNoise
To ascertain the extent to which the presumed drifting of the static magnetic
    field is the cause of the phase cycling noise
    in \cref{fig:PhCycleNoiseBefore},
    the iterative maximization of
    \cref{eq:CorrelOptMulti}
    (subjected to a Gaussian mask of width
    $\sigma=20$ and nonzero only for the coherence
    pathways $\Delta p_1=0$, $\Delta p_2=-1$ and
    $\Delta p_1=0$, $\Delta p_2=0$~\cref{fig:Mask})
    determines the frequency shifts required to align
    the individual transients,
    leading to the result shown in
    \cref{fig:PhCycleNoiseAfter}.
The alignment
    mitigates the phase cycling noise
    while improving the consistency of the signal in
    the desired \gls{ct} pathway.
Thus, the signal alignment
    of \cref{eq:2DcorrelationWithMask}
    proves a viable means to
    address the experimental
    complications owed to randomly varying offset.
\subsubsection{Example: Inversion Recovery}\label{sec:exampleIR}
\figStepByStep
\figStepByStepEp
\paragraph{inversion-recovery example}
\cref{fig:StepByStepInv}
    shows the data for an inversion recovery
    experiment at various stages of processing for
    an aqueous $500\;\text{μM}$ TEMPOL solution.
First,
    \cref{fig:StepByStepInvRawSig}
    shows the ``raw'' data
    with two separate phase cycling
    ($\varphi_1$, $\varphi_2$) and direct
    ($\nu_2$) dimensions.
This figure has been subjected only to a time-domain
    Fourier transform and is included here as a conceptual aid:
    note that viewing data in the phase domain rarely
    proves more diagnostic than viewing it in the coherence domain.
Without requiring any phase correction,
    \cref{fig:StepByStepInvRawSig,fig:StepByStepInvPhaseCorr}
    illustrate
    a 180° phase change (sign inversion) as the
    signal passes through the signal null
    of the inversion recovery curve
    ($M(\tau) = 0$ at $\tau = -T_{1}\ln(\frac{1}{2})$).
A unitary Fourier transformation subsequently converts
    the phase cycling
    dimensions into the respective coherence
    domains,
    gathering
    the majority of the signal into the correct coherence
    pathway.
This indicates that the majority of the equilibrium
    magnetization goes toward generating the signal
    of interest --
    as opposed to a situation where pulse misset
    or inhomogeneities expend a significant
    portion of the signal on artefactual
    pathways.
Thus, a quick glance at the \gls{dcct} map 
    reveals the relative success or failure of
    the choice of pulse lengths and inter-pulse delays
    (\eg, the delay between the initial 180°
    pulse and the second 90° pulse)
    without the need
    for phase correction or for intensive data
    processing.
\paragraph{phase correction}
As before,
    the Hermitian symmetry test enables
    an automated means to
    find and apply the timing correction.
Upon shifting $t=0$ to
    the center of the echo,
    much of the first order phase error
    (which appears as a horizontal color variation in
    \cref{fig:StepByStepInvRawSig})
    along the $x$-direction disappears
    (\cref{fig:StepByStepInvPhaseCorr}).
Finally, after alignment
    (\cref{fig:StepByStepInvAligned}),
    the phase cycling noise reduces in intensity.
\paragraph{discussion of alignment}
During the alignment procedure
    here and in the following results,
    five features prove to be essential.
First,
    the algorithm must perform any timing
    (first order phase) corrections before alignment,
    since signals of different phases do not align
    properly.
Second,
    a long acquisition
    (corresponding to a well-resolved, if noisy,
    frequency domain)
    capable of capturing sharp features yields the most
    dramatic improvements.
Third, to allow for high resolution in the calculated
    $\nu_{shift,max}$,
    a significant
    zero-filling of the time domain,
    together with a simple (exponential or Tukey)
    apodization
    precedes Fourier
    transformation to the frequency domain.
Fourth,
    \cref{eq:CorrelOptMulti}
    requires a cross-correlation function for
    each transient cross-correlated
    against all others,
    which permits higher \gls{snr}
    than, \eg,
    calculating only cross-correlations for
    nearest-neighbor transients, as might be implied by
    \cref{eq:CorrelOptFull}.
Finally, as noted in \cref{sec:crossCorrel},
    optimizing the masked energy/norm of the signal,
    rather than the full energy/norm of the signal,
    proves to be essential.
Once the echo signal is aligned,
    the \gls{fid} is isolated \cref{eq:s_FID},
    and its Fourier transform
    is shown in
    \cref{fig:StepByStepInvReal}.
\subsubsection{Example: Progressive Enhancement}\label{sec:exampleEp}
Following the same data treatment applied to the
    inversion recovery dataset shown in
    \cref{fig:StepByStepInv} and discussed in
    \cref{sec:exampleIR},
    \cref{fig:StepByStepEp}
    presents data for a progressive enhancement
    experiment for an aqueous $6\;\text{mM}$ TEMPOL
    sample.
Notable features of this data include an increase
    in signal intensity along the indirect dimension,
    $p$ (\gls{mw} power)
    as well as an inversion of the data
    following the first indirect step
    (corresponding to the point at which \gls{odnp}
    polarization transfer exceeds the thermal
    polarization). 
The raw data in the frequency domain
    shows the characteristic banding expected of an
    echo detected without a timing correction,
    in \cref{fig:stepOneEp}.
The timing correction 
    (\cref{fig:StepByStepEpCohDom}$\rightarrow$\cref{fig:stepTwoEp})
    makes the signal phase uniform across all
    frequencies and coherent (one color) along
    the indirect dimension.
The correlation alignment
    (\cref{fig:stepTwoEp}$\rightarrow$\cref{fig:stepThreeEp})
    removes the scan-to-scan shifting of the signal
    and also relieves the slight phase cycling noise,
    as shown
    in~\cref{fig:stepThreeEp}.
Lastly, the data with integration bounds is shown
    in~\cref{fig:AlignEpOverlay}.
\pdfcommentJF{Your wording here makes me realize I should check -- none of
    the inversion recovery data uses a clock correction?
    Should I briefly mention the clock correction?}
\pdfcommentAB{We are using it for the step by step
    IR data. I could have sworn an earlier version
    said something about it, but I cannot find it}
\pdfcommentJF{we might have regarded this as a
    hardware issue and removed when we sliced out the
    coherence stuff -- I think it is specific to the
    hardware, and am OK to leave it out in these last
    days}
\subsubsection{Lineshape Improvement and Choice of Integration Bounds}\label{sec:lineshapeDCCT}
\figGaussianApoReal
\gls{odnp} and other low-field \gls{mr} techniques require
    care when filtering and apodizing the data.
Therefore, this section investigates the interplay
    of the \gls{dcct} technique with these methods.
\paragraph{Filtering both Frequency and Time Domains (needs to precede Gaussian apodization)}
To filter out unwanted frequencies in an
    experiment,
    the operator inspects
    the frequency-domain \gls{dcct} map
    and slices out
    the portion of the frequency axis
    containing significant signal,
    thus filtering out off-resonance noise and reducing the
    memory footprint of the signal.
Domain coloring significantly assists in this process,
    since incoherent noise appears as a multi-color
    scatter that can appear grey at a distance,
    while the weaker shoulders of peaks,
    \etc,
    trend toward a common color,
    as exemplified in \cref{fig:RevMicBefore}.
The pySpecData library accepts slices given in
    frequency units (not requiring an array index)
    in a compact notation (\cref{lst:frqSlice})
    and automatically recalculates the new time axis
    (since the slicing operation increases the spacing
    between the time-domain datapoints).
\paragraph{}
\cref{fig:GaussianApodizationRealBefore}
    presents
    preliminary data from
    an echo-detected inversion recovery experiment
    conducted on a sample
    of $150\;\text{μM}$ TEMPOL in toluene;
    this particular inversion recovery was conducted
    with no microwave power, thus without \gls{odnp} signal
    enhancement.
For compactness, it shows only the coherence pathway of
    interest ($\Delta p_1=+1$, $\Delta p_2=-2$).
This dataset presents a few challenges:
    \gls{snr} is limited,
    the acquisition time is relatively short --
    reflected here 
    as a pixelation along the direct dimension
    (which has not been zero-filled),
    and the resonance frequency varies slightly with
    the magnetic field of the electromagnet.
Despite these issues,
    two distinct peaks
    (here vertical red bands of color)
    appear in the correct block of the
    \gls{dcct} map.
Furthermore, despite the fact that,
    the echo time is relatively short,
    \cref{eq:realAbsCost} can determine the center of
    the echo to enable facile phasing.
The peaks in
    \cref{fig:GaussianApodizationRealBefore}
    (blue-green at lower scan number and red at higher scan number) fade off into a
    constant yellow-green and purple
    color to either side of the direct frequency
    dimension.
When properly phased,
    large portions of
    the dispersive tails
    of peaks
    present a constant imaginary
    phase
    that does not vary in phase with frequency
    (because the phase of a Lorentzian
    follows an arctan function as a function of frequency).
Therefore, the yellow-green and purple regions
    are the dispersive tails of the resonance
    and the consistency of their coloration
    indicates that the \gls{fid} has been
    correctly sliced from the echo with no
    further phasing correction required.
\paragraph{}
Most standard \gls{nmr} data-processing routines
    involve filtering in the time domain (apodization),
    with the specific goal of reducing or eliminating
    the contribution of noise from points
    where the signal has decayed to near zero.
However, this step requires more care than
    frequency domain filtering.
Inhomogeneities frequently lead to
    sharp peaks in otherwise broad lineshapes,
    meaning that signal in the direct time domain
    does not decay with a single $T_2^*$ time constant.
Optimal processing of \gls{odnp} data, therefore, requires
    acquisition with a long direct dimension 
    (corresponding to more detailed frequency resolution)
    followed by
    apodization that filters
    out unnecessary noise.
In particular,
    tests on the results acquired on customized
    instrumentation (where the acquisition length can
    sometimes be limited)
    indicate that proper treatment of the time domain
    signal is required for
    attempts at phasing by \cref{eq:realAbsCost}
    or at alignment (as will be discussed later).
Specifically, the acquisition length
    must either exceed $5/\pi\times$ the linewidth of the
    finest feature
    or else must be multiplied in the time domain
    by an exponential decay (\ie, apodized)
    whose time constant is at most $1/5^{th}$ of the
    total acquisition length.
For example,
    the proper application of \cref{eq:realAbsCost}
    to generate
    \cref{fig:GaussianApodizationRealBefore}
    requires such apodization.
While workarounds based on non-Fourier
    processing
    methods~~\cite{MandelshtamPNMRS,Hoch2017,Zambrello2018,Hyberts2010,Stern2002}
    exist,
    attempts to ignore this reality using Fourier-based
    algorithms lead either to issues with peaks
    described by relatively few points
    (pixels in the \gls{dcct} representation)
    or with peaks subjected to dramatic sinc
    interpolation.
\paragraph{}
Ideally, knowledge of the time-domain decay of the
    signal envelope guides the choice of a
    well-matched apodization function.
The varying signal phase and frequency
    in \cref{fig:GaussianApodizationRealBefore} indicate a preference for
    finding the signal envelope
    by summing the absolute value of the individual
    scans in the time domain.
A simple least squares fit of the result to
    \cref{eq:foldedNormal,eq:FWHMdefLorentz}, in general,
    properly determines both the peak amplitude
    ($A$) and
    noise level ($\sigma_n$) of the signal.
However, the least squares fitting algorithm adjusts
    the parameters associated with linewidth
    ($\lambda_L$)
    in a manner inconsistent with tracing the outer
    edge of the signal envelope -- instead passing
    through the middle of any oscillations.
Therefore,
    keeping the same functional form,
    the code scans from a minimum $\lambda_L$
    (here, typically 10~Hz)
    up to the least-squares fit value of $\lambda_L$,
    and determines the norm of the data exceeding the
    fit function at each linewidth.
The point where this norm is a fifth of the way from
    its lowest value to its highest value
    (at the least squares $\lambda_L$)
    is chosen as the $\lambda_L$ of the envelope.
The dotted line in \cref{fig:GaussianApodizationRealEnvelope}
    shows the envelope calculated for the toluene data.
\paragraph{}
Armed with this choice of an enveloped-matched filter
    (\cref{fig:GaussianApodizationRealEnvelope}),
    the results of \cref{fig:GaussianApodizationReal}
    explore the impact of some common apodization
    functions.
Some instances of \gls{odnp} benefit from the ability to
    resolve chemical shifts,
    as demonstrated recently in several
    contexts~\cite{levienSpin2021,Keller2020}.
In \cref{fig:GaussianApodizationRealAfter}
    an equal-linewidth \gls{l2g}
    transformation (\cref{eq:L2GequalWidth})
    aids the resolution of two peaks in the \gls{nmr}
    spectrum without noticeably degrading the
    \gls{snr}.
Notably,
    the apodization emphasizes
    a faint white region between the peaks with
    neither dispersive/imaginary
    (yellow-green or purple)
    nor absorptive/real
    (red) signal.
\cref{fig:GaussianApodizationRealOneDRe}
    demonstrates the same effect in 1D format
    and also the impact of an equal-energy
    \gls{l2g} transformation
    (\cref{eq:L2GequalE}),
    which slightly degrades resolution, but
    significantly reduces the noise.
It is also interesting to note that
    (not shown here),
    by emphasizing particular regions of time,
    the apodization function can have a notable effect
    on the inactive pathways in the coherence domain,
    reducing some phase cycling noise and/or reducing
    the amplitude of artefacts that occupy different
    portions of the time domain \vs the signal.
The equal-energy transformation
    improves the \gls{snr} of the
    peak,
    without significantly broadening the range over
    which it is non-zero.
A procedure to automate the choice of integration
    bounds can apply \cref{eq:L2GequalE},
    and also use the $\sigma_n$ fit from the envelope
    equation
    (\cref{eq:foldedNormal,fig:GaussianApodizationRealEnvelope})
    to determine the integration bounds from the points
    where the (equal energy \gls{l2g}) apodized
    peak intersects with 0.5 the standard deviation of
    the noise of the original spectrum.
The grey lines in \cref{fig:GaussianApodizationRealOneDRe} provide an
    example of the procedure for the present data.
Overall, the results shown and discussed here emphasize
    that a well-chosen implementation of the standard
    technique of
    \gls{l2g} apodization provides
    significant benefit to low-field \gls{odnp} data
    and that these effects contribute cooperatively
    to the benefits of the \gls{dcct} map.
\subsubsection{Example: Integration of Inversion Recovery and Progressive Enhancement}\label{sec:integration}
\figAlignIR
\figAlignEp
Every rigorous \gls{odnp} dynamics analysis following the
    current standard protocol~\cite{FranckMethEnz2018}
    requires a series of inversion recovery ($R_1(p)$) and
    progressive enhancement ($E(p)$) measurements
    and
    demands a means to easily process the raw data and
    unambiguously determine the relevant parameters
    (see \cref{sec:ODNPtheory})
    without significant user input.
An important component of this challenge involves
    converting spectra along the direct dimension
    ($\nu_2$)
    into signal intensities.
\cref{fig:AlignIR,fig:AlignEp} compare
    data
    that has been integrated,
    both with and without first implementing the
    correlation alignment of \cref{sec:alignDCCT}
    and shown in
    \cref{fig:stepThreeEp,fig:StepByStepInvAligned}.
\paragraph{inv rec}
First, we consider the inversion recovery data.
A significant improvement in the signal alignment is observed in
    \cref{fig:StepByStepInvAligned},
    indicated by the consistent color and intensity
    of the pixels along the $\tau$ axis,
    which is not observed in
    \cref{fig:StepByStepInvPhaseCorr} before alignment.
In addition, an inspection of the \gls{dcct} map shows
    much weaker signal intensity in the inactive
    coherence transfer pathways,
    corresponding to reduced phase cycling noise.
Integrals, of course,
    prove somewhat forgiving of the resulting errors in the
    unaligned signal.
The expected noise variance of an integral over a given bandwidth
    can be determined from standard error propagation
    formulae to be $\sigma_\nu^2 N \Delta \nu^2$,
    where $\sigma_\nu^2$ gives the variance of the
    frequency-domain datapoints,
    $\Delta \nu$ gives the spacing between them,
    and $N$ is the number of points.
Thus, the errors (standard deviation) scale with 
    the square root of the size of the integration
    window.
\cref{fig:AlignIR} shows modest improvement
    in the integrals after
    alignment:
    tighter integration bounds slightly reduce the error
    of the aligned signal.
In keeping with the smaller error bars,
    as well as the less ambiguous choice of the
    integration bounds,
    alignment also corrects the two datapoints at
    longer recovery delay
    with both showing a better fit to
    the model (\cref{eq:invRec}) after alignment.
\paragraph{}
The signal intensities for progressive enhancement
    ($E(p)$) data were considered somewhat more carefully.
Here, after slicing out the \gls{fid}
    from the data of \cref{fig:stepThreeEp},
    the code applies an equal-energy \gls{l2g}
    apodization~\cref{eq:L2GequalE} to improve the \gls{snr} of the signal.
Because this data comes largely from highly enhanced
    signal,
    the result gives a controlled example where the
    main concerns focus on issues of field
    stability and misalignment.
The stability of the signal is somewhat
    run-dependent, with the data in
    \cref{fig:AlignEpOverlay} an example of fairly stable
    progressive enhancement signal.
While the lineshape of  the aligned
    signal (green) remains relatively consistent, the
    lineshape of the unaligned signal (blue)
    slightly varies --
    not only in the position of the central peak
    but also in terms of subtle variations of the
    shoulder.
While the unaligned signal demands integration over fixed
    bounds, the aligned signal proves amenable to a
    weighted sum over the lineshape, resulting in
    tighter error bars (propagated from the noise of
    the datapoints) that produce a tight match
    to the model
    ($R^2$ 0.9987 for unaligned 0.9995 for aligned).
The overall scaling of the error bars for
    the unaligned integration proves very sensitive to
    the choice of integration bounds (data not
    shown); however, no such
    ambiguity exists for the weighted average.
This offers promise for not only the precise
    quantification of important reference standards
    for \gls{odnp}, as done here, but also for a complete
    \gls{odnp} analysis of samples with very low
    concentrations of  either spin label or water.
\subsubsection{Example: Thermal Scan on Reverse Micelles}\label{sec:exampleRM}
\figRevMicBefore
\figRevMicAfter
\figRevMicZoom
\paragraph{RM motivation}
Typical \gls{odnp} studies rely on the fact
    that water comprises most of
    the sample, enabling the use of relatively
    small sample volumes ($\sim 4.8\;\text{μL}$).
Studying the water inside reverse micelles proves
    challenging,
    as the proton spectra present
    $\sim 1/30 \times$ the \gls{snr} of aqueous samples.
However, with longitudinal
    relaxation rates, $R_1$, that are significantly faster
    (\eg $\sim 7.7\;\text{s}^{-1}$)
    than those of aqueous solutions
    ($\sim 0.4\;\text{s}^{-1}$),
    reverse micelles provide an opportunity for rapid
    signal averaging not otherwise possible in
    typical \gls{odnp} samples.
\gls{odnp} studies of solutions of organic solvents
    typically employ larger
    sample volumes
    (even in recent studies
    \cite{levienNitroxide2020}),
    following the rationale that much of
    the sample is made of low-loss dielectric solvent
    that can extend into the electric
    field of the cavity with less substantially detrimental
    heating effects.
However,
    water and dielectrically active surfactant
    comprise an uncomfortable intermediate fraction of the sample,
    resulting in a lower water
    \gls{nmr} signal density,
    but still (based on changes to cavity $Q$)
    presenting a significant concern
    \textit{w.r.t.} sample heating.
Thus,
    even though initial
    reverse micelle studies used large probes and
    subsequently larger sample volumes~\cite{valentine_reverse_2014},
    studies of the dynamics of water inside the reverse
    micelles still require
    a small ($\le 0.6\;\text{mm}$) sample radius
    in order to minimize heating effects.
\paragraph{}
\cref{fig:RevMicBefore} shows a portion of the \gls{dcct} map for
    thermally polarized spin echo \gls{nmr} signal for a reverse
    micelle sample.
To compensate for the very low $^{1}$H content of this system,
    the pulse sequence cycles through the 8-step spin echo phase cycle
    200 times, resulting in 1600 individual
    transients.
The 1D plots at the top of \cref{fig:RevMicBefore} show
    the spectrum from one
    complete phase cycle (\ie one scan) in blue,
    as well as the
    average over all 200 scans
    in red.
While individual transients have insufficient \gls{snr}
    to clearly identify signal, acquiring many
    transients can lead to broadening  of the averaged
    signal in the presence of slow field drift,
    rendering characterization of thermal signal
    particularly problematic.
In contrast to either of the 1D
    plots,
    the \gls{dcct} map shows an obvious red band of signal
    in the expected
    coherence channel at +75~Hz offset.
It also shows that the signal appears to
    randomly drift over 50-75~Hz throughout the
    course of the experiment.
\paragraph{after correlation}
Despite the low \gls{snr} of the experiment,
    the signal-averaged correlation function
    (\cref{eq:CorrelOptMulti})
    should still be able to align the
    various transients,
    since the correlation functions used to determine
    the frequency shifts are each averaged from 200
    separate correlation functions
    (one for each pair of transients in the experiment).
Indeed, after the correlation alignment
    (\cref{fig:RevMicAfter}), signal again appears in the
    expected coherence channel, but now as an even more
    distinct band of a solid color that spans the 200
    phase-cycled transients -- namely
    the intense red band at 50 Hz,
    with blue side bands arising from
    truncation of the rising portion of the echo.
The 1D plot shows that, after the correlation
    alignment, signal is observed after a single phase
    cycle (blue), and improved and reduced in
    noise significantly after averaging over all 200 cycles through
    the phase cycle (red).
This signal exhibits a significantly reduced
    linewidth of $\sim 125\;\text{Hz}$,
    explained by the alignment of the consecutive
    transients shown in \cref{fig:RevMicZoom}.
\paragraph{}
Even though the net signal energy,
    averaged across repeats,
    remains the same,
    aligning the signal improves the phase coherence of
    the signal from transient to transient.
\cref{fig:RevMicBefore} \vs
    \cref{fig:RevMicAfter} demonstrates a visual
    explanation of this process
    that improves the \gls{snr} slightly within each
    phase cycled scan.
A visual effect
    arising from the improved phase coherence between
    adjacent scans
    also likely contributes to the increased visibility of the
    signal in \cref{fig:RevMicAfter}.
Overall, the signal-averaged alignment makes the signal
    much more visible in not
    only the averaged signal
    but also within the individual scans displayed in
    the \gls{dcct} map.
This is a striking and somewhat unexpected result.
\section{Discussion}\label{sec:discussion}
\paragraph{what makes up DCCT}
The results here advocate for
    a non-standard approach to utilizing
    coherence pathways.
The \gls{dcct} schema comprises 4 interdependent features:
(1) storage of all transients
    (\ie, not phase cycling the receiver and/or
    averaging on-board),
(2) multidimensional organization of data,
(3) object-oriented code that assists in data
    manipulation,
    and 
(4) visualization of data as a \gls{dcct} map at a
    relatively early stage of the data processing.
\paragraph{value of coherence domain}
The visualization of all separable
    coherence pathways along an additional, short
    dimension,
    enables efficient diagnosis and mitigation
    of effects as simple as 
    instrument miscalibration and as subtle as
    the phase cycling noise caused by unstable
    fields.
In fact, as shown in \cref{sec:data_on_scope},
    less specialized equipment -- \ie, hardware
    without the capability to phase cycle the
    receiver -- 
    requires this approach in order
    to separate the desired signal from undesired artefacts.
\paragraph{combo of domain color and CT is fortuitous}
One notable observation is
    the fact that
    domain coloring combines synergistically with the
    simultaneous presentation of multiple \gls{ct}
    pathways.
In particular,
    domain coloring
    enables the straightforward,
    simultaneous visualization of all \gls{ct} pathways
    by bypassing complexities
    that arise
    when different pathways have
    different phases or timings and/or varying
    resonance frequencies.
By offering a direct, compact visualization of
    phase coherence as pixels of matching or slowly
    varying color,
    the \gls{dcct} map also serves as
    an excellent tool for detecting signals that might
    otherwise appear to be only noise.
In contrast,
    a lack of coherence,
    \eg, between
    echo centers of subsequent scans
    in the presence of field instabilities (\cref{fig:ModCoil}),
    also becomes quite obvious.
Observation of low-\gls{snr} \gls{dcct} maps
    motivated the development and implementation
    of the signal-averaged mean-field correlation
    alignment presented here,
    which results in a very
    clear improvement in the signal linewidth,
    \gls{snr},
    and the means for quantifying signal.
\paragraph{DCCT is comprehensive and immediate}
While traditional techniques involve discarding
    at least some data
    (undesired \gls{ct} pathways,
    the imaginary part of data, \etc),
\gls{dcct} maps afford a comprehensive overview of all
    acquired data in one image,
    at a very early stage of data processing.
Importantly, the \gls{dcct} map can yield these results 
    and other informative comparisons between
    datasets without
    first requiring detailed
    phasing or other manipulation of the raw
    datasets (\cref{fig:DomainColorIntro,fig:ScopeData,fig:NutationResults}).
Thus, the \gls{dcct} schema enables deterministic and rapid
    progress toward acquisition and optimization of
    \gls{nmr} signal.
The studies here, in particular, demonstrate that \gls{dcct} maps
    prove invaluable in the initial stages of instrument
    setup, as well as in diagnostic efforts for failed
    experiments -- \eg, the incorporation of
    new probes or receiver chain components --
    thus providing significant support for
    instruments with modular capability.
These results also showcase the ability of the
    \gls{dcct} map to highlight three important
    features of the data:
    (1) the desired signal in the desired
    \gls{ct} pathway,
    (2) artefacts arising from artefactual
    pathways and potentially contaminating the
    desired pathway,
    and (3) phase cycling noise
    spread across multiple pathways.
\paragraph{history: data processing lagged}
An argument can be made that standard data processing
    technology frequently
    lags behind the established understanding of
    coherence pathways by jumping
    straight to the selection of the desired coherence
    pathway.
Even though researchers were aware
    detailed analysis of coherence pathways could enrich
    experimental design~\cite{jerschowEfficient1998},
    and one could infer similar gains
    by extending such an analysis to
    the development of data-processing algorithms,
    it was frequently necessary to skip the development of such
    algorithms due to the
    resulting complexity and incompatibility with
    standard acquisition techniques
    and data processing software.
As has repeatedly been noted in the
    literature~\cite{ivchenko_multiplex_2003,gan_enhancing_2004,schlagnitweitMQD2012,baltisbergerCommunication2012,waudbyTwodimensional2020},
    the days before
    direct detection and digital filtration of rf
    signals
    (when
    limited memory space requirements
    and slow data transfer rates were a primary
    concern)
    have left many historical artefacts that continue
    to impact typical current approaches to \gls{nmr}
    data acquisition and processing.
In particular,
    the rather stringent constraints of minimizing
    the size of temporarily-stored time domain data
    through onboard averaging of transients acquired
    with different phases
    motivated the design of what has become the
    traditional methodology for treating coherence
    pathways in \gls{nmr}~\cite{keeler_understanding_2011}.
While the traditional schema still proves optimal in
    many cases,
    equipment available today welcomes virtually
    cost-free storage of all
    transients in a phase cycle.
The \gls{dcct} schema outlines a straightforward means to
    profit from these benefits
    \textit{via} object-oriented treatment of the resulting
    multi-dimensional data
    and robust multi-color
    domain coloring plots.
Importantly, it does not add any
    time cost to a more
    traditionally-acquired NMR experiment.
\paragraph{historical context}
The \gls{dcct} approach is not without historical precedent.
Very early
    examples within the context of
    multiple quantum
    spectroscopy~\cite{wokaunSelective1977,drobnyFourier1978,Bodenhausen1984}
    explicitly Fourier transform
    along a dimension where phases were cycled in
    a procedure formalized as the ``Phase Fourier
    Transform.''
More recently, separate storage of the transients to perform this
    transformation has proven advantageous in 2D
    multi-quantum experiments,
    such as in multiplex phase
    cycling~\cite{ivchenko_multiplex_2003} and
    MQD~\cite{schlagnitweitMQD2012},
    as well as in echo train
    experiments~\cite{baltisbergerCommunication2012}.
It is therefore important to note that many modern
    laboratories utilize a methodology
    similar to a subset of the \gls{dcct} schema,
    although the authors are unaware of an extant
    formal description of this very useful procedure
    in the literature,
    nor of a consistent and convenient
    accompanying data-visualization technique,
    like the one presented here.
As observed in the current contribution,
    the general concept garners the widest utility
    when also used as a means to explicitly
    consider signal
    in regions of the coherence domain that are
    \textit{not} selected by
    the pulse sequence.
\paragraph{new visual context for understanding}
The \gls{dcct} schema, further, offers a means for visually
    representing several known
    but important properties of phase cycling.
For example,
    Plancheral's theorem emphasizes that the noise from
    different transients in the phase cycle domain will
    spread equally across the coherence domain.
Therefore, the selection of a single coherence pathway
    yields the same \gls{snr} benefit as (not phase cycled) signal averaging.
The low field experiments in the results here typically
    benefit from the added \gls{snr} of additional phase cycling,
    since these studies tend to be \gls{snr}-limited.
However, for the implementation of more elaborate pulse
    sequences,
    one desires a straightforward strategy for choosing an optimal
    phase cycle~\cite{jerschowEfficient1998}.
The \gls{dcct} schema offers such a strategy.
Specifically,
    a \gls{dcct} map of signal from
    a comprehensive phase cycle
    (\eg, where all pulses are phase cycled by 4 or more
    steps)
    would clearly show where
    in the coherence domain
    (for which values of $\Delta p_1$, $\Delta p_2$, \etc)
    artefacts occur and where they do not occur;
this clearly indicates which parts of the phase cycle
    can be reduced and which cannot.
Eliminating or reducing the number of phase cycle steps
    for pulses whose coherence domain
    shows only noise
    along one or more dimensions only decreases \gls{snr}.
In contrast,
    eliminating or reducing the phase cycle for pulses
    where the coherence domain
    shows artefactual signal
    will alias those artefacts~\cite{Bodenhausen1984},
    potentially into the desired coherence pathway.
\paragraph{example of how to improve phase cycling}
As a simple example, in
    \cref{fig:NutationResults}, elimination of the phase
    cycle of the first pulse
    would add noise of the
    $\Delta p_1 = \pm 1$, $\Delta p_2 = \pm 1$ pathway to
    that of the
    $\Delta p_1 = 0$, $\Delta p_2 = - 1$
    pathway and the noise of the
    $\Delta p_1 = 0$, $\Delta p_2 = 0$
    pathway to the $\Delta p_1 = + 1$, $\Delta p_2 = -2$
    pathway.
While this decreases the \gls{snr} by $\sqrt{2}$, the
    artefact
    ($\Delta p_1=0$, $\Delta p_2=-1$)
    remains separated (no significant DC
    artefacts at $\Delta p_1 = 0$, $\Delta p_2 = 0$
    are present here),
    and the number of scans reduces
    by a factor of 2.
In contrast, 
   elimination of the phase cycle altogether adds
   all four coherence pathways together,
   superimposing the \gls{fid}-like artefact in $\Delta
   p_1 = 0$, $\Delta p_2 = -1$ onto the
   actual echo signal in $\Delta p_1 = + 1$, $\Delta p_2 = -2$. 
Similarly, the CPMG
    results in \cref{fig:CPMGsignal} indicate 
    when a 4-step \vs 2-step phase cycle of the
    excitation pulse proves useful.
These are simple examples,
    but provide a demonstration of an empirical method
    that can be extended to more complex pulse
    sequences.
\paragraph{overall summary of how to acquire data}
For application to experiments in electromagnets and
    other non-ideal fields,
    the results here advocate a strategy of
    acquisition lengths spanning many times
    the $T_2^*$
    (with ideal timescales governed rather by the
    sharpest features of the lineshape),
    with phasing identified from the Hermitian symmetry
    of the echo signal (\cref{sec:firstorderEcho}),
\pdfcommentJF{note to self -- I need to address whether
    the apodized traditional version also works well}
    followed by alignment (\cref{sec:crossCorrel}) and
    apodization (\cref{sec:GaussianApodization}).
The benefits of echo-based
    detection complement the \gls{dcct} schema,
    as echo-based signal follows a distinctive
    coherence pathway.
Thus, phase cycled echoes offer particular advantages over
    90° pulse-acquire schemes when
    acquiring signal on a system
    for the first time.
Aside from offering a maximum signal intensity
    that is not sensitive to field inhomogeneity,
    echo signals are also cleanly isolated from any ringdown
    arising from the pulses.
The signal appears in 1 out of the 8 separate pathways
    in the coherence domain that are
    resolved by a typical phase cycle,
    and it can be compared to the amplitude of the noise
    and artefacts of signal in these other pathways.
Furthermore,
    echo-based signals do
    not require post-processing
    (\eg, linear prediction or polynomial fitting)
    to yield baseline-free spectra,
    which here proves ideal for the quantitative \gls{nmr}
    necessary for \gls{odnp}.
\paragraph{}
Beyond these specifics,
    however,
    the most important overarching result reported
    here is the unexpected synergy between domain
    coloring for the visualization of signal phase,
    the simultaneous visualization of all coherence
    pathways,
    and modern object-oriented programming that enables
    organization and plotting of the data.
\section{Conclusion}\label{sec:conclusion}
\paragraph{extending modular design}
The \gls{dcct} schema
    offers a powerful new standard for visualizing data
    (the \gls{dcct} map)
    that proves unexpectedly useful 
    and versatile.
The results presented here highlight the
    \gls{dcct} schema as
    critical for quickly identifying and
    optimizing signal in the presence of obstacles
    such as inhomogeneous fields
    or reduced \gls{snr}, as in low field \gls{nmr}.
As repeatedly noted in the literature,
    eschewing on-board signal averaging offers a range of
    practical benefits,
    but beyond this
    the \gls{dcct} schema
    provides a detailed yet comprehensible view of all of the data
    acquired during the course of an \gls{nmr} experiment,
    leading to a significant degree of confidence
    in the quality of the results and subsequent
    data processing.
This technique also serves as a means
    to elucidate experimental errors that
    otherwise
    would be overlooked, serving to improve the
    overall execution of \gls{nmr}.
Further, the use of the \gls{dcct} map to visualize the
    complete suite of coherence transfer pathways
    attainable for a given pulse sequence offers
    insight into empirical phase cycle optimization for a given
    experiment.
The success of these methods in improving the
    practicability of the modular \gls{odnp} system here
    indicates future successes in other fields of
    customized \gls{nmr}.
Due to these benefits, the \gls{dcct} map
    will serve as an excellent tool for detecting signal
    whether in routine spectroscopy of biomolecules,
    in a spectrometer subjected to appreciable field
    drift, or in samples with challenging low proton
    concentrations that would otherwise produce only
    noise.
The ability of the \gls{dcct} map to aid in
    detecting signal despite significant noise has
    applications to low field systems as well as when
    dealing with simple but costly human errors
    on well-configured high
    field systems.
Lastly, the \gls{dcct} map is expected to contribute to
    aid in the development of data processing routines
    and algorithms, such as the signal-averaged
    mean-field correlation alignment presented here,
    given its ability to highlight shortcomings in
    the execution of \gls{nmr} experiments.
The authors forsee particular applications to pulsed
    \gls{esr}~\cite{Saxena1997,Dzikovski2020} and 2D NMR at low field.
However, the target of future applications ranges
    broadly,
    from
    experiments as closely related as portable \gls{mr} to
    those as distantly related as two-dimensional
    coherent laser spectroscopy.
\begin{acknowledgments}
This material is based in part upon work supported by the
National Science Foundation under Grant No.~2146270.
This work is also supported in part by
the Syracuse University College of Arts + Sciences
in the form of by startup funds,
as well as the Graduate School Summer Dissertation
Fellowship received by AAB.
The authors would like to thank all the members of
ACERT and Prof.~Jack Freed's lab, as well as members of
Prof.~Songi Han's lab for fruitful discussions and the
opportunity to develop nascent stages of the concepts presented
here.
The authors would also like to thank the ODNP team at Bridge12
and Dr. Tom Casey for fruitful discussions
and to note that several features of the pySpecData
library have also been ported to the DNPLab Python package
\url{http://dnplab.net/}
for
broader dissemination to the community.
\end{acknowledgments}
\section*{Author Declarations}
The authors have no conflicts to disclose.
\section*{Data Availability}
Raw datasets underlying all figures are stored in
HDF5 format and are freely available from the
authors upon request.


\appendix
\renewcommand\thefigure{Apdx.\arabic{figure}}
\setcounter{figure}{0}
\section{Code Snippets}\label{sec:codeSnippets}
\paragraph{FID slicing}
The \gls{fid} is sliced from the echo by subtracting the
delay between the start of acquisition and the center
of the echo, $t_{delay}$ from the original $t_2$ axis
coordinates,
then placing the axis in register (so that one of the
axis coordinates occurs at exactly $t_2=0$),
slicing out only values $t_2>0$,
and finally setting the first datapoint to $1/2$ its
original value.
\begin{lstlisting}[language=python,label=lst:FID,caption=FID slice]
s["t2"] -= t_delay # set back the t2 axis
                   # by time t_delay
s.register_axis({"t2":0}) # ensure that the t2
#                           axis contains a
#                           time point 0
s=s["t2":(0, None)] # discard time points
#                    before t=0
s["t2",0] *= 0.5 # multiply first point by 0.5
\end{lstlisting}
\paragraph{bruker ppg}
Pulse programs supplied from Bruker
    implement the `traditional' schema for phase
    cycling--namely, cycling of the receiver phase and
    signal averaging on board.
In order to store all the
    phase cycled transients separately,
    slight modifications to existing pulse programs are
    needed.
The following demonstrates a template that has been
    successfully used to accomplish this task in a
    Bruker pulse program:
\lstdefinelanguage{bruker}
{
    keywords={lo,wr,times,loopcounter,st,st0,ipp0,ipp1,ipp2,if,to},
    morecomment=[l]{;},
    morecomment=[s][\color{brown}]{"}{"},
}
\begin{lstlisting}[language={bruker},label=lst:bruker,
    caption={Separately saved phase cycling in Bruker
    pulse program}]
"l21=4" ; length of phase program ph1
"l22=2" ; length of phase program ph2
loopcounter num_ph
"num_ph = l21*l22"
; td1 set to total number of transients
; (including the phase cycle)
"l20 = td1/num_ph" ; calculates indirect
                   ; dimension, if it exists
"nbl = num_ph"

1     st0 ; reset buffer pointer to 0 when
          ; phase cycle is complete
2     p1 ph1 ; apply 90 pulse according
             ; to phase program ph1
      ; (rudimentary template showing only key
      ; features -- delays, etc are omitted.
      ; A standard spin echo would have a
      ; delay here.)
      p2 ph2 ; apply 180 pulse according
             ; to phase program ph2
      goscnp ph31 ;acquisition of data,
                  ;rx phase always 0
      st
      ipp1 ; increment phase of first pulse
      lo to 2 times l21 ; controls loop over
                        ; phase program ph1.
                        ; By convention, we
                        ; use loop counter
                        ; l2X for phase
                        ; cycle X
      ipp2 ; increment phase pointer
           ; to program ph2, as outermost
           ; phase cycling loop
    lo to 2 times l22
  wr #0 if #0 id0 ; even for experiments that
  ; don't require d0, id0 is required for
  ; standard behavior
lo to 1 times l20 ; if indirect dimension
                  ; employed, this loop
                  ; generates the indirect
                  ; dimension

; phase programs
ph1 = 0 1 2 3
ph2 = 0 2
ph31=0 ; receiver phase, always 0
;note through loops and phase pointers,
;we avoid:
;ph1 = 0 1 2 3 0 1 2 3
;ph2 = 0 0 0 0 2 2 2 2
\end{lstlisting}
\paragraph{freq slicing}
The following demonstrates the ease with which one
    can filter along a given axis (here, frequency
    filtering) in pyspecdata, referring to values
    along the axis itself, and returning data along
    that slice.
\begin{lstlisting}[language=python,label=lst:frqSlice,caption={Frequency
        slicing}]
s.ft("t2", shift=True) # Fourier Transform
#                        (shifted to center
#                        about 0 Hz) from the
#                        time to frequency
#                        domain along the
#                        direct dimension.
s = s["t2":(-400, 400)] # slice out the signal
#                         within ±400 Hz of
#                         the carrier
\end{lstlisting}
\paragraph{setting time axis}
The following pySpecData python easily sets the
    time axis of the data to center the echo at $t=0$.
The data instance \texttt{s} is either directly
    supplied by the spectrometer
    (utilizing an in-house library for running the
    SpinCore transceiver)
    or by loading the relevant Bruker data into
    pySpecData.
Here, we assume that \texttt{s} was previously
    moved from the direct time to frequency domain
    (\eg, in order to slice out a relevant range of
    spectral frequencies):
\begin{lstlisting}[language=python,label=lst:TimeSetting,caption={Setting
time axis}]
s.ift("t2")  # move to the time domain
echo_center = hermitian_function_test(
    s
)  # runs the Hermitian function test and
#    returns the center of the echo
s.setaxis(
    "t2", lambda x: x - echo_center
).register_axis(
    {"t2": 0}
)  # applies this best shift to the time axis
\end{lstlisting}
\paragraph{scope acquisition}
Acquisition with the digital oscilloscope requires
    digital demodulation and filtering,
    which typically takes the following form.
(Note how the time domain name for the axis is
    typically used to refer to the axis,
    regardless of whether it is currently in
    the time or frequency domain.)
\begin{lstlisting}[language=python,label=lst:oscope,caption={Oscilloscope-based
    acquisition}]
with GDS_scope() as g:
    pulse = g.acquire_waveform(
        ch=1
    )  # typically, split off and capture the
    #    pulse waveform for reference
    s = g.acquire_waveform(
        ch=2
    )  # signal after duplexer and LNA
    pulse.ft(
        "t", shift=True
    )  # move the pulse waveform into the
    #    frequency domain
    center_frq = abs(
            pulse["t":(0, None)]
            ).argmax("t")  # estimate the
    #                        carrier frequency
    #                        from the max
    s.ft(
        "t", shift=True
    )  # move the signal into the time domain
    s = s[
        "t" : (center_frq + r_[-10e3, 10e3])
    ]  # filter out a 20 kHz bandwidth
    s.ift("t")  # move s back to the time
    #             domain
\end{lstlisting}
\paragraph{chunk indirect, FT ph cycle, vis DCCT}
Storing each phase cycled transient separately
    results, in the simplest case scenario,
    in one long concatenated sequence of complex data
    (1-dimensional array).
One can easily reshape this long
    1-dimensional array into the appropriate
    dimensions of choice,
    an operating referred to in
    pySpecData as `chunk'-ing.
\begin{lstlisting}[language=python,label=lst:chunkExample,caption=Example of chunking data for DCCT]
s.chunk(
    "t", ["ph2", "ph1", "t2"], [2, 4, -1]
)  # here we break data into phase cycle
#    steps → assuming that the leftmost
#    dimension is iterated by the outermost
#    loop of the pulse sequence while the
#    rightmost label is iterated by the
#    innermost loop of the pulse sequence
#    (typically the counter for the direct
#    acquisition datapoints)
s.setaxis(
    "ph1", r_[0:4] / 4
)  # label the phase cycling dimension, in
#    units of cycles -- the first pulse here
#    is cycled in 4 steps
s.setaxis(
    "ph2", r_[0:2] / 4
)  # label the phase cycling dimension, in
#    units of cycles -- the second pulse here
#    is cycled in 2 steps
s.ft(
    ["ph1", "ph2"], unitary=True
)  # Fourier Transform to move from the phase
#    cycling domain to the coherence domain →
#    this example employs a vector-unitary
#    Fourier Transformation (which preserves
#    the vector norm)
s.ft(
    "t2", shift=True
)  # Fourier Transformation from the time to
#    frequency domain along the direct
#    dimension
DCCT(
    s["t2":(-500, 500)]
)  # generate a DCCT plot over the
#    range ±500 Hz
\end{lstlisting}

\bibliographystyle{new}
\bibliography{references}
\end{document}